\documentclass[11pt]{article}
\usepackage{amsmath}
\usepackage{amsthm}
\usepackage{amsfonts}
\usepackage{amssymb}
\usepackage{mathtools}
\usepackage{enumitem}
\usepackage{algorithm}
\usepackage{algcompatible}
\usepackage{latexsym} 
\usepackage{graphicx}
\usepackage{caption}
\usepackage{float}
\usepackage{natbib}
\usepackage[table]{xcolor}
\usepackage{diagbox}
\usepackage{changepage}
\usepackage{listings}
\usepackage{dsfont}
\usepackage{setspace}
\usepackage{subfigure}
\usepackage{multirow}
\usepackage{scalerel}

\usepackage[left=3cm, right=3cm, top=2.5cm, bottom=2.5cm]{geometry}

\usepackage[matrix,tips,graph,curve]{xy}
\newcommand{\indep}{\rotatebox[origin=c]{90}{$\models$}}

\newcommand{\defeq}{\vcentcolon=}

\makeatother
\makeatletter 
\@addtoreset{equation}{section}
\makeatother

\usepackage{hyperref}
\hypersetup{
            colorlinks={true}, 
            linkcolor={blue}, 
            citecolor=blue,
            pdfencoding=auto, 
            psdextra
            }

\usepackage[capitalise,compress]{cleveref} 
\newlist{assumpenum}{enumerate}{1} 
\setlist[assumpenum]{label=(\arabic*), ref=\theassumption~(\arabic*)}
\crefname{assumpenumi}{Assumption}{assumption}

\newlist{lemmaenum}{enumerate}{1}  
\setlist[lemmaenum]{label=(\arabic*), ref=\thelemma~(\arabic*)}
\crefname{lemmaenumi}{Lemma}{lemma}

\newlist{propenum}{enumerate}{1}  
\setlist[propenum]{label=(\arabic*), ref=\theproposition~(\arabic*)}

\usepackage{xr-hyper}

\newtheoremstyle{plain}     
  {3pt}{3pt}{}{}{\bfseries}{.}{ }{}    

\theoremstyle{plain}
\newtheorem{assumption}{Assumption}[section]
\newtheorem{theorem}{Theorem}[section]
\newtheorem{lemma}{Lemma}[section]

\newtheorem{remark}{Remark}[section]

\theoremstyle{definition}
\newtheorem{definition}{Definition}[section]

\newtheorem{example}{Example}

\newcommand{\argmax}{\operatorname{argmax}}
\DeclareMathOperator*{\essinf}{ess\,inf}

\begin{document}

\title{Policy Learning with $\alpha$-Expected Welfare\footnote{We thank Gregory M. Duncan, Sukjin Han, Toru Kitagawa, Alex Luedtke, Hyeonseok Park, Jing Tao, and participants of seminars at University of California San Diego, University of California Irvine, Institute for Advanced Economic Research at Dongbei University of Finance and Economics, Optimal Transport and Distributional Robustness in Banff, and INFORMS Annual Meeting in Seattle for feedback on an earlier version of this paper titled: ``Targeted Policy Learning.'' }}
\author{Yanqin Fan\\
\href{mailto:fany88@uw.edu}{\texttt{fany88@uw.edu}}
\and Yuan Qi \\
\href{mailto:ayqi@amazon.com}{\texttt{ayqi@amazon.com}}
\and Gaoqian Xu \\
\href{mailto:gaoqian-xu@uiowa.edu}{\texttt{gaoqian-xu@uiowa.edu}}}

\date{\today}

\onehalfspacing
\maketitle
\begin{abstract}
    This paper proposes an optimal policy that targets the average welfare of the worst-off $\alpha$-fraction of the post-treatment outcome distribution. We refer to this policy as the $\alpha$-Expected Welfare Maximization ($\alpha$-EWM) rule, where $\alpha \in (0,1]$ denotes the size of the subpopulation of interest. The $\alpha$-EWM rule interpolates between the expected welfare ($\alpha=1$) and the Rawlsian welfare ($\alpha\rightarrow 0$). For $\alpha\in (0,1)$, an $\alpha$-EWM rule can be interpreted as a distributionally robust EWM rule that allows the target population to have a different distribution than the study population. Using the dual formulation of our $\alpha$-expected welfare function, we propose a debiased estimator for the optimal policy and establish its asymptotic upper regret bounds. In addition, we develop asymptotically valid inference for the optimal welfare based on the proposed debiased estimator. We examine the finite sample performance of the debiased estimator and inference via both real and synthetic data. 
    \vspace{12pt}
    

\end{abstract}

\newpage
\section{Introduction}
\label{sec:intro}

\subsection{Motivation}

Targeted/personalized policy rules assign treatments to individuals based on their observable characteristics. 
Learning treatment assignment policies that benefit the relevant population in a desirable way often require careful consideration. The fact that treatment effects tend to vary with individual observable characteristics prompts policy makers to design policies that determine treatment statuses based on individual characteristics. Examples include deciding which patients should receive medical treatment, assigning unemployed workers to training programs, and selecting which students to offer financial aid. Using experimental or observational data from a sample that represents the relevant population, the mean-optimal policy maximizes the sum of individual welfare in the sample. The empirical welfare maximization (EWM) approach in \cite{kitagawa2018a} provides a solution in this regard.

As noted in \cite{kitagawa2021equality}, EWM overlooks distributional impacts. This motivates \cite{kitagawa2021equality} to introduce an equality-minded rank-dependent social welfare function that places greater emphasis on individuals with lower-ranked outcomes. When the policy class is restricted due to considerations such as implementability, cost, and interpretability, EWM may even hurt those who are disadvantaged in the population. For example, if welfare is measured as the (negative) mean blood sugar level of individuals at risk of diabetes and the treatment is a new medication, an EWM policy may prescribe the medication to most individuals because it can substantially benefit the low-risk individuals, who form the majority of the sample, but high-risk individuals who receive the medication may be hurt and end up in even worse situations.\footnote{Throughout this discussion, we follow the standard empirical welfare maximization literature \citep{kitagawa2018a} in treating the observed 
outcome as the welfare-relevant object. This corresponds to assuming that individuals' 
utility is linear in the outcome (risk neutrality).} Similarly, if welfare is evaluated by the average post-training income, a mean-optimal policy is more inclined to select individuals who are high school graduates and have experienced relatively short periods of unemployment to participate in the training program, while overlooking those with lower educational attainment or longer unemployment durations who might also benefit substantially from the training; see Section 5.1 in \cite{athey2021policy}.

Taking a group-agnostic and risk-averse point of view, this paper proposes to learn an optimal policy that favors individuals on the lower tail of the outcome distribution. 
Specifically, for any $\alpha\in(0,1)$, we introduce the $\alpha$-expected welfare function as the expected outcome among the worst-affected $(\alpha\times100)\%$ of the population, i.e., a lower-tail conditional average. We study non-randomized binary policies which maximize the  $\alpha$-expected welfare and refer to such policies as \textit{$\alpha$-expected welfare maximization} ($\alpha$-EWM) policies.
The choice of $\alpha$ is problem-specific and should be based on domain knowledge. A smaller $\alpha$ means that the policy is tailored for the more disadvantaged, whereas a larger $\alpha$ generates a policy that considers a broader less-advantaged subpopulation but those who are most disadvantaged receive less attention. From a philosophical standpoint, when $\alpha$ is small, our $\alpha$-EWM objective aligns with John Rawls' \textit{difference principle}, which aims to maximize the welfare of the least-advantaged group to maintain social stability and fairness \citep{41549aa6-42b1-36d1-ab6a-84fdd10b1f93}. Indeed, the $\alpha$-expected welfare converges to the essential infimum of the outcome random variable as $\alpha$ approaches zero. We note that the definition of the $\alpha$-expected welfare function also applies to $\alpha = 1$, in which case it reduces to  the standard expected welfare underlying the empirical welfare maximization studied in \cite{kitagawa2018a} and \cite{athey2021policy}. Consequently, we use $1$-EWM and EWM interchangeably throughout the rest of this paper.

To further motivate our $\alpha$-EWM for $\alpha \in (0,1)$, we provide a simple numerical comparison with the $1$-EWM criterion from \cite{kitagawa2018a}, the equality-minded welfare criterion from \cite{kitagawa2021equality}, and quantile maximization from \cite{wang2018quantile}. Section \ref{section:relation} discusses the relationship between our $\alpha$-EWM and these criteria in more detail. We use a simple data generating process (DGP) similar to the motivating example in \cite{wang2018quantile}:
\begin{align}
\label{Spec: illustrative}
    Y=20+3 A+X-5 A X+\left(1+A+2 A X\right) \epsilon,
\end{align}
where the covariate $X \sim \operatorname{Unif}[0,1] $, the binary treatment $A\sim\operatorname{Bernoulli}(0.5)$, and $\epsilon \sim N(0,1)$. We assume that the propensity score $e_o(\cdot)=0.5$ is known, and the policy class is defined as $\Pi_{\mathrm{c}} = \mathds{1} \{ X \leq c \}$ for the policy parameter $c \in [0,1]$.                                                                                                      

\begin{figure}[H]
    \centering
    \subfigure[Different welfare types]{
        \includegraphics[width=0.485\textwidth]{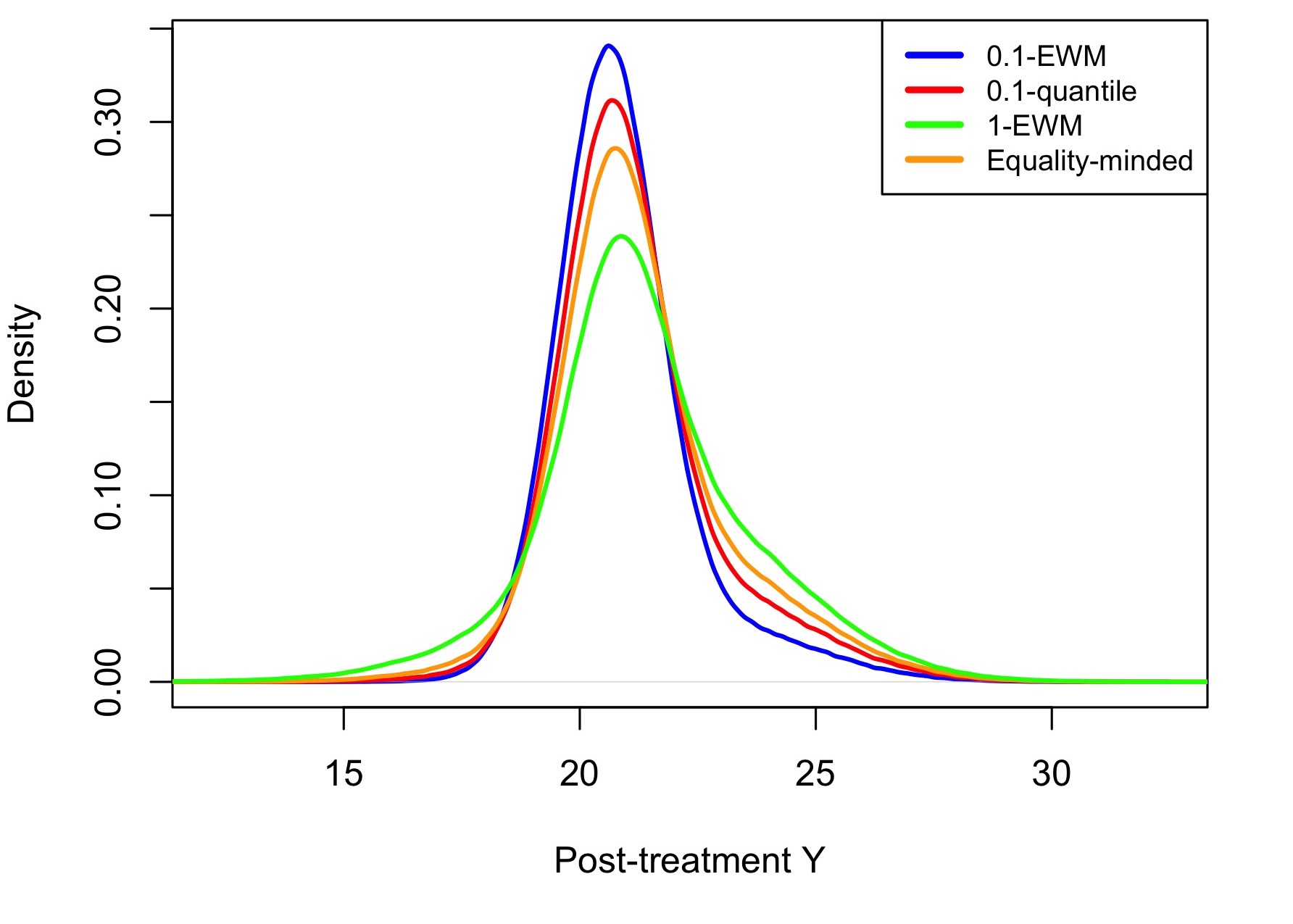}}
    \subfigure[EWM with different $\alpha$ targets]{
        \includegraphics[width=0.485\textwidth]{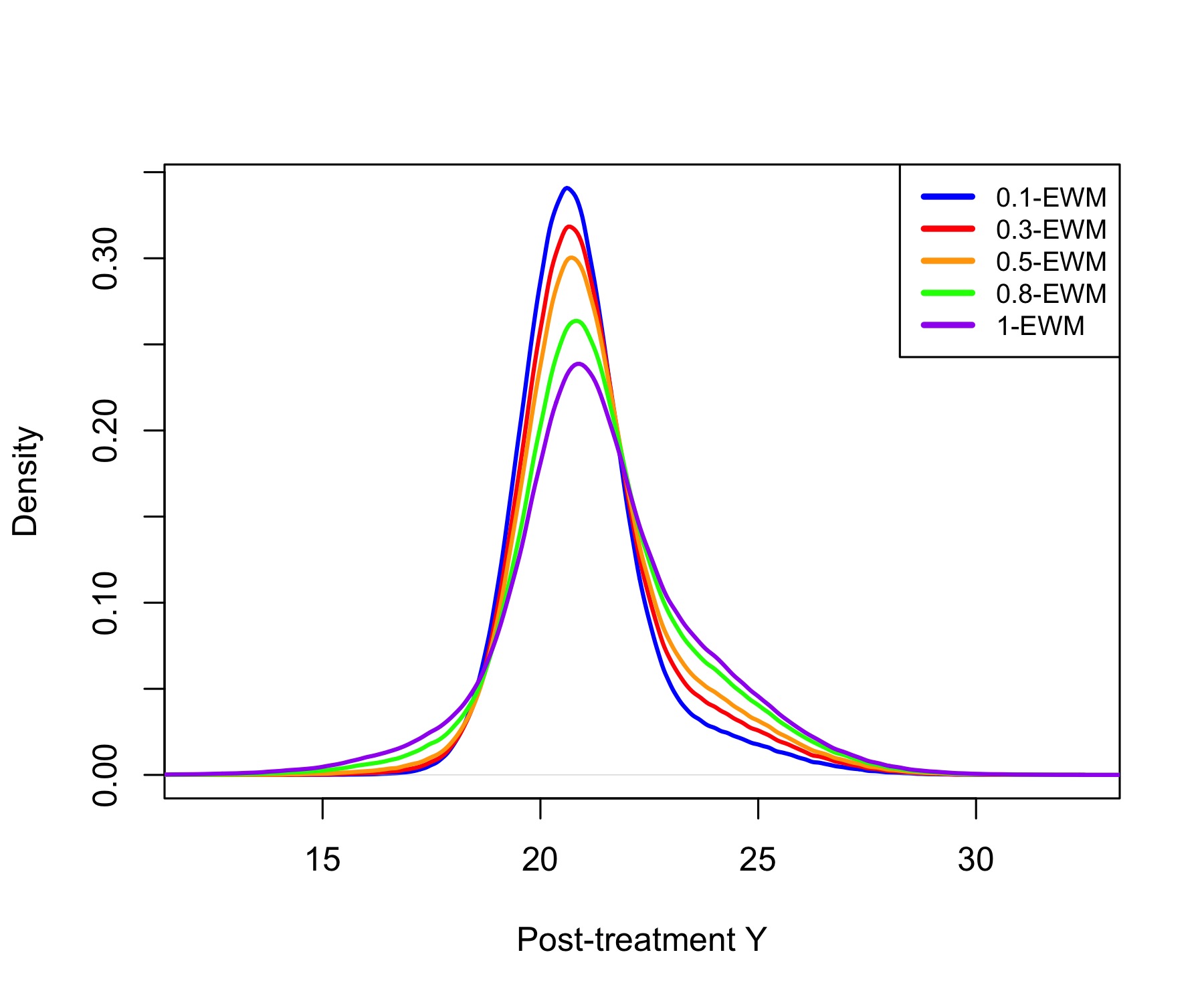}}
    \caption{Distributions of post-treatment outcomes induced by the optimal policies under different welfare criteria.}
   \label{Figure: illustrative density}
\end{figure}

We create a superpopulation of size one million. Since we can generate $Y_i$ for both $A_i = 0$ and $A_i = 1$, we have full knowledge of the true outcome distribution induced by any policy $c$. For comparison, we select values of $c$ that maximize the following: the 0.1-expected welfare, the standard Gini social welfare, the 0.1-outcome quantile, and the mean outcome. These correspond to the 0.1-EWM, equality-minded, 0.1-quantile-optimal, and 1-EWM policies, respectively. Figure \ref{Figure: illustrative density}(a) displays the resulting post-treatment outcome densities. We observe a gradual tightening of the outcome distribution as we move from the 1-EWM policy to the equality-minded policy, then to the 0.1-quantile-optimal policy, and finally to the 0.1-EWM policy, which yields the most concentrated distribution with the thinnest tails on both sides.

To further illustrate this pattern within the EWM family, Figure \ref{Figure: illustrative density}(b) shows the distributions for the EWM policies under different $\alpha$ targets. As $\alpha$ decreases, the distributions become increasingly concentrated, with less probability mass in both tails. This pattern reflects that targeting smaller lower-tail subpopulations reduces the probability of very poor outcomes, and it also reduces the occurrence of extremely large outcomes. Taken together, the two panels illustrate that welfare criteria emphasizing the lower tail, including EWM with smaller $\alpha$, generate outcome distributions that are more equal and less dispersed.

\subsection{Main Contributions}
This paper makes several contributions to the literature on policy learning. First, 
under the assumption of unconfoundedness,\footnote{The assumption of unconfoundedness is not essential and can be replaced with any assumption that identifies the conditional marginal distributions of the potential outcomes.} we show that the $\alpha$-expected welfare function is identified and propose a debiased estimator. Our debiased estimator utilizes cross-fitted nuisance estimators and the orthogonal moment function based on the dual form of the $\alpha$-expected welfare function. Optimizing the $\alpha$-expected welfare poses noticeable challenges compared with $1$-EWM. Adopting a group-agnostic perspective, the worst-off subpopulation being targeted changes dynamically with different policies. Consequently, estimating the $\alpha$-expected welfare requires the estimation of the $\alpha$-quantile of the welfare, which serves as a ``cutoff" for computing the tail average (see \cref{section: Average Value-at-Risk Welfare Function} for details). 

Second, we establish theoretical guarantees of our $\alpha$-EWM for any $\alpha\in (0,1)$ by deriving asymptotic upper regret bounds with an explicit expression for the constant. This complements similar regret bounds for $1$-EWM in \cite{kitagawa2018a} and \cite{athey2021policy}.

Third, we develop asymptotically valid inference for the optimal $\alpha$-expected welfare. When the optimal policy is unique, Wald-type inference is asymptotically valid. When the optimal policy is not unique, we develop inference by applying
the numerical delta method introduced by \cite{hong2018numerical}.

Fourth, we demonstrate that more comprehensive policy evaluations can be performed by consistently estimating the welfare of the worst-off $(\alpha\times100)\%$ of the population for \text{any} $\alpha\in(0,1)$ and policy. Put differently, even if a policy does not specifically target the worst-affected $(\alpha\times100)\%$, we can still assess its performance at $\alpha$ to gain insights into the associated trade-offs. We illustrate our $\alpha$-EWM method using experimental data from the National Job Training Partnership Act (JTPA) Study, as analyzed by \cite{bloom1997benefits}. We find that targeting smaller subpopulations—such as the bottom 25\% or 30\% of the outcome distribution—leads to more robust welfare performance across a range of welfare objectives. In contrast, targeting broader groups (e.g., the bottom 80\% or the expected welfare) can result in substantial welfare losses for the bottom 25\%, indicating that policies aimed at broader groups may come at the expense of welfare among the most disadvantaged.

Lastly, we conduct simulation studies based on synthetic JTPA data generated using Wasserstein Generative Adversarial Networks (WGANs) developed by \cite{athey2024using}, to evaluate the performance of our estimator and compare policy outcomes. In the WGAN-JTPA setup, both the $0.25$-EWM and equality-minded policies enhance the welfare of lower-ranked individuals while reducing that of higher-ranked individuals relative to the $1$-EWM policy, with the $0.25$-EWM policy placing much greater emphasis on these adjustments. Additional simulation studies based on stylized DGPs from \cite{athey2021policy} are provided in Section H.3 of the Online Supplement \citep{fan2025policy_supplement}. Across all simulation setups, the debiased estimator and Wald inference perform satisfactorily for all $\alpha$ values considered.

The rest of the paper is organized as follows. Section \ref{sec:literature} provides an overview of the related literature. Section \ref{section: Average Value-at-Risk Welfare Function} introduces our model preliminaries, including the $\alpha$-expected welfare measure and its identification under the selection-on-observables assumption. We also point out relations and differences between four welfare measures: the 1-expected welfare, equality-minded welfare, quantile welfare, and our $\alpha$-expected welfare. Section \ref{Section: debiased} reviews the dual form of the $\alpha$-expected welfare function and presents its debiased estimator, and \cref{Section: theory} establishes an asymptotic upper regret bound for our debiased optimal policy. \cref{section:inference} constructs asymptotically valid inference for the optimal $\alpha$-expected welfare. \cref{Section: empirical} presents numerical results, including an empirical application based on experimental data from the JTPA Study and a simulation study using WGAN-generated JTPA data. Section \ref{Section: conclusion} concludes. In the Appendix, Section \ref{section: First_best_policy} establishes the first-best $\alpha$-EWM policy. Section \ref{section: improved rate} derives an improved rate for the regret bound under the margin assumption. Section \ref{section: inference for the optimal value} presents a uniform inference procedure. An online supplement contains technical proofs and additional numerical results. 

\subsection{Related Literature}
\label{sec:literature}

Our work builds on existing literature on policy learning from experimental and observational data, as well as statistical inference for the mean outcome under the optimal policy. In the following, we provide a brief discussion of related work. 

\paragraph{Mean-optimal Policy Learning}
Existing research on policy learning in economics and statistics has mainly focused on mean-optimal policy under unconfoundedness \citep{qian2011performance,zhao2012estimating,zhang2012estimating,bhattacharya2012inferring,luedtke2016statistical,kallus2018balanced,luedtke2020performance,athey2021policy}. Most work on policy learning focuses on establishing theoretical guarantees by deriving regret bounds. The  paper by \cite{kitagawa2018a} explores the mean-optimal policy learning from experimental data in a nonparametric framework. When propensity scores are known and the policy class  has a finite VC dimension, they employ inverse propensity weighting to estimate the welfare function, achieving $n^{-1/2}$-rate regret bounds, where $n$ is the sample size. 
\cite{athey2021policy} extend this setup to observational studies where propensity scores are unknown and the policy class $\Pi_n$ may vary with $n$. They estimate the objective function using doubly robust scores, a method that is shown to be efficient in the sense of \cite{newey1994asymptotic}. The resulting policies achieve regret bounds of the order $\sqrt{\mathrm{VC}(\Pi_n)/n}$. Notably, their regret bound depends on the convergence rate of nuisance parameter estimation and the semiparametric efficient variance for evaluating an optimal policy. Finally, under mild conditions, \cite{luedtke2020performance} show that regret can decay faster than $n^{-1/2}$ for a fixed data distribution. 

Several studies have examined statistical inference for the mean-optimal welfare associated with the first-best policies. For example, \cite{luedtke2016statistical} propose an online one-step estimator that is $\sqrt{n}$-consistent for the optimal value function, where the estimated policy and value function are recursively updated using new observations. Similarly, \cite{shi2020breaking} conduct inference for optimal welfare through subsample aggregation and cross-validation. In contrast, \cite{rai2018statistical} studies the inference of optimal mean welfare under a restricted policy class. The author utilizes bootstrap and numerical delta methods, e.g. \cite{hong2018numerical}, to approximate the limiting distribution of the estimator. We apply the same set of tools to develop inference for the optimal $\alpha$-expected welfare associated with a pre-specified policy class when the optimal policy may not be unique.

\paragraph{Fairness and Robustness of Policy Learning.}
In many applications, objectives beyond the mean are used to incorporate fairness and distributional concerns. \cite{kitagawa2021equality} study rank-dependent social welfare functions that place greater weight on lower-ranked outcomes, and \cite{wang2018quantile} consider policies that maximize a target quantile of the induced outcome distribution. A complementary approach is to learn mean-optimal policies subject to explicit fairness constraints; see, e.g., \cite{fang2023fairness,viviano2024fair,kim2023fair}.

A related strand studies robustness and external validity by optimizing welfare under the least favorable distribution in an ambiguity set, thereby delivering performance guarantees under distribution shift. Examples include Wasserstein-based uncertainty sets in \cite{adjaho2022externally,kido2022distributionally,fan2023quantifying} and KL-divergence-based sets in \cite{si2023distributionally}. Robustness has also been studied with respect to unobserved confounding or selection \citep{kallus2021minimax,lei2023policy}, and to strategic or equilibrium uncertainty in allocation problems \citep{wang2024robust}. As discussed in Section \ref{Section: relation EWM}, our $\alpha$-expected welfare admits a distributionally robust representation and thus provides a robustness check for mean-based EWM in the presence of distribution shift.

The paper closest to ours is \cite{qi2023robustness}, which uses the average value-at-risk (AVaR) criterion to learn robust individualized decision rules. Since AVaR coincides with our $\alpha$-expected welfare (\cref{equation: AVaR as DRO}), the key differences are methodological and inferential. Relative to \cite{qi2023robustness}, we allow unknown propensity scores and estimate the objective via doubly robust, cross-fitted scores; we study general (possibly $n$-dependent) VC-type policy classes $\Pi_n$; we provide explicit upper regret bounds; and we develop asymptotically valid inference for the optimal welfare value under both experimental and observational designs. On the computational side, \cite{qi2023robustness} rely on a non-convex surrogate and difference-of-convex optimization, whereas we use derivative-free optimization for the original policy objective. 

We close this section by summarizing the notation used in this paper.
We use $O, o, O_{P}, o_{P}, \asymp,  \gtrsim, \lesssim$ in the following sense: $a_n=O\left(b_n\right)$ if $\left|a_n\right| \leq$ $C b_n$ for $n$ large enough; $a_n = o(b_n)$ if $a_n/b_n \rightarrow 0$; $X_n=O_{P}\left(b_n\right)$, if for any $\delta>0$, there exist $M, N>0$, such that $\mathbb{P}\left|\left|X_n\right| \geq\right.$ $\left.M b_n\right] \leq \delta$ for any $n>N ; X_n=o_{P}\left(b_n\right)$, if $\mathbb{P}\left[\left|X_n\right| \geq \epsilon b_n\right] \rightarrow 0$ for any $\epsilon>0 ; a_n \asymp b_n$ if there exist $k_1, k_2>0$ and $n_0$, such that for all $n>n_0, k_1 a_n \leq b_n \leq k_2 a_n$;  $a_n \gtrsim b_n$ if $b_n=O\left(a_n\right) ; a_n \lesssim b_n$ if $a_n=O\left(b_n\right)$. Furthermore, we write $f(n) = \widetilde{O}(g(n))$ if there is a function $h$ that grows poly-logarithmically such that $f(n) \leq h(g(n))g(n)$. The notation $f(n) = \Omega(g(n))$ means that there is a universal constant $c_o >0$ such that $f(n) \geq  c_og(n)$ uniformly in $n$. We use the shorthand $[n] = \{1, \ldots, n\}$, $a \vee b = \max\{a,b\}$ and $a \wedge b = \min\{a,b\}$.
The abbreviation i.i.d. stands for  independent and identically distributed. In the sequel, $c_o$ and $C_o$ denote generic positive constants, whose values may vary from line to line.

\section{$\alpha$-Expected Welfare Function and Optimal Policy}
\label{section: Average Value-at-Risk Welfare Function}

Suppose that we have a random sample $\left(X_i, Y_i, A_i \right)_{i=1}^n$, where $X_i\in\mathcal{X} \subseteq \mathbb{R}^p$ denotes the observable characteristics of individual $i$ (continuous or discrete), $Y_i\in\mathcal{Y}\subseteq\mathbb{R}$ represents the outcome of individual $i$ (or utility/welfare), and $A_i\in\{0,1\}$ denotes the treatment status of individual $i$, for $i\in [n]$. Without loss of generality, larger values of $Y_i$ are assumed to be preferable. To simplify notation, we define $Z_i\defeq(X_i,Y_i,A_i)\in\mathcal{Z}$ and $\mathcal{Z}=\mathcal{X}\times \mathcal{Y}\times \{0,1\}$.  Let $Y_{i}(0)$ and $Y_{i}(1)$ denote the potential outcomes that would have been observed if $A_i=0$ and $A_i=1$, respectively. Then $Y_i=A_iY_{i}(1)+(1-A_i)Y_{i}(0)$ is the realized outcome under the Stable Unit Treatment Value Assumption \citep{rubin1978bayesian, rubin1990comment}. Further, we denote by $P$ the
distribution of $Z_i$, and by $\mathbb{E}_P$ and $\mathrm{Var}_P$ the expectation and variance under $P$, respectively.

\subsection{\( \alpha \)-Expected Welfare Function and Identification}

We study non-randomized binary policy/rule $\pi: \mathcal{X} \rightarrow \{0,1\}$. Throughout this paper, we let $\Pi_o$ denote the policy class that contains all Borel measurable functions from $\mathcal{X}$ to $\{0,1\}$ and $\Pi\subseteq \Pi_o$ denote a pre-specified policy class.
As discussed in \cite{kitagawa2018a} and \cite{athey2021policy}, practitioners  may adopt a pre-specified policy class $\Pi\subseteq \Pi_o$ that incorporates constraints relevant to the problem context, such as budgetary limitations, specific functional forms, fairness considerations, and other pertinent factors. The policy class $\Pi$ may depend on the sample size $n$ and if needed,  we write it explicitly as $\Pi_n$.

For any policy $\pi\in \Pi_o$, let $Y_i(\pi):=Y_i(\pi(X_i) )$,  the outcome of individual $i$ when $\pi$ is implemented. Further, let $F_{\pi}(y)$, $y\in \mathcal {Y}$, denote the distribution function of $Y_i(\pi)$ and
$F^{-1}_\pi(\alpha):=\inf \left\{y \in\mathbb{R}:F_\pi(y)\geq\alpha \right\}$, $\alpha\in (0,1]$, denote the quantile function of $Y_i(\pi)$.

\begin{definition}[$\alpha$-Expected Welfare and Optimal Policy]\label{definition: Expected Welfare}
    Given a policy class $\Pi$ chosen by the policymaker, we define   the $\alpha$-expected welfare of $Y_i(\pi)$ as the expected welfare of the worst-off subpopulation of size $\alpha\in (0,1]$, i.e.,
\begin{equation}\label{equation: Expected Welfare given policy}
\mathbb{W}_\alpha(\pi):=\frac{1}{\alpha}\int_0^{\alpha} F_\pi^{-1}(t) \mathrm{d} t \quad  \mbox{ for } \pi \in \Pi.
\end{equation}
An $\alpha$-expected welfare maximization ($\alpha$-EWM) policy is defined as
\begin{equation}\label{equation:alpha-EWM}
\pi_\alpha^*\in\argmax_{ \pi \in \Pi  }\mathbb{W}_\alpha(\pi).
\end{equation}
\end{definition}

As discussed in \cref{sec:intro}, $\lim_{\alpha \rightarrow 0}\mathbb{W}_\alpha(\pi)=\essinf Y_i(\pi) $ and $\mathbb{W}_1(\pi)=\mathbb{E} \left[Y_i(\pi) \right]$. Our welfare function $\mathbb{W}_\alpha(\pi)$ therefore flexibly interpolates between the expected welfare and infimum welfare of the  target population by varying $\alpha\in (0,1]$, where $\alpha=1$ gives the expected welfare of the target population adopted in \cite{kitagawa2018a} and \cite{athey2021policy}.

\begin{remark}
(i) For a given $\pi$, $\mathbb{W}_\alpha(\pi)$ is the  Expected Shortfall of $Y_i(\pi)$, a commonly used coherent risk measure in finance and risk management. When the distribution function of $Y_i(\pi)$ is continuous at $F_{\pi}^{-1}(\alpha)$, $\mathbb{W}_\alpha(\pi)$ is also the same as the Conditional Value at Risk (CVaR) of $Y_i(\pi)$, defined as $\mathrm{CVaR}_\alpha(\pi):=\mathbb{E}\left[Y_i(\pi)\mid Y_i(\pi) \leq F_{\pi}^{-1}(\alpha) \right]$,  see \cite{rockafellar2000optimization,shapiro2021lectures}.

(ii)  For a given $\pi$, 
$\mathbb{W}_\alpha(\pi)$ is also closely related to the generalized Lorenz function, a popular tool for measuring and comparing inequality, see \cite{greselin2018classical} and \cite{shorrocks1983ranking}. Specifically, 
let $L_\alpha^{\mathrm{gen}}(Y_i(\pi))$ denote the generalized (unnormalized) Lorenz function at level $\alpha$: $L_\alpha^{\mathrm{gen}}(Y_i(\pi))\defeq\int_0^{\alpha} F_\pi^{-1}(t)dt$. Then 
$\mathbb{W}_{\alpha}(\pi)=\frac{1}{\alpha }L_\alpha^{\mathrm{gen}}(Y_i(\pi))$. 
\end{remark}

To identify $\mathbb{W}_\alpha(\pi)$ as defined in \cref{equation: Expected Welfare given policy}, we note that $$Y_i(\pi)=\pi(X_i)Y_i(1)+[1-\pi(X_i)]Y_i(0).$$ The conditional (given $X_i = x$) and unconditional distribution functions of $Y_i(\pi)$ are
\begin{equation} \label{equation: identification of cdf_pi}
F_{\pi}( y | x) = \pi(x) F_1(y | x)+(1-\pi(x)) F_0(y | x) \ \mbox{ and } \ F_{\pi}(y) = \int_{\mathcal{X}}  F_{\pi}( y | x)  \mathrm{d} P_X(x),
\end{equation}
where $F_1(y |x)$ and $F_0(y|x)$ are the conditional distribution functions of $Y_i(1)$ and $Y_i(0)$ given $X_i=x$, respectively. 

\cref{equation: Expected Welfare given policy} and \cref{equation: identification of cdf_pi} imply that $\mathbb{W}_\alpha(\pi)$ is a function of policy $\pi(\cdot)$ and conditional distribution functions  $F_1( \cdot | \cdot)$ and $F_0( \cdot| \cdot)$. Consequently, for any $\pi \in \Pi_o$, $\mathbb{W}_\alpha(\pi)$ is identified as long as  $F_1(  \cdot| \cdot)$ and $F_0( \cdot| \cdot)$ are identified. Hence, any assumption that ensures the identification of $F_{1}(\cdot|\cdot)$ and $F_{0}(\cdot|\cdot)$ also identifies $\mathbb{W}_\alpha(\pi)$.  

In the rest of this paper, we adopt the assumption of selection-on-observables, which includes unconfoundedness and common support, as detailed in Assumption \ref{Assumption: Selection-on-observables}.

\begin{assumption} \label{Assumption: Selection-on-observables}
Assume that  $\mathbb{E}\left|Y_i(a)\right| < \infty$ for $a \in \{0,1\}$, and that the following conditions hold:
\begin{assumpenum}
\item {\it Unconfoundedness:} \label{assumption: Unconfoundedness}$\left(Y_i(0), Y_i(1) \right)\indep A_i \mid X_i$.  
\item \label{assumption: Strict Overlap}{\it Strong overlap}: Let  $e_o(x) := \mathbb{P}\left[A_i =1 \mid X_i=x \right]$ denote the propensity score. There is a constant $\kappa \in \left(0,\frac{1}{2} \right)$ such that $e_o(x)\in [\kappa, 1- \kappa]$ for all $x \in \mathcal{X}$.
\end{assumpenum}
\end{assumption}

Assumption \ref{assumption: Unconfoundedness} states that the potential outcomes are independent of the treatments after conditioning on the observed covariates. Heuristically, it requires that all confounders that affect both treatments and potential outcomes simultaneously be observed. For identification, Assumption \ref{assumption: Strict Overlap} can be relaxed to the weaker condition that $ e_o(x) \in (0, 1)$ for all $x \in \mathcal{X}$, but the regret bounds and inference developed in later sections of this paper rely on it. 

Under Assumption \ref{Assumption: Selection-on-observables}, the distribution functions $F_{a}(\cdot|x)$, $a\in\{0,1\}$, for all  $x \in \mathcal{X}$ are point-identified: 
\[
F_{a}(y|x)\defeq\mathbb{P}\left[Y_i(a)\leq y|X_i=x\right]=\mathbb{P}\left[Y_i\leq y|X_i=x, A_i=a\right].
\]
Consequently, $\mathbb{W}_\alpha(\pi)$ is identified  for any $\pi \in \Pi_o$.

\subsection{Relations with Other Welfare Maximization Criteria}
\label{section:relation}
In this subsection, we compare our $\alpha$-expected welfare $\mathbb{W}_\alpha(\pi)$, for $\alpha \in (0,1)$, with three welfare functions commonly used in the literature: the expected welfare, the equality-minded welfare, and the quantile welfare functions.

\subsubsection{Expected Welfare Maximization}
\label{Section: relation EWM}

EWM in \cite{kitagawa2018a} and \cite{athey2021policy} coincides with $1$-EWM with the mean outcome \(\mathbb{E}[Y_i(\pi)]\) as the population welfare function. Let $F_{\pi}$ denote the distribution of the outcome under policy $\pi: \mathcal{X} \rightarrow \{0,1\}$ in the study population. \citet{kitagawa2018a} and \citet{athey2021policy} assume 
that the study population (from which the data $(X_i, A_i, Y_i)$ is 
collected) and the target population (to which policy $\pi$ will be 
applied) share the same marginal distribution $F_\pi$ of the potential 
outcome $Y_i(\pi)$.

For $\alpha\in (0,1)$, our $\alpha$-expected welfare $\mathbb{W}_\alpha(\pi) $  represents a distributionally robust version of the $1$-expected welfare function. To see this, consider the uncertainty set centered at probability distribution $F_{\pi}$ of the outcome under policy $\pi: \mathcal{X} \rightarrow \{0,1\}$:
\[
\begin{aligned}
\mathcal{U}_{\alpha} (F_\pi) & = \left\{Q: D_\infty(Q \| F_\pi ) \leq  \log \frac{1}{\alpha}  \right\},
\end{aligned}
\]
where $D_\infty ( Q\| F_\pi) = \mathrm{ess \ sup} \log \frac{d Q}{d F_\pi}$. In other words, $\mathcal{U}_{\alpha} (F_\pi)$ is a $D_\infty$-divergence ball with radius $\log \alpha^{-1}$. As $\alpha$ decreases, this ball expands, corresponding to a greater degree of distributional robustness. From \cite{ rockafellar2002deviation} and \cite{duchi2023distributionally}, it follows that 
\begin{equation} \label{equation: AVaR as DRO}
\mathbb{W}_\alpha(\pi) = \inf_{Q \in \mathcal{U}_{\alpha}(F_\pi ) } \mathbb{E}_{Z\sim Q} \left[  Z\right]. 
\end{equation}

The uncertainty set $\mathcal{U}_\alpha(F_\pi)$ has an alternative representation:
\[
\begin{aligned}
\mathcal{U}_{\alpha} (F_\pi) 
& = \left\{Q  : \exists \ P \in \mathcal{P}(\mathcal{Y}) , t \in[\alpha, 1] \text { s.t. } F_\pi  = t Q+(1-t) P\right\}.
\end{aligned}
\]
Consequently, it includes all distributions $Q$ that could appear as a component of $F_\pi$ in the study population with weight at least $\alpha$ in a mixture. Taking the infimum over $Q \in \mathcal{U}_\alpha(F_\pi)$ gives the minimum expected welfare across all such sub-populations, so $\pi^*_\alpha$ maximizes the worst-case expected welfare over perturbations of $F_\pi$ within $\mathcal{U}_\alpha(F_\pi)$. As $\alpha$ decreases, $\mathcal{U}_\alpha(F_\pi)$ expands, yielding stronger protection against adverse shifts in the outcome distribution.

\subsubsection{Equality-Minded Welfare Maximization}
\label{Section: relation EM}

\cite{kitagawa2021equality} propose equality-minded policies by maximizing rank-dependent social welfare functions (SWFs), which assign greater weights to lower-ranked individuals.\footnote{Following \cite{kitagawa2021equality}, we restrict attention in this subsection to \(Y_i(0)\geq0\) and \(Y_i(1)\geq0\) almost surely.}  Given a decreasing function $\Lambda: [0,1] \rightarrow [0,1]$ with $\Lambda(0) = 1$ and $\Lambda(1) = 0$, the equality-minded welfare under policy $\pi$ is defined as
\begin{equation}
W_\Lambda(F_\pi)\defeq\int_0^\infty \Lambda \left( F_\pi(y) \right) \mathrm{d}y=\int_0^1  F^{-1}_\pi \left(t\right) \omega(t)\mathrm{d}t,
\end{equation}
where $\omega(t) \coloneqq -\frac{\mathrm{d}}{\mathrm{d}t} \Lambda(t)$ is the associated  weight function.
When $\Lambda$ is strictly convex, the associated SWF, $W_\Lambda$, upholds the Pigou-Dalton Principle of Transfers, as rank-preserving transfers from higher-ranked individuals to lower-ranked individuals are preferred under the welfare $W_\Lambda$.  The function $\Lambda$, chosen by practitioners, captures the degree of inequality aversion through its level of complexity.
An important class of rank-dependent SWFs is the extended Gini SWFs, where $\Lambda (t) =\Lambda_k(t) = (1-t)^{k-1}$ for some $k \geq 2$, and the weight function is $\omega(t)=\omega_k(t) = (k-1)(1-t)^{k-2}$. 
The expected welfare and the standard Gini SWFs correspond to $k = 2$ and $k = 3$, respectively. 

Extended Gini SWFs can, in fact, be expressed in terms of our $\alpha$-expected welfare $\mathbb{W}_\alpha(\pi)$. For example, when $k > 2$,  
\begin{equation}\label{equation: generalized Gini welfare}
W_\Lambda(F_\pi) = \left(k-2\right) \left(k-1\right) \int_0^1 \mathbb{W}_{\alpha}(\pi)
\alpha (1-\alpha)^{k-3}\mathrm{d}\alpha.     
\end{equation}
On the other hand, our $\alpha$-expected welfare can be written as
\begin{equation}\label{equation: weighted}
\mathbb{W}_\alpha(\pi)= \frac{1}{\alpha} \int_0^\alpha F_\pi^{-1}(t) \mathrm{d}t = \int_0^1  F^{-1}_\pi \left(t\right) \omega_{\alpha}(t)\mathrm{d}t,
\end{equation}
where $\omega_{\alpha}(t) = \frac{1}{\alpha} \mathds{1} \{ 0 \leq  t \leq \alpha \}$ with the associated $\Lambda_{\alpha}(t)  =  \left(1 - t/\alpha\right) \mathds{1}\{ 0 \leq t \leq \alpha \}$. However, for any $\alpha\in(0,1)$, $\Lambda_{\alpha}(t)$ is not strictly convex. Consequently, our $\alpha$-expected welfare $\mathbb{W}_\alpha(\pi)$ does not belong to the equality-minded class of \cite{kitagawa2021equality} and does not satisfy the Pigou-Dalton Principle of Transfers.
This principle is satisfied only if the rank-preserving transfer happens across the probability level $\alpha$, i.e., from an individual ranked above $\alpha$ to an individual ranked below $\alpha$. Transfers on the same side do not affect $ \mathrm{W}_\alpha(\pi)$ since all the individuals involved have the same weight. 

\subsubsection{Quantile Welfare Maximization}
\label{Section: relation QW}
To prioritize the lower tail of population welfare over the (weighted) expected welfare, \cite{wang2018quantile} propose a quantile-optimal policy, defined as
\[
\argmax_{\pi \in \Pi} \mathrm{VaR}_\alpha(Y_i(\pi)) = F_\pi^{-1}(\alpha) ,
\]
where $\alpha \in (0,1)$ is the quantile level of interest. For the class of linear policies with a fixed number of covariates $\Pi$,  \cite{wang2018quantile} establish the cube root asymptotics for the estimator of the parameter that defines the optimal linear policy. 

Compared with quantile welfare $F_\pi^{-1}(\alpha)$ that overlooks the welfare of the population with outcomes below it, our $\alpha$-expected welfare function $ \mathbb{W}_\alpha(\pi)$ integrates $F_\pi^{-1}(t)$ over the range $[0, \alpha]$, thereby accounting for welfare levels below the $\alpha$-quantile and providing a more comprehensive assessment of the lower tail of the welfare distribution.

\section{Debiased Estimation and Practical Implementation}
\label{Section: debiased}

The $\alpha$-expected welfare function $\mathbb{W}_\alpha(\pi)$ has an equivalent representation, which we will use to construct a debiased estimator of $\mathbb{W}_\alpha(\pi)$.

\subsection{An Equivalent Representation of $\mathbb{W}_\alpha(\pi)$}

Let $(u)_-\defeq\min{(u, 0)}$ and  $(u)_+ \defeq \max{(u, 0)}  $. Further, let $\theta = (\pi , \eta)$ and
$$
\mathbb{V}_\alpha(\theta) =  \frac{1}{\alpha} \mathbb{E}\left[\left(Y_i(\pi)-\eta\right)_{-}\right]+\eta.
$$

\begin{lemma}[Dual Representation of $\mathbb{W}_\alpha(\pi)$]
\label{lemma: AVaR dual}
For any $\alpha \in (0,1]$ and $\pi \in \Pi$,
\[
\label{dual}
\mathbb{W}_\alpha(\pi)=\sup_{\eta \in \mathbb{R} }\mathbb{V}_\alpha(\pi, \eta).
\]
Furthermore, for $\alpha \in (0, 1)$, the supremum is attained on the interval $[t^\star, t^{\star\star}]$, where $t^\star = \sup\{ y\in \mathbb{R}: F_\pi (y) \leq \alpha \}$ and $t^{\star \star} =F_\pi^{-1}(\alpha)$. When $\alpha = 1$, if the support of $Y_i(\pi)$ is bounded, then the supremum is attained on $\left[ F_\pi^{-1}(1), \infty \right)$. Otherwise, the supremum is unattainable and $\sup_{\eta \in \mathbb{R} }\mathbb{V}_1(\pi, \eta) = \lim_{\eta \rightarrow \infty} \mathbb{V}_1(\pi, \eta)$.
\end{lemma}

Lemma \ref{lemma: AVaR dual} extends the dual representation established in \cite{rockafellar2000optimization} for $\alpha \in (0,1)$; see also Theorem 6.2 of \cite{shapiro2021lectures}, to $(0,1]$ so it covers the boundary case $\alpha=1$ as well. This extension is crucial as it allows us to unify the mean-based and tail-based objectives within a single framework. 

Let
\[\mu_a(x, \eta)\defeq \mathbb{E}\left[\left(Y_i(a)-\eta  \right)_- |X_i = x\right] \mbox{ for } a\in \{0,1\},\] 
and $\tau(x, \eta) \defeq\mu_1(x, \eta)- \mu_0(x, \eta)$ for any $x \in \mathcal{X}$ and $\eta \in \mathbb{R}$. Under Assumption \ref{Assumption: Selection-on-observables}, $\tau(x, \eta)$ is identified for any given $\eta$.

\begin{theorem}\label{theorem: Value function identification}
Under Assumption \ref{Assumption: Selection-on-observables}, for any $0 <\alpha \leq 1$ and any $\theta = (\pi, \eta) \in \Pi_o \times \mathbb{R}$, it holds that 
\begin{equation}\label{equation: V_identification}
    \begin{aligned}
\mathbb{V}_\alpha(\theta) 
& = \frac{1}{\alpha}\left\{  \mathbb{E} \left[  \pi (X_i) \mu_1(X_i, \eta)\right] + \mathbb{E} \left[ \left(1-\pi(X_i ) \right) \mu_0(X_i, \eta) \right]\right\}  +\eta, \\
& = \frac{1}{\alpha}\left\{  \mathbb{E} \left[  \pi (X_i) \tau(X_i, \eta)\right] + \mathbb{E} \left[ \mu_0(X_i, \eta) \right]\right\}  +\eta, \\ 
& = \frac{1}{\alpha} \mathbb{E} \left[  w(X_i, A_i ,\pi) (Y_i - \eta)_{-} \right] + \eta, \\
\end{aligned}
\end{equation}
where the function $w: \mathcal{X} \times \{0,1\} \times \Pi_o \rightarrow [0, \infty)$ is defined as
\[
 w(x, a ,\pi)\defeq 
\frac{a \pi (x)}{e_o(x)}  + \frac{ (1- a) \left(1- \pi(x) \right)}{1- e_o(x)}.
\]
\end{theorem}

Lemma \ref{lemma: AVaR dual} and Theorem \ref{theorem: Value function identification} provide a theoretical foundation for our debiased estimation procedure. 

\begin{remark}\label{remark: compact feasible set}
(i) When $0<\alpha<1$, the feasible set in the dual representation of $\mathbb{W}_\alpha(\pi)$ in Lemma \ref{lemma: AVaR dual} can be restricted to a compact set. Since $|Y_i(\pi)| \leq |Y_i(0)| + |Y_i(1)|$ for all $\pi \in \Pi_o$, the $\alpha$-quantile of $|Y_i(\pi)|$ is no greater than the $\alpha$-quantile of $|Y_i(0)| + |Y_i(1)|$, while the $\alpha$-quantile of $-|Y_i(\pi)|$ is no less than the $\alpha$-quantile of $-|Y_i(0)| - |Y_i(1)|$. Therefore, the solution to $\sup_{\eta \in \mathbb{R}} \mathbb{V}(\pi, \eta)$ is $\mathrm{VaR}_\alpha(Y_i(\pi))$, which satisfies the bounds
\[
- \mathrm{VaR}_{1-\alpha} \left( |Y_i(0)| + |Y_i(1)|  \right) \leq   \mathrm{VaR}_\alpha(Y_i(\pi)) \leq   \mathrm{VaR}_\alpha \left(|Y_i(0)| + |Y_i(1)|  \right).
\]
Thus, we can express $\mathbb{W}_\alpha(\pi)$ as $\sup_{\eta \in \mathcal{B}_Y } \mathbb{V}_\alpha(\pi, \eta)$ for some compact set $\mathcal{B}_Y \subset \mathbb{R}$.

(ii) When $Y_i$ has a bounded support, the claim in (i) holds for $\alpha=1$ as well.
\end{remark}

\subsection{Debiased Estimator of $\mathbb{W}_\alpha(\pi)$}\label{section: welfare_Debiased_Estimator}

As noted in the previous sections, the 1-expected welfare function  $\mathbb{W}_1(\pi) $ is the same as the expected welfare \(\mathbb{E}[Y_i(\pi)]\) in \cite{kitagawa2018a} and \cite{athey2021policy}. In the rest of this paper, we focus on estimation and asymptotic theory for an $\alpha$-EWM rule for $\alpha\in (0,1)$.

Theorem \ref{theorem: Value function identification} suggests two plug-in methods for estimating $\mathbb{V}_\alpha(\theta)$ or the welfare function $\mathbb{W}_\alpha(\pi)$: IPW and outcome equation estimation. It is known that the IPW estimator is sensitive to the estimator of the propensity score and may suffer from severe bias. The outcome equation estimator may be sensitive to the estimators of $\tau$ ($\mu_1$ and $\mu_0$). This motivates the debiased estimator proposed in this section. 

\Cref{theorem: Value function identification} implies that under Assumption \ref{Assumption: Selection-on-observables}, the function $\mathbb{V}_{\alpha}(\theta)$ is identified for any fixed $\theta = (\pi, \eta)$.   Following \cite{robins1994estimation} and \cite{robins1995analysis}, we build our doubly robust score for $\mathbb{V}_\alpha(\theta)$   by introducing the augmentation term. 
Given any $\theta = (\pi, \eta)$, and for  any function $\check{e}: \mathcal{X} \rightarrow (0,1)$ and  $\check{\mu}_a :\mathcal{X} \times \mathbb{R} \rightarrow\mathbb{R}$ with $a \in \{0,1\}$, define 
\begin{equation} \label{equation: g_theta}
\begin{aligned}
g_\theta (z;  \check{\mu}, \check{e} ) & =\frac{1}{\alpha} \left[ \left(1-\pi(x)  \right)\check{\mu}_0 (x,\eta )+ \pi(x) \check{\mu}_1 (x,\eta ) \right]  +\eta \\
&+ \frac{1}{\alpha} \left[\frac{(1-\pi(x))(1-a)}{1-\check{e}\left(x \right)}( (y-\eta)_{-}  - \check{\mu}_0(x,\eta))\right] \\
& + \frac{1}{\alpha} \left[\frac{\pi(x) a}{\check{e}(x)}\left( (y-\eta)_{-}   - \check{\mu}_1(x,\eta)\right)\right] ,
\end{aligned} 
\end{equation}
where $\check{\mu} = (\check{\mu}_0, \check{\mu}_1)$ and the augmentation term  is defined as the sum of the last two components in \eqref{equation: g_theta}. The augmentation term has mean zero and the Neyman orthogonality condition holds:
\[
\partial_{\mu} \mathbb{E}_{P} [ g_\theta(Z_i; \mu_o, e_o) ] [ \check{\mu} - \mu_o] = 0,\quad \text{ and }\quad \partial_{e} \mathbb{E}_{P} [ g_\theta(Z_i; \mu_o, e_o) ] [  \check{e}- e_o] = 0,
\]
where $\mu_o = (\mu_0, \mu_1)$. To simplify notation, we let $g_\theta(\cdot) = g_\theta(\cdot; \mu_o, e_o)$, where the function $g_\theta(\cdot)$ indexed by $\theta$ is referred to as the (doubly robust) score function for estimating $\mathbb{V}_\alpha(\theta)$.
It is clear that for any given $\theta$,  the function $g_{\theta} - \mathbb{E}_P[ g_{\theta} (Z_i)]$ is the efficient influence function for $\mathbb{V}_{\alpha} (\theta)$; see \cite{luedtke2016statistical,kennedy2016semiparametric} for more detailed discussion.

Building upon \cite{chernozhukov2018double} and \cite{chernozhukov2022locally}, we construct our doubly robust score \( \widehat{g}_\theta(Z_i) \) for \( \mathbb{V}_{\alpha}(\theta) \) based on \( K \)-fold cross-fitting, a sample-splitting method used to validate asymptotic properties and leverage high-level conditions concerning the predictive accuracy of nuisance estimation methods.  The estimation procedure is
described below.
\begin{enumerate}
\item[(a)] 
Randomly partition the sample into $K$ folds $\cup_{k=1}^K \mathcal{I}_k$ such that $|\mathcal{I}_k | = n/K$.
\item[(b)] For each \( k \), define \( \mathcal{I}_k^c = [n] \setminus \mathcal{I}_k \). Fit estimators for the nuisance parameters \( e_o(\cdot) \) and \( \mu_a(\cdot, \cdot) \) for \( a \in \{0,1\} \) using the observations in the remaining \( K-1 \) folds, specifically, \( (Z_i)_{i \in \mathcal{I}_k^c} \). Denote these estimators as \( \widehat{e}^{(-k)}(\cdot) \) and \( \widehat{\mu}^{(-k)}_a(\cdot, \cdot) \).

\item[(c)] The doubly robust score is 
\begin{equation} \label{equation: Gamma_theta}
\begin{aligned}
\widehat{g}_\theta(Z_i; \widehat{\mu}^{-k(i)}, \widehat{e}^{-k(i)}) & =\frac{1}{\alpha} \left[ \left(1-\pi(X_i)  \right)\widehat{\mu}^{-k(i)}_0 (X_i,\eta )+ \pi \left(X_i \right) \widehat{\mu}^{-k(i)}_1 (X_i,\eta ) \right] +\eta  \\
&+ \frac{1}{\alpha} \left[\frac{\left(1-\pi(X_i)\right)\left(1-A_i \right)}{1-\widehat{e}^{-k(i)} \left( X_i \right)} \left[ ( Y_i -\eta)_{-}  - \widehat{\mu}_0^{-k(i)}(X_i,\eta)    \right ]  \right] \\
& + \frac{1}{\alpha} \left[\frac{\pi(X_i) A_i}{\widehat{e}^{(-k(i))} \left( X_i \right) }\left[ (Y_i-\eta)_{-}   - \widehat{\mu}_{1}^{-k(i)}(X_i,\eta)\right]\right] ,
\end{aligned} 
\end{equation} 
where $k(i)$ is the index in $[n]$ such that $i \in \mathcal{I}_k$.
\item[(d)] For each $\theta=(\pi, \eta)$, $\mathbb{V}(\theta)$ and $\mathbb{W}_\alpha(\pi)$  can be estimated by 
\[
\widehat{\mathbb{V} }_{n} (\theta) = \frac{1}{n} \sum_{i = 1}^n  \widehat{g}_\theta(Z_i) \quad \text{and} \quad 
\widehat{\mathbb{W} }_{n}(\pi) = \sup_{\eta \in \mathcal{B}_Y  } \widehat{\mathbb{V} }_{n}  (\pi, \eta),
\]
where $\mathcal{B}_Y$ is introduced in Remark \ref{remark: compact feasible set}.\footnote{If the support of $Y_i$ is bounded, i.e., Assumption \ref{Assumption: Bounded support} holds, then one can take $\mathcal{B}_Y$ as the support of $Y_i$. In our numerical work, we took $\mathcal{B}_Y$ as the closed interval with lower and upper bounds as the minimum and maximum statistics of $Y_i$ respectively.  } 

\item[(e)] The debiased estimator $\widehat{\theta}_{n}  = (\widehat{\pi}_{n}, \widehat{\eta}_{n} )$ is the maximizer of $\widehat{\mathbb{V} }_{n} (\theta)$.  
\end{enumerate}

\begin{remark} 
Since \cref{lemma: AVaR dual} implies that $\mathbb{W}_\alpha(\pi)=\mathbb{V}_\alpha(\pi, F_\pi^{-1}(\alpha))$, a debiased estimator of $\pi^*$ can also be constructed from the expression  $\mathbb{V}_\alpha(\pi, F_\pi^{-1}(\alpha))$. Noting that $F_\pi^{-1}(\alpha)$ is an optimal solution to $\sup_{\eta\in\mathbb{R}}\mathbb{V}_\alpha(\theta)$, the orthogonal moment function based on  $\mathbb{V}_\alpha(\pi, F_\pi^{-1}(\alpha))$ is equal to 
\[
g_{\left(\pi,F_{\pi}^{-1}(\alpha)\right)} (z;  \check{\mu}, \check{e} ) - \mathbb{V}_\alpha(\pi, F_\pi^{-1}(\alpha)) .
\]

A cross-fitting estimator can be constructed from $g_{(\pi,F_{\pi}^{-1}(\alpha))} (z;  \check{\mu}, \check{e} )$, but it requires an estimator of the quantile function $F_\pi^{-1}(\alpha)$. We leave a detailed comparison between these two estimators in future work. 
\end{remark}

\section{Asymptotic Upper Regret Bounds}
\label{Section: theory}

In this section, we establish asymptotic regret bounds on the debiased $\alpha$-EWM policy proposed in Section \ref{Section: debiased} for any fixed $\alpha\in(0,1)$. They complement similar regret bounds for the $1$-EWM and equality-minded policies established in \cite{kitagawa2018a}, \cite{athey2021policy}, and  \cite{kitagawa2021equality}. 

\subsection{Policy Class and Examples}
For each $n$, let $\Theta_n = \Pi_n \times \mathcal{B}_Y$, where $\Pi_n$
is the class of pre-specified policies and $\mathcal{B}_Y \subset \mathbb{R}$ is a compact set introduced in \cref{remark: compact feasible set}. For brevity, we write $\mathbb{V}(\theta) \equiv \mathbb{V}_\alpha(\theta)$, omitting the subscript $\alpha$. 

The following assumption restricts the complexity of the policy class $\Pi_n$. 

\begin{assumption}\label{policy size}
There exists a constant $b_o \in \left(0,1/2\right)$ such that the VC-dimension of $\Pi_n$ is bounded as $\mathrm{VC}(\Pi_n) \leq n^{b_o}$ for all $n \in \mathbb{N}^+$.
\end{assumption}

A policy is a classifier that assigns the covariate $X_i$ to a binary treatment status. Any machine learning classification model can serve as a candidate policy class. In the following, we list three examples of policy classes and their VC dimensions.

\begin{example}[Linear Rules]  \label{example: Linear Rules}
The linear policy class can be  characterized by 
\begin{equation}\label{equation: Linear policy Class}
\Pi_n = \left\{ \mathds{1}\{  x^\prime \beta > 0 \}: \beta \in B \right\},
\end{equation}
where $B$ is a compact subset of $\mathbb{R}^{p_n}$, where the dimension of the covariates $p_n$ is allowed to grow with the sample size $n$. Although the eligibility score is linear in $\beta$, it can include intercepts, interaction terms, higher-order terms, and other transformations of the original covariate $X_i$.  The VC-dimension of $\Pi_n$ is $p_n + 1$.
\end{example}

\begin{example}[Decision Trees] \label{example: Decision Tree}
A decision tree is a predictor $\pi: \mathcal{X}\subset \mathbb{R}^p \rightarrow \{0,1\}$ that recursively partitions the feature space $\mathcal{X}$ into a set of rectangles and assigns a label to each resulting partition. Following \cite{bertsimas2017optimal} and \cite{zhou2023offline}, we define a decision tree recursively. A decision tree of depth $L$ consists of $L$ levels, with the first $L-1$ levels containing branch nodes and the final $L$-th level comprising exclusively of leaf nodes.  For any branch node, we choose the split-point $b$ and the variable $x(j)$ that is a single component of $x$ . If $x(j) < b$, the path taken is towards the left; if not, the decision leads to the right branch. Each path will end with a leaf node that is assigned a unique label. \cite{zhou2023offline} show that the  VC-dimension  of the class $\Pi$  of decision trees of depth-$L$ over $\mathbb{R}$ is $\mathrm{VC}(\Pi) = \widetilde{\mathcal{O}}( 2^L \log p)$. Choosing the tree depth $L_n = O(\log n)$ ensures that the VC dimension grows polynomially in $n$, thereby satisfying Assumption \ref{policy size}.
\end{example}

\begin{example}[ReLU Neural Networks]\label{example: ReLU_DNN}
Formally, a Rectified Linear Unit (ReLU) neural network is characterized by the ReLU activation function, $\sigma(x)=\max (0, x)$, and structured as a directed acyclic graph with weights and biases. Let $W$ denote the total number of parameters and $L$ the depth. Let $\Pi$ denote the policy class of deep ReLU networks. Bartlett et al. (2019) establish that $\mathrm{VC}(\Pi) = O\left(WL \log W \right)$ and $\mathrm{VC}(\Pi) = \Omega\left(WL \log(W/L)\right)$. Provided that
$W L\log W=O \big(n^{1/2-\epsilon}\big)$
for some $\epsilon>0$, the VC dimension satisfies
Assumption \ref{policy size}.
\end{example}

\subsection{Assumptions on Nuisance Estimators and a Preliminary Lemma}

In this section, we establish a fundamental lemma showing that the estimation error of the nuisance parameters can be ignored when $\mathbb{V}(\cdot)$ is estimated using the doubly robust score with cross-fitting. Before presenting the lemma, we introduce additional assumptions.  

Let $ \mathbb{V}_{n}(\theta)= \mathbb{P}_n g_\theta$, where $g_\theta (z):= g_\theta(z; \mu_o, e_o)$ is defined in \eqref{equation: g_theta}. We assume that the nuisance parameter estimators $\widehat{\mu}_a\left(\cdot, \cdot\right)$ and $\widehat e(\cdot)$ converge to their true values at sufficiently fast rates.
 \begin{assumption}\label{assumption: nuisance parameter convergence rate}
 \begin{assumpenum}
 \item   $\sup_{ (x,\eta) \in \mathcal{X} \times \mathcal{B}_Y }   \left | \widehat{\mu}_a(x, \eta) - \mu_a(x, \eta) \right|=o_P(1)$ for $a\in \{0,1\}$, and \\
 $\sup_{ x  \in \mathcal{X}  }   \left| \widehat{e} \left(x\right) - e_o(x) \right|  = o_P(1)$.
 \item There exist constants $\zeta_\mu, \zeta_e > 0$ satisfying $\zeta_\mu+ \zeta_e > 1/2$ such that for $a \in \{0,1\}$,
 \[
\sup_{\eta \in \mathcal{B}_Y }  \left[ \mathbb{E} \left|  \widehat{\mu}_a(X_i, \eta) - \mu_a(X_i, \eta)  \right|^2  \right]^{1/2}   = O (n^{-\zeta_\mu})  \ \ \text{and} \ \
\left[  \mathbb{E} \left| \widehat{e}\left(X_i \right) - e_o(X_i)  \right|^2  \right]^{1/2} = O( n^{-\zeta_e}).
\]
 \item $b_o < 2 \left(\zeta_\mu \wedge \zeta_e\right)$, where $b_o$ is defined in Assumption \ref{policy size}.
\end{assumpenum}
\end{assumption}

\begin{remark} The regression function $\mu_a(x, \eta)$ can be estimated by regressing the transformed outcome $\{ (Y_i - \eta)_{-}: A_i = a \}$ on $\{ X_i: A_i = a \}$ using standard nonparametric methods.
For example, under Assumption \ref{Assumption: Selection-on-observables} and standard regularity conditions (e.g., Hölder smoothness), kernel-based estimators can achieve minimax optimal convergence rates, both uniform and in $L^2$ \citep{gine2002rates,gine2021mathematical}. Similarly, sieve-based approaches can be applied; optimal uniform convergence rates are established by \cite{chen2015optimal, belloni2015some}. Furthermore, deep neural networks can  also be used to estimate nuisance parameters, with \cite{schmidt2020nonparametric} and \cite{kohler2021rate} establishing nearly optimal $L^2$-convergence rates.

\end{remark}

We conclude this subsection by demonstrating that $ \widehat{\mathbb{V}}_{n}(\theta)$ is a good approximation to $\mathbb{V}_{n}(\theta) = \mathbb{P}_n g_\theta$ with convergence rate faster than $n^{-1/2}$. Consequently, we can ignore the nuisance parameter estimation errors in subsequent asymptotic analysis. 

\begin{lemma}\label{Lemma: M_error}
Suppose Assumptions \ref{Assumption: Selection-on-observables}, \ref{policy size} and \ref{assumption: nuisance parameter convergence rate} hold. Then,
\[
\mathbb{E}_{P} \left[ \sup_{\theta \in \Theta_n} \left|\widehat{\mathbb{V}}_{n}(\theta) - \mathbb{V}_{n}(\theta) \right | \right] = o \big(n^{- 1/2} \big).
\]
\end{lemma}

\subsection{Asymptotic Upper Regret Bound}
In this subsection, we study the regret upper bound of implementing $\widehat{\pi}_{n}$ under the following assumption.

\begin{assumption}\label{assumption: L^2 boundedness}
 $Y_i(a)$ is $L^2(P)$-bounded, i.e.,  $\mathbb{E}_P \left[ |Y_i (a) |^2 \right] < \infty$ for $a \in \{0,1\}$.
\end{assumption}

For any policy class $\Pi_n$, which may depend on $n$, the regret of deploying a policy $\pi \in \Pi_n$  relative to the best policy in $\Pi_n$, is defined as
\[
\mathrm{Reg}(\pi, \Pi_n) = \max_{\pi^\prime \in \Pi_n} \mathbb{W}_\alpha(\pi^\prime) - \mathbb{W}_\alpha(\pi).
\]
When $\Pi_n$ is clearly understood from the context, we write $\mathrm{Reg}(\pi)  =\mathrm{Reg}(\pi, \Pi_n)$ for notational simplicity. Our primary result regarding the asymptotic regret of our $\alpha$-EWM policy incorporates the following two key quantities:
\[
\begin{aligned}
\Xi := \sup_{\eta \in \mathcal{B}_Y } \mathbb{E} \left| \gamma_\eta(Z_i ) \right|^2  \quad \text{and} \quad  \Xi^\dagger := \sup_{\eta \in \mathcal{B}_Y } \mathbb{E} \big| \gamma^\dagger_\eta(Z_i ) \big|^2 ,
\end{aligned}
\]
where 
\[
\begin{aligned}
\gamma_\eta(z) & :=   \tau(x, \eta) +  \frac{a - e_o(x)}{e_o(x) \left( 1 - e_o(x) \right)}  \left\{ (y-\eta)_{-}   - \mu_{a}(x,\eta)\right\} \mbox{ and }\\
\gamma^\dagger_\eta(z ) & :=  \mu_0(x, \eta)  +  \frac{(1-a)}{1-e_o \left(x \right)} \left\{ (y-\eta)_{-}  -\mu_0(x,\eta) \right\}.
\end{aligned}
\]

\begin{theorem}\label{theorem: regret bound with semiparametric efficiency score}
Suppose Assumptions \ref{Assumption: Selection-on-observables}, \ref{policy size}, \ref{assumption: nuisance parameter convergence rate},  and \ref{assumption: L^2 boundedness} hold. If $\inf_{\eta \in \mathcal{B}_Y}\mathbb{E}|\gamma_\eta(Z_i)|^2 > 0$, then for $\alpha\in (0,1)$, 
\begin{equation}\label{equation: semiparametric efficiency score regret bound}
\limsup_{n \rightarrow \infty} \frac{ 
 \mathbb{E}\left[   \mathrm{Reg} \left( \widehat{\pi}_{n} \right )    \right] }{   \sqrt{\mathrm{VC}(\Pi_n)  / n}   }   \leq \frac{30}{\alpha}  \sqrt{ \Xi + \Xi^\dagger } + 72 \sqrt{ \left(\bar{K}/ \alpha + 1\right)^2 + \Xi /\alpha^2 },
\end{equation}
where $\bar{K} = 3 + 2/\kappa$.
\end{theorem}

\cref{theorem: regret bound with semiparametric efficiency score} complements Theorem 1 in \cite{athey2021policy} for $1$-EWM policy.  The constant in 
\cref{theorem: regret bound with semiparametric efficiency score} depends on $\alpha$: it increases as $\alpha$ decreases, partly due to estimation error. Specifically, estimating the average welfare of the $\alpha$-worst-affected group makes use of only an $\alpha$-fraction of the total sample, leading to greater instability in welfare estimation.

\begin{remark}\label{remark: upper bound_characterization}
Suppose that $\mathcal{B}_Y \subseteq \left[-\eta_B, \eta_B \right]$ for some $\eta_B > 0$. Under the strict overlap condition in Assumption \ref{assumption: Strict Overlap}, it follows that $\Xi$ and $\Xi^\dagger$ can be upper bounded as
\[
\begin{aligned}    
\Xi  \leq  \left(1 + \frac{2}{\kappa}  \right) \left( \mathbb{E} |Y_i(0)|^2 + \mathbb{E} |Y_i(1)|^2  + \eta_B^2 \right) \quad \text{and} \quad
\Xi^\dagger   \leq  \left(1 + \frac{2}{\kappa}  \right) \left( \mathbb{E} |Y_i(0)|^2  + \eta_B^2 \right).
\end{aligned}
\]
\end{remark}

\begin{remark}\label{remark: near-optimal solution}
Recall that we learn the optimal policy by simultaneously solving $\widehat{\pi}_{n}$ and $\widehat{\eta}_{n}$ from $\max_{(\pi, \eta)\in \Pi_n \times \mathcal{B}_Y } \widehat{\mathbb{V}}_{n}( \pi, \eta)$. Like \cite{athey2021policy}, we also allow for an approximate optimal policy. Specifically, 
let $\widetilde{\theta} \equiv (\widetilde{\pi}, \widetilde{\eta}) \in \Pi_n \times \mathcal{B}_Y$ denote any near-optimal solution satisfying
\[
\widehat{\mathbb{V}}_{n}(\widetilde{\theta}) \geq \sup_{\theta \in \Theta_n }\widehat{\mathbb{V}}_{n}(\theta)  - o_P\left( r_n \right),
\]
where $r_n =  \sup_{\theta \in \Theta_n} \left|\widehat{\mathbb{V}}_{n}(\theta) - \mathbb{V}_{n}(\theta) \right | $.
\cref{theorem: regret bound with semiparametric efficiency score} holds if the exact optimizer  $\widehat{\pi}_{n}$ is replaced by any near-optimal welfare maximizer $\widetilde{\pi}$ and $r_n = o_P(n^{-1/2})$.   
The term $o_P(r_n)$ enables us to find an approximate solution to $\max_{\theta \in \Theta_n} \widehat{\mathbb{V}}_{n} (\theta) $, which is particularly useful when the optimization is non-concave. 
\end{remark}

\paragraph{Technical Comparisons with \cite{kitagawa2018a, athey2021policy}.}

The regret bounds in \cref{theorem: regret bound with semiparametric efficiency score} and those in \cite{kitagawa2018a, athey2021policy}  are all of the same order $\sqrt{\mathrm{VC}(\Pi_n) / n}$. Additionally,  \cref{theorem: regret bound with semiparametric efficiency score}  and Theorem 1 in \cite{athey2021policy} provide explicit expressions for the constants which require more delicate technical proofs than  \cite{kitagawa2018a, kitagawa2021equality}. Technically, establishing \cref{theorem: regret bound with semiparametric efficiency score} is more challenging than Theorem 1 in \cite{ athey2021policy} due to the nonlinear score function. To elaborate on the challenges and  how we overcome them, we first provide a sketch of the argument for the rate in  \cref{theorem: regret bound with semiparametric efficiency score}. 

Standard empirical process arguments alongside \cref{Lemma: M_error} would suffice to establish the rate. However, even in this simpler case, we face additional technical challenges due to the nonlinear score function. To see this, we
follow \cite{kitagawa2018a, kitagawa2021equality} and establish the order of the regret bound for $\widehat{\pi}_n$ based on the following lemma.

\begin{lemma}\label{lemma:REG} Suppose Assumptions \ref{Assumption: Selection-on-observables}, \ref{policy size} and \ref{assumption: nuisance parameter convergence rate} hold. Then for $r_n=o_P(n^{-1/2}),$
    \[\mathrm{Reg}(\widehat{\pi}_{n}) \leq 2 \sup _{\theta \in \Theta_n}\left|\left(\mathbb{P}_n-P\right) g_\theta\right| + r_n.\]
\end{lemma}

\cref{lemma:REG} implies that it is sufficient to study the concentration of the empirical process: $$\mathbb{V}_{n}(\theta) - \mathbb{V}(\theta) = (\mathbb{P}_n - P) g_\theta \
\mbox{ over } \ \theta \in \Theta_n. $$ 
In contrast to \cite{kitagawa2018a,kitagawa2021equality} and \cite{athey2021policy}, the score function for the $\alpha$-expected welfare  $g_\theta$ is nonlinear in $\theta$ rendering the VC dimension of the function class $\mathcal{G}_{\Theta_n} :=\{g_\theta : \theta \in \Theta_n\}$ difficult to derive. Instead of exploiting the VC dimension of the corresponding function classes as in \cite{kitagawa2018a,kitagawa2021equality} and \cite{athey2021policy}, we directly upper bound the covering number of $\mathcal{G}_{\Theta_n}$ and then apply the classic empirical process maximal inequality, such as Theorem 2.14.1 in \cite{vaart1998empirical}.  
\begin{lemma}\label{lemma: covering number of G_theta}
If Assumption \ref{assumption: L^2 boundedness} holds,  then there is an envelope function $G$ for $\mathcal{G}_{\Theta_n}$ and constant $c_o > 0$ not depending on $n$ and $p$ such that 
\[
N\left( \epsilon  \|G \|_{Q,2} , \mathcal{G}_{\Theta_n}, L^2(Q)\right) \leq\left(c_o / \epsilon\right)^{24 \mathrm{VC}(\Pi_n)+25}, \quad \forall \epsilon > 0,
\]
for all finite discrete probability measures $Q$ on $\mathcal{Z}$.
\end{lemma}
Assumption \ref{assumption: L^2 boundedness} and Assumption \ref{assumption: Strict Overlap} ensure the existence of an envelope function that is bounded in $L^2(P)$.
Applying Theorem 2.14.1 in \cite{vaart1998empirical} and \cref{lemma: covering number of G_theta}, we conclude that there is a constant $C_o > 0$ not depending on $n$ such that
\begin{equation}\label{equation:order}
\mathbb{E}_{P} \left[ \sup_{\theta \in \Theta_n}  \left| (\mathbb{P}_n - P) g_{\theta} \right| \right] \leq  C_o\sqrt{ \mathrm{VC}(\Pi_n)/ n}.
\end{equation}

\cref{theorem: regret bound with semiparametric efficiency score} further provides an explicit expression for the constant $C_o$ in Eq. (\ref{equation:order}). To derive it,  we follow \cite{athey2021policy} and employ a classical chaining argument to derive an upper bound for the Rademacher complexity of the score function class. However, due to the nonlinearity of score function $g_\theta$ with respect to $\theta = (\pi, \eta)$, the slicing technique used in \cite{athey2021policy} is difficult to implement. Instead, we introduce a new conditional semi-metric  and apply Dudley's chaining argument to directly bound the Rademacher complexity of $\mathcal{G}_{\Theta_n}$. We refer interested readers to Section D.5 of the Online Supplement for details.

\section{Inference for the Optimal Welfare}
\label{section:inference}

In this section, we develop asymptotically valid inference for the optimal $\alpha$-expected welfare. Compared with regret bounds, inference on optimal welfare is lacking even for EWM except for the first-best EWM policy; see  \cite{luedtke2016statistical,luedtke2018parametric,shi2020breaking}, and Appendix B of the Online Supplement to \cite{kitagawa2018a}.

We first construct Wald-type inference assuming uniqueness of the optimal policy and then summarize a general inference procedure without the uniqueness assumption. A detailed treatment of the general inference procedure is postponed to Section \ref{section: inference for the optimal value} of the Appendix. 

For simplicity, we assume that the policy class does not change with the sample size $n$, i.e., $\Pi_n = \Pi$ for all $n$, and write $\Theta = \Pi \times \mathcal{B}_Y$.  We define a metric space $(\Theta, \| \cdot \|)$, where 
\[
\left\| \theta_1 - \theta_2 \right\| \equiv  |\eta_1- \eta_2|+ \|\pi_1 - \pi_2\|_{P, 2}  = |\eta_1- \eta_2| + \sqrt{ \mathbb{E} |\pi_1(X_i) - \pi_2(X_i)|^2}.
\]
for any $\theta_1, \theta_2 \in \Theta$. This premise will be upheld throughout the subsequent analysis.

We adopt the following assumption in this section.

\begin{assumption}\label{assumption: assumption for inference}
\begin{assumpenum}
\item \label{Assumption: Bounded support}
The outcome $Y_i = Y_i(A_i)$ has bounded support, i.e., $\mathbb{P}\left( | Y_i | \leq c_o \right)  = 1$ for some constant $c_o > 0$.
 \item  \label{Assumption: fixed policy class}  The policy class $\Pi$ has finite VC-dimension, i.e.,  $\mathrm{VC}(\Pi) < \infty$.
\end{assumpenum}
\end{assumption}

Assumption \ref{assumption: assumption for inference} is widely adopted in policy learning research, see, e.g., \cite{kitagawa2018a, kitagawa2021equality, rai2018statistical, kallus2018confounding, luedtke2016statistical, luedtke2020performance}.\footnote{Although studies like \cite{athey2021policy} do not adopt this assumption for regret bounds, it substantially simplifies the technical analysis for statistical inference. } 
Assumption \ref{Assumption: Bounded support} implies that the feasible set $\mathcal{B}_Y$ of the dual reformulation of $\mathbb{W}_\alpha(\pi)$ can be restricted to $[-c_o, c_o]$  and the regression functions  $|\mu_a(x, \eta)| \leq 2c_o$ for all $\eta \in \mathcal{B}_Y$ and $a \in \{0,1\}$. Moreover, the functions $g_\theta(\cdot)$ are also uniformly bounded, i.e., $\sup_{\theta \in \Theta}\| g_{\theta} \|_{\infty} < \infty$.

\subsection{Asymptotic Normality and Wald-type Inference}
\label{section: Wald CI}
Wald-type inference relies on the following uniqueness assumption. 
\begin{assumption}[Uniqueness]
\label{Assumption: V_theta smoothness (1)}
There exists a $\theta_o \equiv (\pi_o, \eta_o) \in \Theta$ such that for all $\epsilon > 0$, $   \mathbb{V}(\theta_o) > \sup \{  \mathbb{V}(\theta): \theta \in \Theta,  \| \theta - \theta_o \| > \epsilon   \}$.
\end{assumption}
Assumption \ref{Assumption: V_theta smoothness (1)} is a standard condition in extremum estimation. It ensures that $\theta_o\in \Theta$ is a unique and well-separated point of maximum of $\theta \mapsto \mathbb{V}(\theta)$. Lemma 14.4 in \cite{kosorok2008introduction} gives some sufficient conditions for this assumption. If for all $\epsilon > 0$, $   \mathbb{W}(\pi_o) > \sup_{\pi: \| \pi - \pi_o \| > \epsilon }   \mathbb{W}(\pi)$  and $Y_i(\pi)$ has positive density at $\mathrm{VaR}_\alpha(Y_i(\pi))$ for all $\pi \in \Pi$, then Assumption \ref{Assumption: V_theta smoothness (1)} is satisfied. For policy learning, Assumption \ref{Assumption: V_theta smoothness (1)} is strong, although it is adopted in \cite{wang2018quantile}, Section 2.3 of \cite{kitagawa2018a}, and Section 2.3 of \cite{luedtke2020performance}. 
\begin{remark}
For $1$-EWM, uniqueness of the first-best optimal policy excludes a special class of distributions known as exceptional distributions. For  $\alpha$-EWM with $\alpha\in(0,1)$, we show in \cref{lemma: first best policy} in Section \ref{section: First_best_policy} of the Appendix that the first best policy is given by $\pi_o = \mathds{1}\{ \tau(x, \eta_o) \geq 0 \}$ with $\eta_o = \eta_{\mathrm{FB}}^*$ defined in \cref{lemma: first best policy}.  Assumption \ref{Assumption: V_theta smoothness (1)} excludes the class of exceptional distributions for which $\mathbb{P}\left[\tau(X_i, \eta_o)=0  \right] > 0$. This is because Assumption \ref{Assumption: V_theta smoothness (1)} implies that $\theta_o = (\pi_o, \eta_o)$ is the unique and well separated maximizer. As a result,
\[
 \mathds{1}\{ \tau(X_i, \eta_o) \geq 0 \} =  \mathds{1}\{ \tau(X_i, \eta_o) > 0 \},\quad P\text{-a.s.},
\]
and $\mathbb{P}( \tau(X_i, \eta_o) = 0 ) = 0$. 
\end{remark}

To establish asymptotic normality of $\widehat{\mathbb{V} }_{n} (\widehat{\theta}_{n}  )$,  consider the following decomposition:
\begin{equation}\label{equation: V_DML decomposition}
\begin{aligned}
\widehat{\mathbb{V} }_{n} (\widehat{\theta}_{n}  ) -  \mathbb{V}(\theta_o)  =  \underbrace{ \widehat{\mathbb{V} }_{n} (\widehat{\theta}_{n}  ) - \mathbb{V} _{n} (\widehat{\theta}_{n}  ) }_{= o_P(n^{-1/2})} 
 + \underbrace{ \mathbb{V} _{n} (\widehat{\theta}_{n}  )  -   \mathbb{V}  (\widehat{\theta}_{n}  )   }_{\approx \mathbb{V} _{n} (\theta_o  )  -   \mathbb{V}  (\theta_o )    } +  \mathbb{V}  (\widehat{\theta}_{n}  )   -  \mathbb{V}  (\theta_o) .
\end{aligned}
\end{equation}
Note that the first term on the RHS of \cref{equation: V_DML decomposition} is $o_P(n^{-1/2})$ due to \cref{Lemma: M_error}. In the rest of this section, we will show that 
\begin{enumerate}
    \item[(i)] the second term on the RHS of \cref{equation: V_DML decomposition} is asymptotically equivalent to $\mathbb{V}_{n}(\theta_o) - \mathbb{V}(\theta_o)$;
    \item[(ii)] the third term on the RHS of \cref{equation: V_DML decomposition} is of order $o_P(n^{-1/2})$.  
\end{enumerate}
Consequently,
\[
\begin{aligned}
\sqrt{n} \left[\mathbb{V}_{n} (\widehat{\theta}_{n}  )  -   \mathbb{V}  (\theta_o) \right] & = \sqrt{n} \left[ \mathbb{V} _{n} (\theta_o  )  -   \mathbb{V}  (\theta_o ) \right] + o_P(1)  \\
& = \sqrt{n} (\mathbb{P}_n - P) g_{\theta_o}  + o_P(1).
\end{aligned}
\]
and asymptotic normality follows. 

To show (i), we first prove $\|\widehat{\theta}_{n}  - \theta_o\| = o_P(1)$ in \cref{lemma: consistency} below. Since $\mathcal{G}_\Theta \equiv \{g_\theta : \theta \in \Theta\}$ is $P$-Donsker by \cref{lemma: covering number of G_theta}, (i) follows.

\begin{lemma}\label{lemma: consistency}
Under Assumptions \ref{Assumption: Selection-on-observables}, \ref{assumption: nuisance parameter convergence rate},  \ref{assumption: assumption for inference} and \ref{Assumption: V_theta smoothness (1)}, it holds that $\|\widehat{\theta}_{n} -\theta_o \|=o_P(1)$.
\end{lemma}

To show (ii),  we note that
\[
\begin{aligned}
\mathbb{V}(\theta_o) - \mathbb{V}( \widehat{\theta}_n  ) &= \mathbb{V}(\theta_o) - \widehat{\mathbb{V}}_{n} (\theta_o) + \widehat{\mathbb{V}}_{n} (\theta_o) - \widehat{\mathbb{V}}_{n} (\widehat{\theta} _{n} ) + \widehat{\mathbb{V}}_{ n} (\widehat{\theta}_{n} )  - \mathbb{V}(\widehat{\theta}_{ n} )  \\
& \leq \mathbb{V}(\theta_o) - \mathbb{V}_{n} (\theta_o)  + \mathbb{V}_{ n} (\widehat{\theta}_{ n} )  - \mathbb{V}(\widehat{\theta}_{ n} )  + r_n  \\
& = (\mathbb{P}_n - P) ( g_{\widehat{\theta}_{n}  } - g_{\theta_o} ) + r_n,
\end{aligned}
\]
where the inequality follows from $\widehat{\mathbb{V}}_{n} (\theta_o) - \widehat{\mathbb{V}}_{n} (\widehat{\theta} _{ n} ) \leq 0$, and $r_n = o_P(n^{-1/2})$ by \cref{Lemma: M_error}. Similar to \cite{luedtke2020performance}, one can show that under mild conditions including boundedness and uniqueness, asymptotic equicontinuity arguments ensure that $(\mathbb{P}_n - P) ( g_{\widehat{\theta}_n} - g_{\theta_o} ) = o_P(n^{-1/2})$ for any policy class $\Pi$ satisfying $\text{VC}(\Pi)<\infty$.

Summing up, we obtain asymptotic normality of $\widehat{\mathbb{V}}_{n}(\widehat{\theta}_{n}  )$.

\begin{theorem}\label{Theorem: approximation of DML M_n}
Suppose conditions in \cref{lemma: consistency} hold. Then,
\[
\begin{aligned}
\widehat{\mathbb{V}}_{n}(\widehat{\theta}_{n}  )  -  \mathbb{V}_P (\theta_o) &=  (\mathbb{V}_{n} - \mathbb{V}_P )(\theta_o)  + o_P(n^{-1/2})\\
& = \frac{1}{n} \sum_{i=1}^n \left\{ g_{\theta_o} (Z_i) - \mathbb{E}_{P} [g_{\theta_o}(Z_i)] \right\} + o_P(n^{-1/2})   ,
\end{aligned}
\]
where the function $g_{\theta_o}$, defined in \cref{equation: g_theta}, is evaluated at $\theta = \theta_o$. In particular, 
\[
\sqrt{n} \left[\widehat{\mathbb{V}}_{n}(\widehat{\theta}_{n}  )  -  \mathbb{V}(\theta_o)  \right] \rightsquigarrow N\left( 0, \sigma_o^2 \right),
\]
where $\sigma_o^2 = \mathrm{Var}\left[g_{\theta_o}(Z_i) \right]$.
\end{theorem}
\begin{remark}
Drawing on \cite{newey1994asymptotic, luedtke2016statistical}, and under the assumptions stated in \cref{Theorem: approximation of DML M_n} and other mild conditions, our optimal welfare estimator achieves semiparametric efficiency bound.
\end{remark}
The next theorem presents a consistent estimator of the asymptotic variance $\sigma_o^2$.
\begin{theorem} \label{theorem: variance estimation}
Consider the following estimator of $\sigma_o^2$:
\[
\widehat{\sigma}_o^2 =   \frac{1}{n} \sum_{i=1}^n \left[ g_{\widehat{\theta}_{n}  }  \big(Z_i ; \widehat{\mu}^{(-k(i))} , \widehat{e}^{(-k(i))} \big )  \right]^2   -   \left[  \frac{1}{n} \sum_{i=1}^n  g_{\widehat{\theta}_{n}  }  \big(Z_i ; \widehat{\mu}^{(-k(i))} , \widehat{e}^{(-k(i))}  \big)  \right]^2. 
\]
Under the conditions of  \cref{lemma: consistency}, 
it holds that $\widehat{\sigma}_o^2 = \sigma_o^2 + o_P(1)$ and 
\[
\sqrt{n} \widehat{\sigma}_o^{-1} \left[\widehat{\mathbb{V}}_{n}(\widehat{\theta}_{n}  )  -  \mathbb{V}(\theta_o)  \right] \rightsquigarrow N\left( 0, 1 \right).
\]
\end{theorem}

\subsection{Uniform Inference}\label{section: uniform inference}

Section \ref{section: inference for the optimal value} of the Appendix applies
the numerical delta method introduced by \cite{hong2018numerical} to develop uniform inference for the optimal welfare without requiring the optimal policy to be unique, i.e., Assumption \ref{Assumption: V_theta smoothness (1)}. It improves on the inference proposed in Appendix B of the Online Supplement to \cite{kitagawa2018a}. We provide a summary of the procedure here and refer the interested reader to Section \ref{section: inference for the optimal value} of the Appendix for technical details. 

Define the supremum functional  $\psi: \ell^{\infty} (\Theta) \rightarrow \mathbb{R}$ as $\psi: h \mapsto \sup_{\theta \in \Theta}  h (\theta)$. 
Consider the multiplier bootstrap $ \widehat{\mathbb{G}}_{n}^*: \Theta \rightarrow \mathbb{R}$ defined as 
\begin{equation}\label{equation: multiplier boostrap process}
\widehat{\mathbb{G}}_{n}^*: \theta \mapsto  n^{-1/2} \sum_{i=1}^n  \xi_i \left [ \widehat{g}_{\theta}(Z_i) - \widehat{\mathbb{V}}_n (\theta) \right ],    
\end{equation}
where $\{\xi_i\}_{i=1}^n$ are i.i.d. random variables independent of $(Z_i)_{i=1}^n$, with $\mathbb{E}( \xi_i) = 0$, $\mathbb{E}(\xi_i^2) =1$ and $\mathbb{E}\left[ \exp |\xi_i| \right] < \infty$. 
For given $\epsilon_n = o(1)$ with $n^{1/2} \epsilon_n \rightarrow \infty$, let
\begin{equation}\label{equation: numerical boostrap}
\begin{aligned}
\widehat{\psi}_n^{\prime} ( \widehat{\mathbb{G}}_n^* )  &= \frac{ \psi \big(\widehat{\mathbb{V}}_n + \epsilon_n  \widehat{\mathbb{G}}_n^* \big )  - \psi(\widehat{\mathbb{V}}_n )    }{  \epsilon_n }.
\end{aligned}
\end{equation}

For any $\gamma \in (0,1)$, let $c_{\gamma}$ denote the $\gamma$-empirical quantile of $\widehat{\psi}_n^{\prime} (\widehat{\mathbb{G}}_n^*)$ which can be obtained from a large number of bootstrap samples. The one-sided confidence interval at the desired confidence level $1-\gamma$ is
\begin{equation}\label{equaton: one-sided_CI}
\left[ \sup_{\theta \in \Theta} \widehat{\mathbb{V}}_n (\theta) - c_{1-\gamma}/\sqrt{n}, \infty \right),     
\end{equation}
with correct asymptotic coverage:
\[
\lim_{n \rightarrow \infty} \inf_{P \in \mathcal{P}_n} \mathbb{P} \left[ \mathbb{V}_P(\theta_o) \geq \sup_{\theta \in \Theta} \widehat{\mathbb{V}}_n (\theta) - c_{1-\gamma} /\sqrt{n}  \right] \geq 1- \gamma,
\]
where $\mathcal{P}_n$ is a collection of distributions satisfying some regularity conditions specified in Assumption \ref{assumption: Estimation of Nuisance Functions} in Section \ref{section: inference for the optimal value} of the Appendix.
Define $q_{1-\gamma}$ as the $(1-\gamma)$-empirical quantile of $\left|\widehat{\psi}_n^{\prime} ( \widehat{\mathbb{G}}_n^* ) \right|$ for any $\gamma > 0$. The corresponding two-sided confidence interval is
\begin{equation}\label{equaton: two-sided_CI}
\left[ \sup_{\theta \in \Theta } \widehat{\mathbb{V}}_n (\theta) - q_{1- \gamma} / \sqrt{n},  \sup_{\theta \in \Theta } \widehat{\mathbb{V}}_n (\theta) +  q_{1- \gamma} / \sqrt{n}   \right],
\end{equation}
which attains the correct asymptotic coverage for any distribution $P \in \mathcal{P}_n$:
\[
\lim_{n \rightarrow \infty} \inf_{P \in \mathcal{P}_n}  \mathbb{P} \left[ \left| \sup_{\theta \in \Theta} \widehat{\mathbb{V}}_n (\theta) -  \mathbb{V}_P(\theta_o) \right|  \leq q_{1-\gamma} /\sqrt{n} \right] \geq 1-\gamma.
\]

\section{Empirical Application and Simulations}
This section presents extensive numerical results on the finite sample performance of our debiased estimator and proposed inference using both real data and synthetic data.\footnote{
Data and codes for this section can be accessed at
 \href{https://github.com/yqi3/alpha-EWM}{\texttt{https://github.com/yqi3/alpha-EWM}}.}
 
\label{Section: empirical}
\subsection{The JTPA Study}
\label{Section: JTPA}
\cite{kitagawa2018a} apply $1$-EWM method to experimental data from the National Job Training Partnership Act (JTPA) Study. The study randomized whether applicants are eligible to receive training and job-search assistance provided by the JTPA. The pre-treatment covariates included in the data are years of education (\textit{edu}) and pre-program earnings (\textit{prevearn}) and the outcome variable is an applicant's earnings 30 months after the assignment (\textit{earnings}). The sample size is 9,223 and the propensity score is known to be $2/3$. We adopt this data studied by \cite{kitagawa2018a} and, similar to \cite{kitagawa2018a}, we analyze welfare from an intent-to-treat standpoint, considering hypothetically making available the training program to eligible individuals, who may decline it. For detailed data description and evaluation of average program effects, we refer the reader to \cite{bloom1997benefits}.

We consider three policy classes: simple (treat all or none) and linear with and without squared and cubic $edu$. More specifically, the two linear policy classes take the form
\begin{equation}
\label{JTPA LES policy class}
    \Pi_{\mathrm{LES}}\defeq\left\{\{x:\beta_0+\beta_1 edu+\beta_2 prevearn>0\}, (\beta_0,\beta_1,\beta_2)\in\mathbb{R}^3\right\}\text{ and}
\end{equation}
\begin{equation}
\label{JTPA LES policy class 3}
    \Pi^3_{\mathrm{LES}}\defeq\left\{\begin{array}{lr}\{x:\beta_0+\beta_1 edu+\beta_2 prevearn+\beta_3edu^2+\beta_4edu^3>0\}, \\
\quad\quad\quad\quad\quad\quad\quad(\beta_0,\beta_1,\beta_2,\beta_3,\beta_4)\in\mathbb{R}^5 \end{array} \right\}.
\end{equation}
We investigate $\alpha\in\mathcal{A}\defeq\{0.25, 0.3, 0.4, 0.5, 0.8, 1\}$. For each $\alpha \in \mathcal{A}$ and policy class, we estimate $\mu_a(x, \eta)=\mathbb{E}\left[\left(Y_i(a) - \eta\right)_- \mid X_i = x\right] \text{ for } a \in \{0,1\}$ and a given $\eta$, using random forests (RF) developed by \cite{athey2019generalized}. We then apply simulated annealing (SA), proposed by \cite{kirkpatrick1983optimization}, to select the combination of parameters that (approximately) maximizes the objective function.\footnote{We build RF using \texttt{regression\_forest()} in \texttt{R} package \texttt{grf} and implement SA using \texttt{optim\_sa()} in the \texttt{R} package \texttt{optimization} \citep{athey2019generalized, husmann2017r}. We use default tuning parameters for RF. For SA, the specifications are more problem-specific. A good strategy is to plot the loss function and inspect if there is sufficient evidence of convergence.} SA is a derivative-free probabilistic optimization algorithm aiming at finding approximate solutions by iteratively exploring the solution space and gradually decreasing the probability of accepting worse solutions as the algorithm progresses.\footnote{\citet{geman1984stochastic} establish convergence of simulated annealing to the global optimum under idealized conditions, including infinitesimally slow cooling and an infinite number of iterations. Our implementation uses a practical finite-iteration schedule and is therefore intended as an approximate optimization routine. Other derivative-free methods, such as genetic algorithms, could also be employed for this purpose.}

Estimation and inference results for $\mathbb{W}_\alpha(\pi_o)$ are organized in Table \ref{Table: JTPA}. The first two columns consist of simple policies and serve as baselines for $\Pi_{\text{LES}}$ and 
$\Pi^3_{\text{LES}}$ in the third and fourth columns; details on the optimal policies can be found in Section H.1 of the Online Supplement. As $\alpha$ grows, $\widehat{\mathbb{W}}_\alpha(\widehat\pi_{n})$ increases across panels, reflecting an expanding lower-tail subpopulation that includes relatively better outcomes; the percentage of treated individuals increases with $\alpha$ as well. The \(95\%\) CIs from normal inference are in the third row of each panel, and the \(95\%\) CIs from 
uniform inference, obtained via multiplier bootstrap with \(\epsilon = n^{-1/4}\) and \(B = 100\), are in the last row. The uniform 
inference CI is not reported for the first two columns, as these are fixed policies rather than solutions to the optimization problem \eqref{equation:alpha-EWM}, and our uniform inference framework in Section~\ref{section: uniform inference} applies only to the latter. For each combination of $\alpha$ and policy class, the CI from uniform inference is wider than that from normal inference. While uniqueness 
cannot be verified, Section \ref{Section: WGAN simulations} finds that the Wald-type CIs achieve approximately \(95\%\) coverage, supporting their validity in this application.

\begin{table}[H] \centering
  \scriptsize{
\begin{tabular}
{@{\extracolsep{-6.5pt}}lcccc} 
 & \multirow{2}{*}{\textbf{Treat None}} & \multirow{2}{*}{\textbf{Treat All}} & \multirow{2}{*}{\textbf{Linear}} & \textbf{Linear with} \\
 &  &  &  & \textbf{$edu^2$ and $edu^3$} \\
\\[-1.75ex]\hline
\multicolumn{5}{c}{\cellcolor{blue!20}$\textbf{Panel 1: }\boldsymbol{\alpha}\boldsymbol{=0.25}$}
\vspace{.1cm}\\
\multicolumn{1}{l}{$\%$ treated} & 0\% & 100\% & 35\% & 33\%\\
\multicolumn{1}{l}{$\widehat{\mathbb{W}}_\alpha(\widehat\pi_{n})$} & \$377 & \$451 & \$531 & \$546\\
\multicolumn{1}{l}{95\% CI (normal)} & $(\$299, \$455)$ & $(\$373, \$529)$ & $(\$439, \$622)$ & $(\$446, \$646)$\\
\multicolumn{1}{l}{95\% CI ($\epsilon=n^{-1/4}$)} & -- & -- & $(\$146, \$915)$ & $(\$156, \$937)$\\
\\[-2ex]\hline
\multicolumn{5}{c}{\cellcolor{blue!20}$\textbf{Panel 2: }\boldsymbol{\alpha}\boldsymbol{=0.3}$}
\vspace{.1cm}\\
\multicolumn{1}{l}{$\%$ treated} & 0\% & 100\% & 51\% & 33\%\\
\multicolumn{1}{l}{$\widehat{\mathbb{W}}_\alpha(\widehat\pi_{n})$} & \$696 & \$839 & \$918 & \$918\\
\multicolumn{1}{l}{95\% CI (normal)} & $(\$586, \$806)$ & $(\$729, \$949)$ & $(\$794, \$1{,}042)$ & $(\$777, \$1{,}059)$\\
\multicolumn{1}{l}{95\% CI ($\epsilon=n^{-1/4}$)} & -- & -- & $(\$458, \$1{,}378)$ & $(\$507, \$1{,}329)$\\
\\[-2ex]\hline
\multicolumn{5}{c}{\cellcolor{blue!20}$\textbf{Panel 3: }\boldsymbol{\alpha}\boldsymbol{=0.4}$}
\vspace{.1cm}\\
\multicolumn{1}{l}{$\%$ treated} & 0\% & 100\% & 82\% & 82\%\\
\multicolumn{1}{l}{$\widehat{\mathbb{W}}_\alpha(\widehat\pi_{n})$} & \$1{,}648 & \$1{,}947 & \$2{,}038 & \$2{,}039\\
\multicolumn{1}{l}{95\% CI (normal)} & $(\$1{,}469, \$1{,}826)$ & $(\$1{,}768, \$2{,}126)$ & $(\$1{,}846, \$2{,}231)$ & $(\$1{,}840, \$2{,}239)$\\
\multicolumn{1}{l}{95\% CI ($\epsilon=n^{-1/4}$)} & -- & -- & $(\$1{,}477, \$2{,}600)$ & $(\$1{,}516, \$2{,}563)$\\
\\[-2ex]\hline
\multicolumn{5}{c}{\cellcolor{blue!20}$\textbf{Panel 4: }\boldsymbol{\alpha}\boldsymbol{=0.5}$}
\vspace{.1cm}\\
\multicolumn{1}{l}{$\%$ treated} & 0\% & 100\% & 83\% & 83\%\\
\multicolumn{1}{l}{$\widehat{\mathbb{W}}_\alpha(\widehat\pi_{n})$} & \$2{,}981 & \$3{,}419 & \$3{,}525 & \$3{,}527\\
\multicolumn{1}{l}{95\% CI (normal)} & $(\$2{,}746, \$3{,}216)$ & $(\$3{,}185, \$3{,}654)$ & $(\$3{,}274, \$3{,}775)$ & $(\$3{,}269, \$3{,}785)$\\
\multicolumn{1}{l}{95\% CI ($\epsilon=n^{-1/4}$)} & -- & -- & $(\$2{,}952, \$4{,}098)$ & $(\$2{,}898, \$4{,}156)$\\
\\[-2ex]\hline
\multicolumn{5}{c}{\cellcolor{blue!20}$\textbf{Panel 5: }\boldsymbol{\alpha}\boldsymbol{=0.8}$}
\vspace{.1cm}\\
\multicolumn{1}{l}{$\%$ treated} & 0\% & 100\% & 87\% & 79\%\\
\multicolumn{1}{l}{$\widehat{\mathbb{W}}_\alpha(\widehat\pi_{n})$} & \$8{,}672 & \$9{,}522 & \$9{,}662 & \$9{,}691\\
\multicolumn{1}{l}{95\% CI (normal)} & $(\$8{,}327, \$9{,}017)$ & $(\$9{,}177, \$9{,}868)$ & $(\$9{,}293, \$10{,}030)$ & $(\$9{,}310, \$10{,}072)$\\
\multicolumn{1}{l}{95\% CI ($\epsilon=n^{-1/4}$)} & -- & -- & $(\$8{,}940, \$10{,}383)$ & $(\$8{,}984, \$10{,}398)$\\
\\[-2ex]\hline
\multicolumn{5}{c}{\cellcolor{blue!20}$\textbf{Panel 6: }\boldsymbol{\alpha}\boldsymbol{=1}$}
\vspace{.1cm}\\
\multicolumn{1}{l}{$\%$ treated} & 0\% & 100\% & 96\% & 95\%\\
\multicolumn{1}{l}{$\widehat{\mathbb{W}}_\alpha(\widehat\pi_{n})$} & \$15{,}241 & \$16{,}520 & \$16{,}673 & \$16{,}678\\
\multicolumn{1}{l}{95\% CI (normal)} & $(\$14{,}815, \$15{,}668)$ & $(\$16{,}093, \$16{,}946)$ & $(\$16{,}227, \$17{,}118)$ & $(\$16{,}233, \$17{,}123)$\\
\multicolumn{1}{l}{95\% CI ($\epsilon=n^{-1/4}$)} & -- & -- & $(\$15{,}755, \$17{,}590)$ & $(\$15{,}776, \$17{,}581)$\\
\\[-2ex]\hline
\end{tabular}
}\\
\caption{Estimated $\mathbb{W}_\alpha(\pi_o)$ for different $\alpha$'s and policy classes that condition on \textit{edu} and \textit{prevearn}. Baseline results for treating none or all of the individuals are shown in the first two columns. The third and fourth rows of each panel report the $95\%$ CI based on normal and uniform inference, respectively. }
\label{Table: JTPA}
\end{table}

\begin{figure}[H]
    \centering
    \subfigure{
        \includegraphics[width=0.48\textwidth]{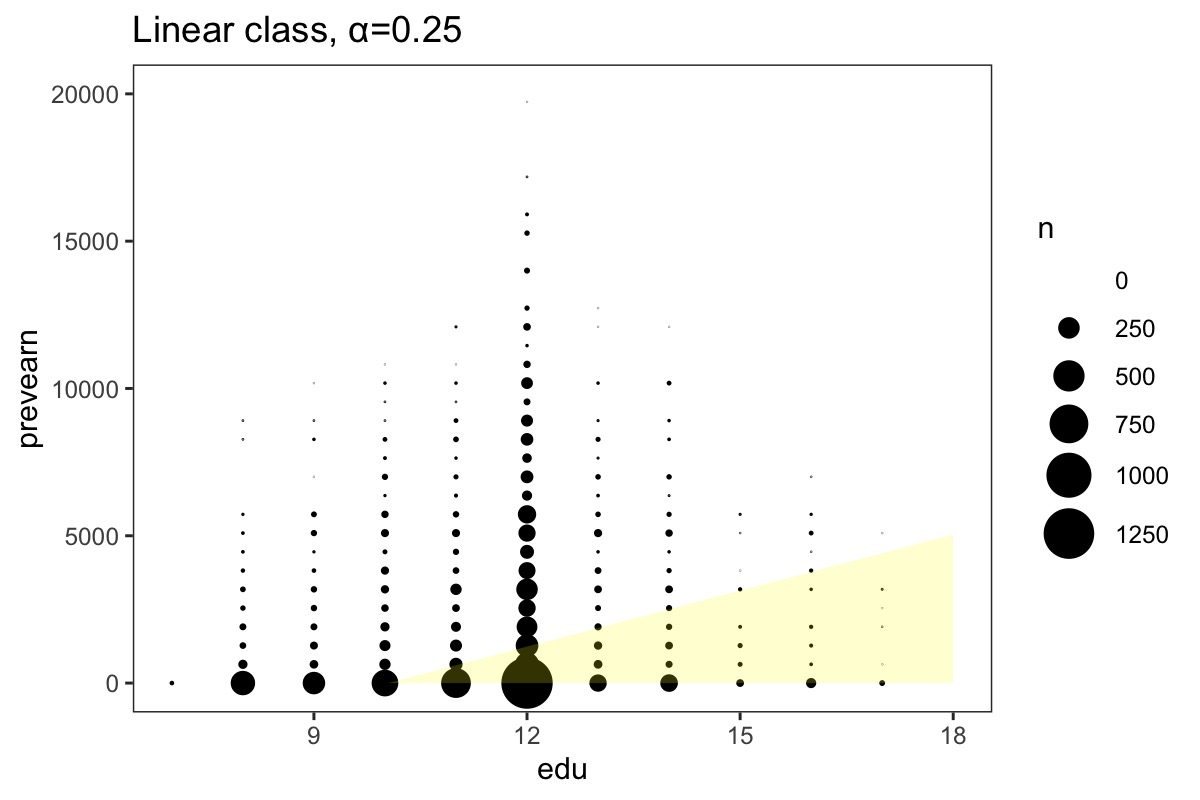}}
    \subfigure{
        \includegraphics[width=0.48\textwidth]{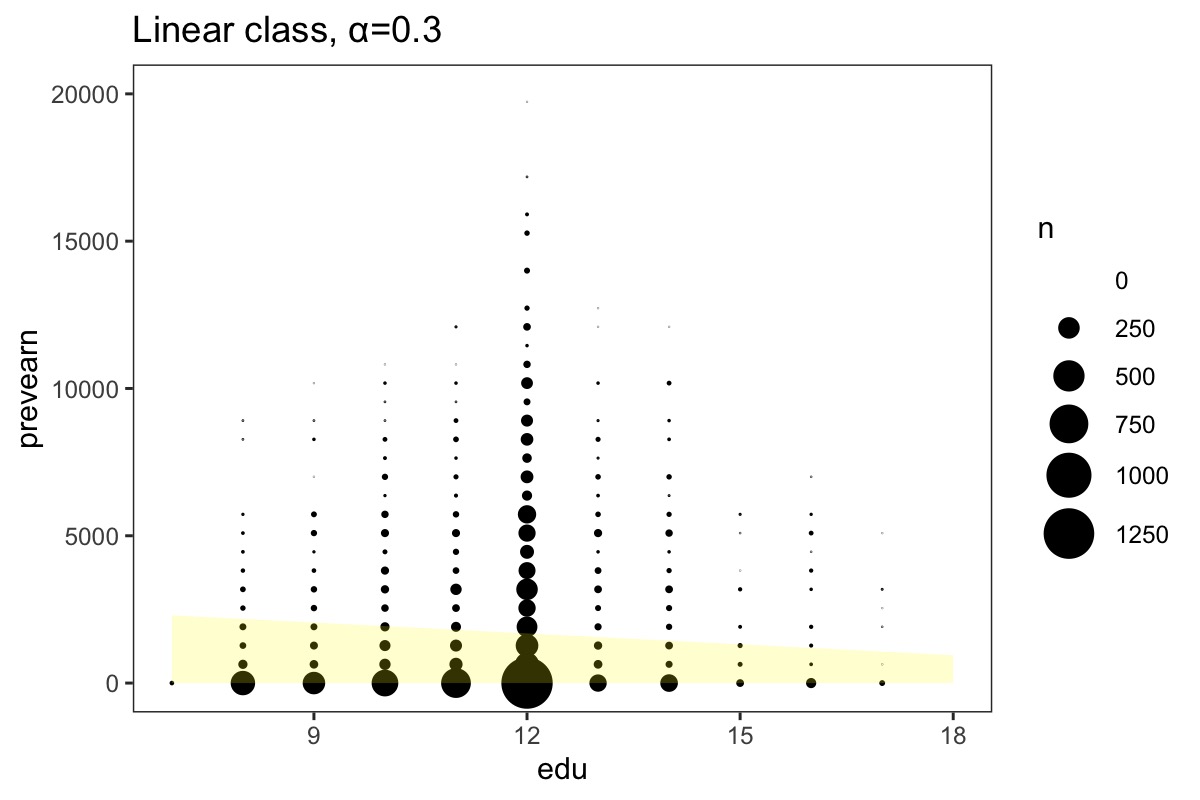}}
    \subfigure{
        \includegraphics[width=0.48\textwidth]{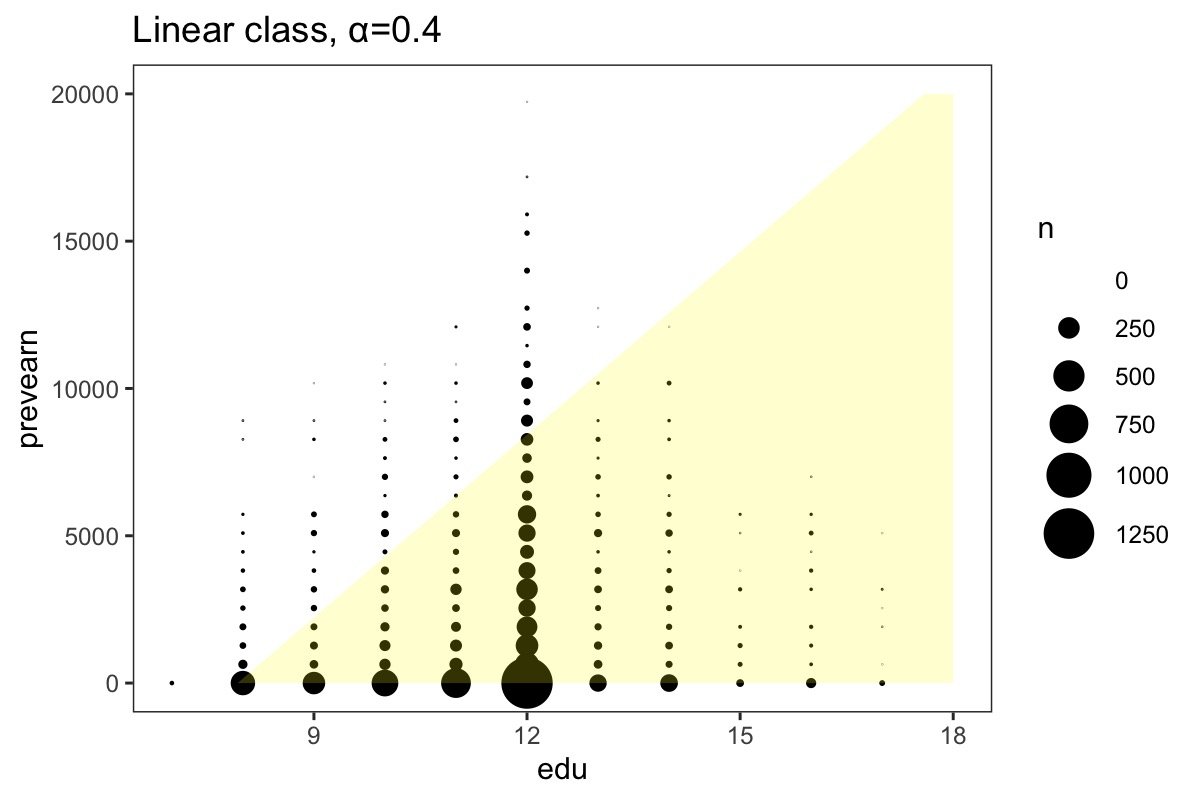}}
    \subfigure{
        \includegraphics[width=0.48\textwidth]{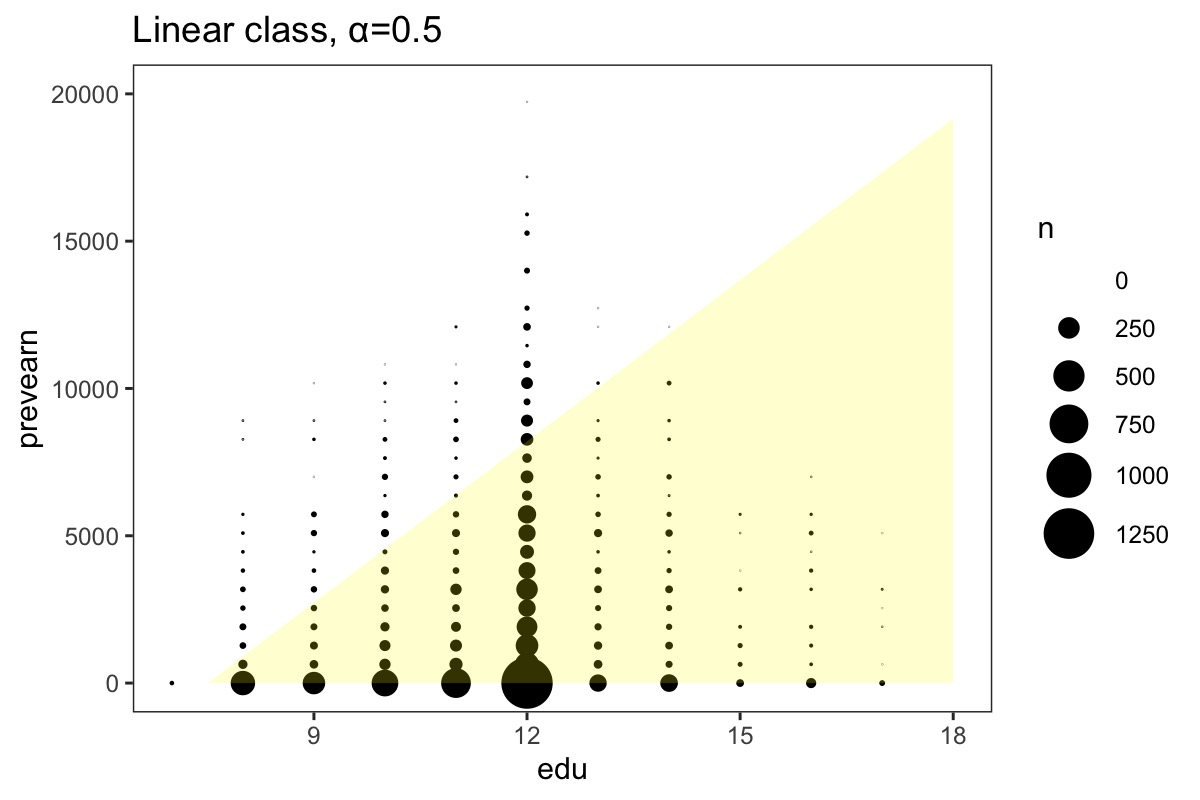}}
        \subfigure{
        \includegraphics[width=0.48\textwidth]{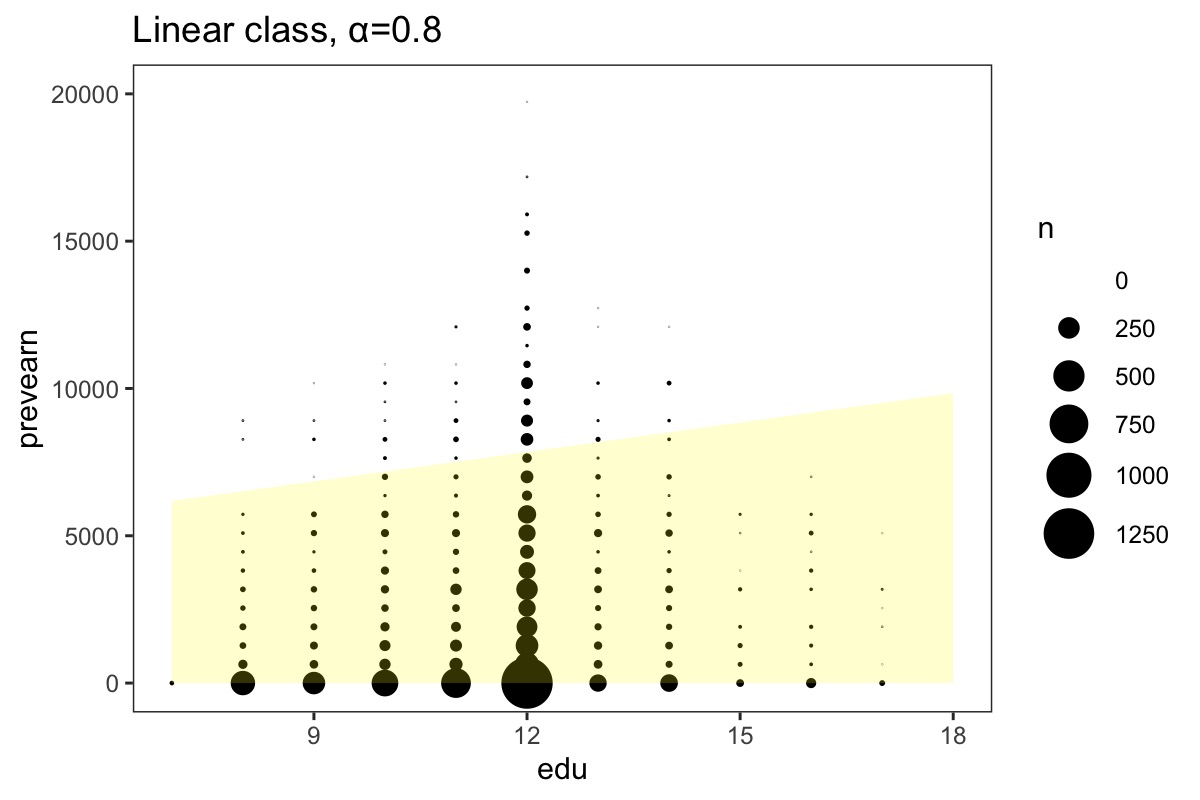}}
        \subfigure{
        \includegraphics[width=0.48\textwidth]{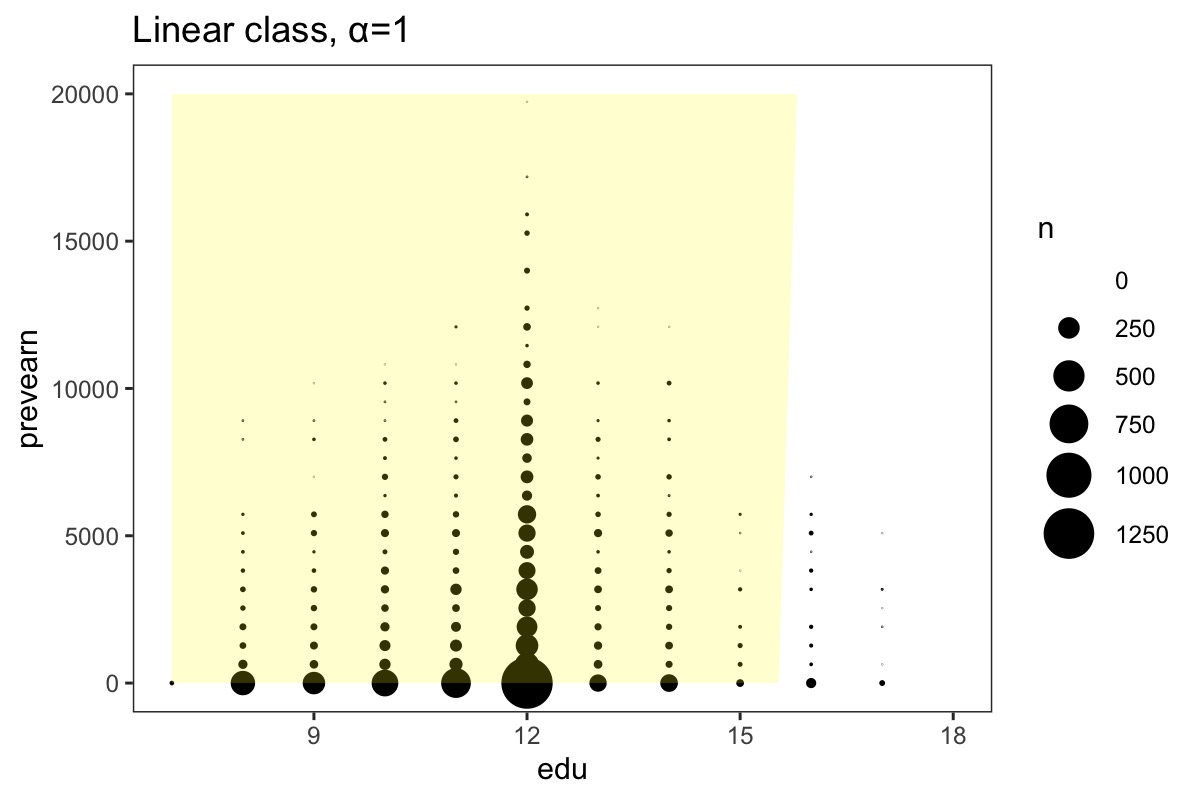}}
        \caption{Optimal policies from the linear class $\Pi_{\mathrm{LES}}$ conditioning on \textit{edu} and \textit{prevearn}. The number of individuals with characteristics closest to each $(edu,prevearn)$ in the grid is represented by the size of the corresponding dot. }
    \label{Figure: JTPA linear}
\end{figure}

\begin{figure}[H]
    \centering
    \subfigure{
        \includegraphics[width=0.48\textwidth]{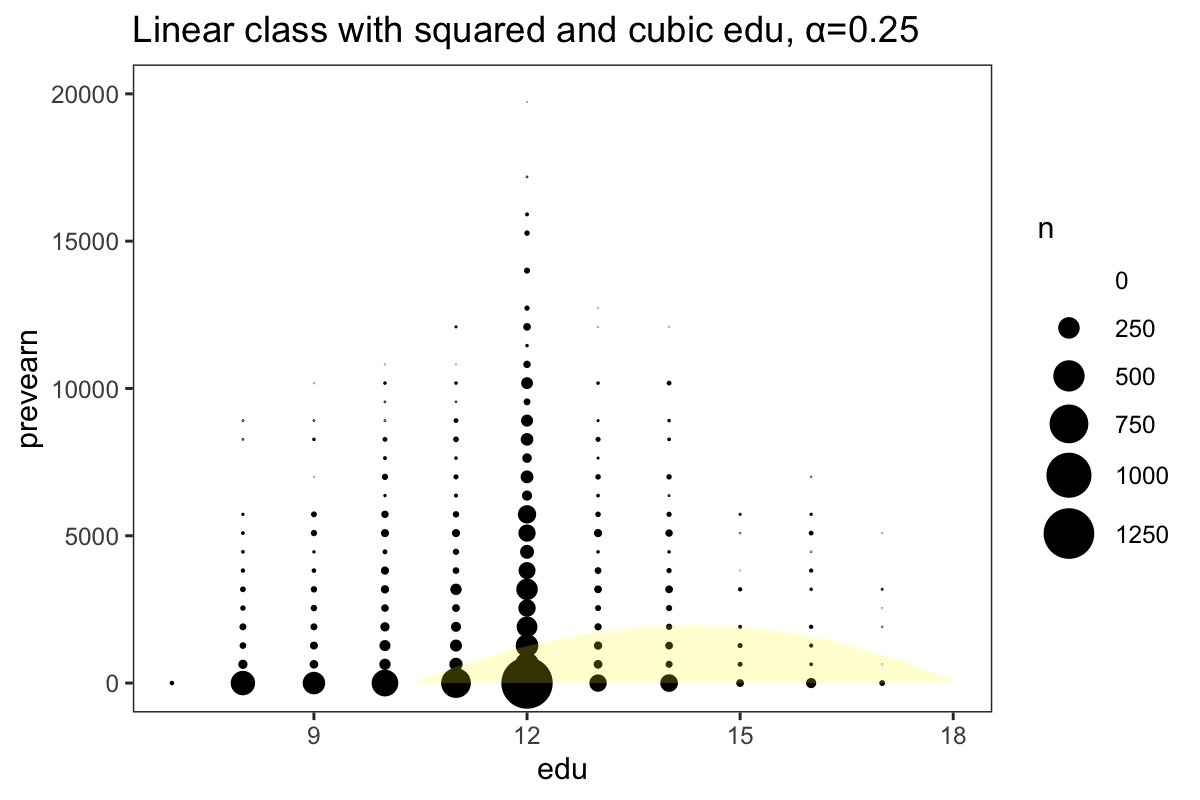}}
    \subfigure{
        \includegraphics[width=0.48\textwidth]{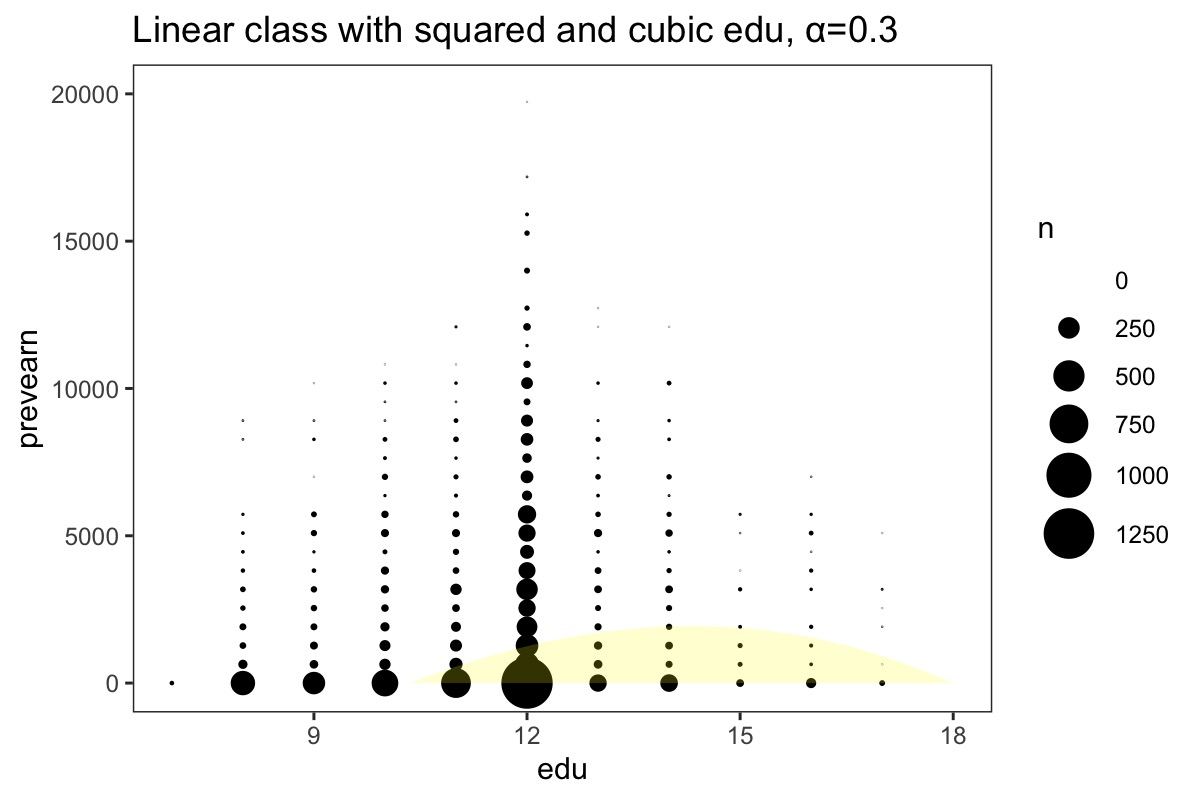}}
    \subfigure{
        \includegraphics[width=0.48\textwidth]{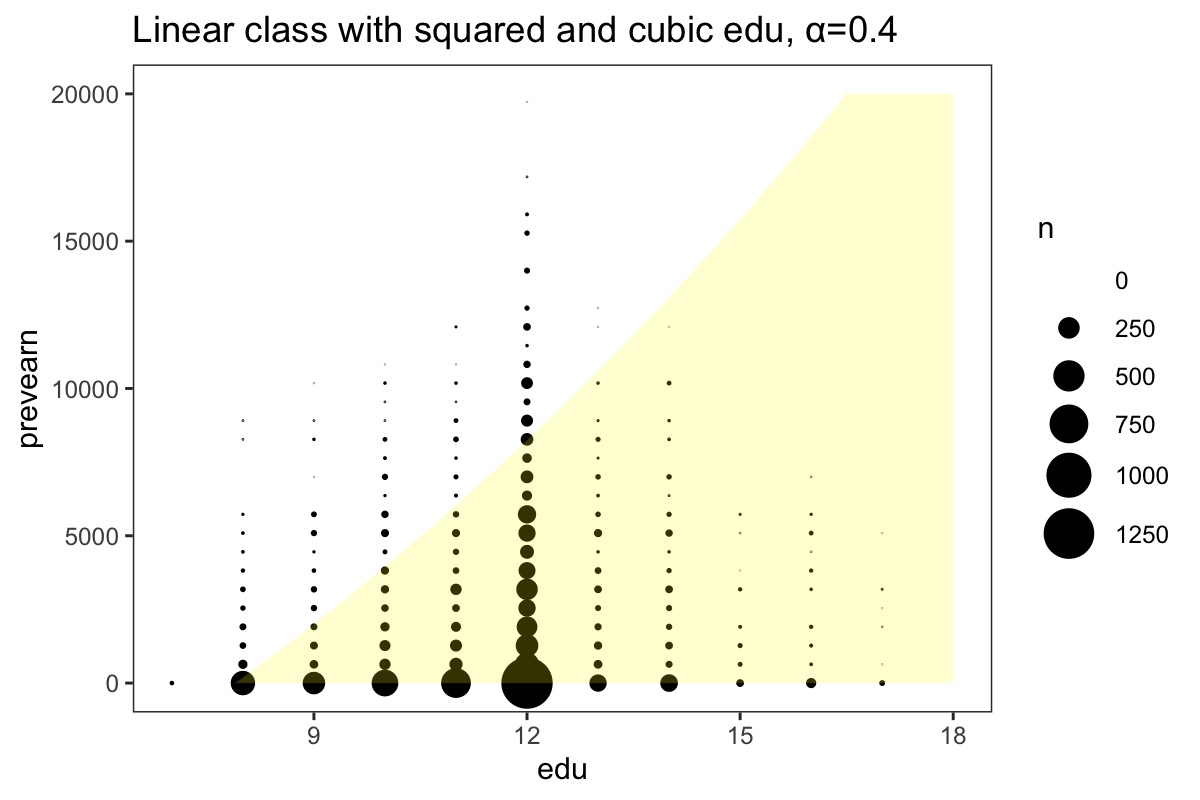}}
    \subfigure{
        \includegraphics[width=0.48\textwidth]{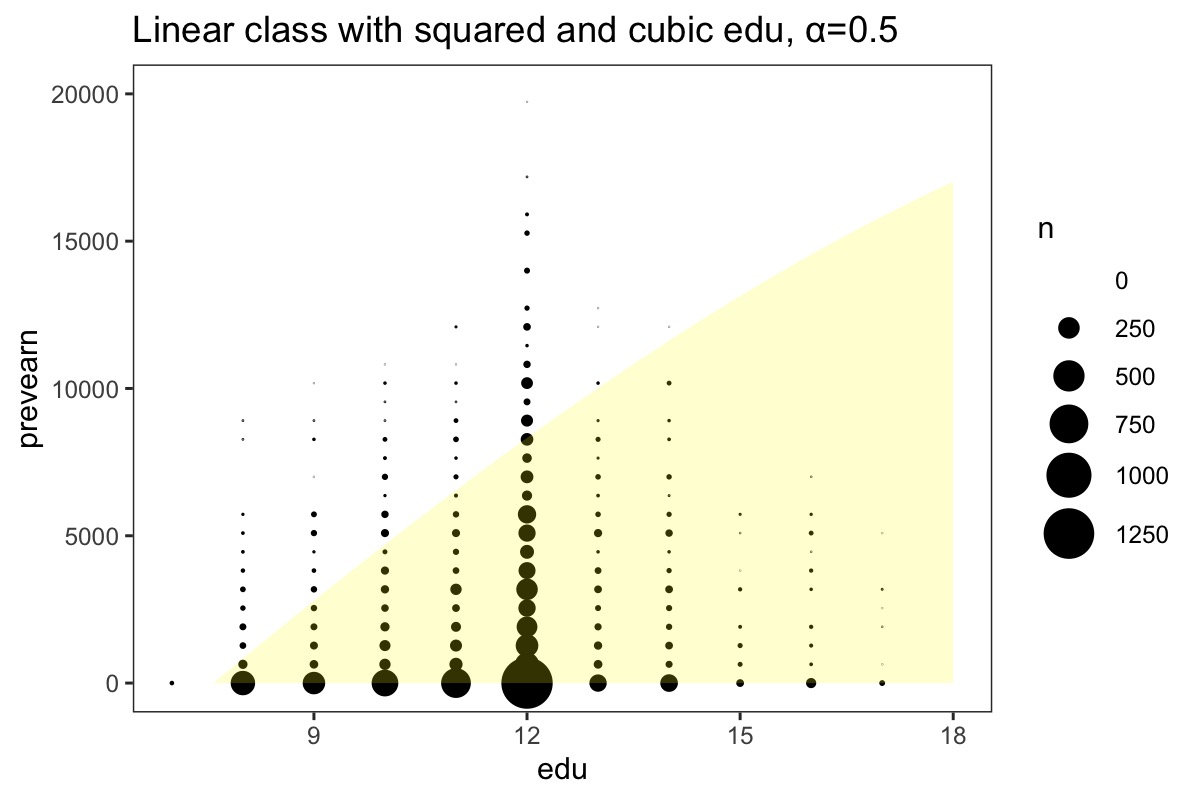}}
        \subfigure{
        \includegraphics[width=0.48\textwidth]{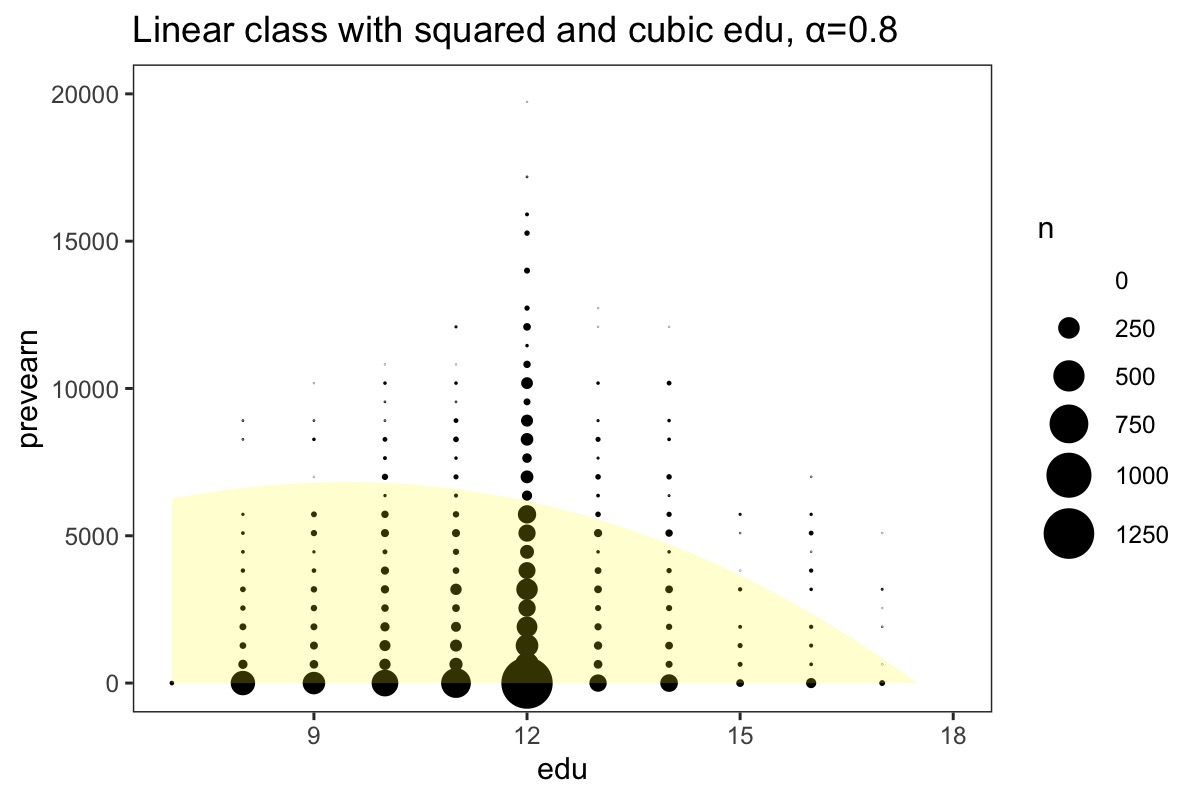}}
        \subfigure{
        \includegraphics[width=0.48\textwidth]{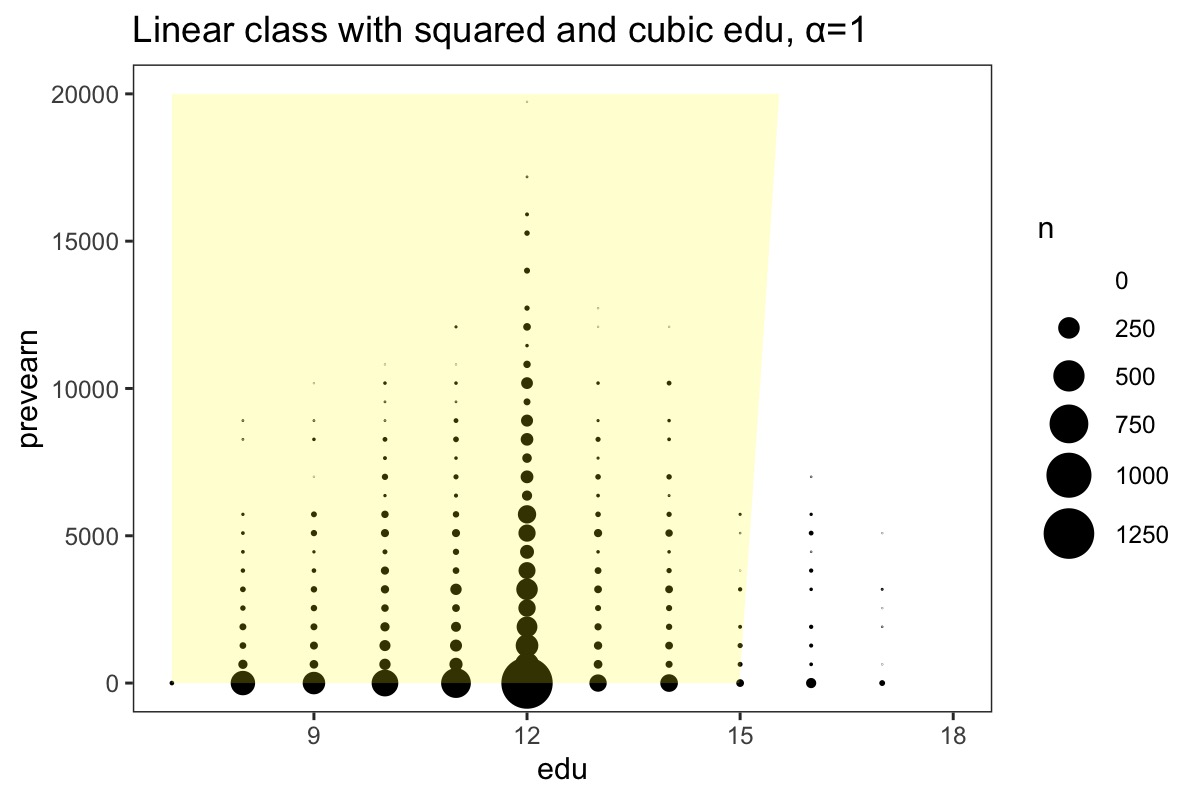}}

    \caption{Optimal policies from the linear class $\Pi_{\mathrm{LES}}^3$ conditioning on \textit{edu}, \textit{prevearn}, $edu^2$, and $edu^3$. The number of individuals with characteristics closest to each $(edu,prevearn)$ in the grid is represented by the size of the corresponding dot. }
    \label{Figure: JTPA cubic}
\end{figure}

Examining the point estimates of welfare, we see that for all $\alpha\in\mathcal{A}$, a simple policy of treating all outperforms treating none. Moreover, relative to treating all, there is a considerable increase in the targeted welfare generated by the optimal policy of class $\Pi_{\mathrm{LES}}$. Linear policies with $edu^2$ and $edu^3$ only bring tiny welfare improvements. Figures \ref{Figure: JTPA linear} and \ref{Figure: JTPA cubic} highlight the optimal treatment regions. Following \cite{kitagawa2018a}, we bin the individuals by $(edu,prevearn)$, and the number of individuals with each combined characteristic is represented by the size of the corresponding dot. When $\alpha = 1$, our setup is comparable in theory to that of \cite{kitagawa2018a}, who also estimate optimal policies within $\Pi_{\mathrm{LES}}$ and $\Pi_{\mathrm{LES}}^3$. Our estimated treatment regions nevertheless differ from theirs, which is not unexpected: \cite{kitagawa2018a} select policies by maximizing a sample analog of the welfare objective, potentially using plug-in estimators of outcome regressions or propensity scores, whereas our approach is based on a doubly robust, orthogonalized welfare criterion implemented via double machine learning. As a result, the corresponding finite-sample optimization problems differ, so the resulting estimated policies (and their implied welfare levels) need not coincide, even under the same policy class.
\vspace{0.5cm}

\vspace{-0.75cm}

Tables \ref{Table: JTPA comparisons linear} and \ref{Table: JTPA comparisons cubic} examine welfare gains and losses as we switch between different targeting policies and estimate the resulting welfare of different targeted subpopulations. For example, the first row in Table \ref{Table: JTPA comparisons linear} shows the estimated welfare of the worst-off $25\%$ of the population when the optimal linear policies are targeted at the worst-off $25\%$, $30\%$, $40\%$, $50\%$, $80\%$, and $100\%$, respectively. The diagonal entries (i.e., the row maxima) are highlighted as these optimal policies are targeted at the actual subpopulations of interest. Tables \ref{Table: JTPA comparisons linear} and \ref{Table: JTPA comparisons cubic} demonstrate a valuable strength of our method, as we are able to conduct rich policy evaluations by estimating the expected welfare at any $\alpha$ for any given policy. In other words, even when a policy is not targeted at the worst-affected $(\alpha\times100)\%$, we can still evaluate its performance at $\alpha$ to obtain a clear picture of the trade-offs, which opens up possibilities for learning policies that promote greater equality across subpopulations.

Based on Tables \ref{Table: JTPA comparisons linear} and \ref{Table: JTPA comparisons cubic}, Tables H.2 and H.3 in the Online Supplement report the percentage welfare loss for every combination of actual $\alpha$ and $\alpha'$ for policy selection. For $\Pi_{\text{LES}}$, adopting the 1-EWM linear policy ($\alpha'=1$) leads to a $14.6\%$ decrease in the average welfare of the worst-affected quarter of the population ($\alpha=0.25$), compared to implementing the optimal linear policy \textit{targeted at} the worst-affected quarter ($\alpha=\alpha'=0.25$). Conversely, adopting the policy targeted at the worst-affected quarter ($\alpha'=0.25$) only leads to a $5.4\%$ decrease in the 1-expected welfare ($\alpha=1$) relative to implementing the 1-EWM policy ($\alpha=\alpha'=1$). Similar patterns emerge with the inclusion of $edu^2$ and $edu^3$ in treatment assignment ($\Pi_{\text{LES}}^3$). For both policy classes, the bottom quarter of the population is particularly vulnerable when the policy targets some $\alpha' \geq 0.4$ instead. Thus, policymakers aspiring for greater equality should prioritize smaller levels of $\alpha$, such as $0.25$ or $0.3$, as evidenced by the small percentage welfare losses in the first two columns of Tables H.2 and H.3 in the Online Supplement, all of which are below $5.5\%$.
\begin{table}[H]
\centering
\fontsize{8}{10}\selectfont
\begin{tabular}{|l||*{6}{c|}}\hline
\backslashbox{$\alpha$ of Interest}{$\alpha'$ for Policy Selection}
&\makebox[2em]{$0.25$}&\makebox[2em]{$0.3$}&\makebox[2em]{$0.4$}&\makebox[2em]{$0.5$}&\makebox[2em]{$0.8$}&\makebox[2em]{$1$}\\\hline\hline

\multirow{2}{*}{$0.25$} & \cellcolor{yellow}531 & 525 & 501 & 495 & 467 & 453\\
& (47) & (48) & (43) & (46) & (41) & (41)\\\hline

\multirow{2}{*}{$0.3$} & 899 & \cellcolor{yellow}918 & 909 & 897 & 862 & 843\\
& (66) & (63) & (63) & (62) & (58) & (58)\\\hline

\multirow{2}{*}{$0.4$} & 1{,}945 & 2{,}021 & \cellcolor{yellow}2{,}038 & 2{,}035 & 1{,}993 & 1{,}964\\
& (109) & (105) & (98) & (100) & (97) & (95)\\\hline

\multirow{2}{*}{$0.5$} & 3{,}331 & 3{,}485 & 3{,}522 & \cellcolor{yellow}3{,}525 & 3{,}493 & 3{,}460\\
& (148) & (141) & (131) & (128) & (132) & (124)\\\hline

\multirow{2}{*}{$0.8$} & 9{,}146 & 9{,}451 & 9{,}552 & 9{,}588 & \cellcolor{yellow}9{,}662 & 9{,}609\\
& (216) & (209) & (186) & (185) & (188) & (179)\\\hline

\multirow{2}{*}{$1$} & 15{,}773 & 16{,}093 & 16{,}232 & 16{,}266 & 16{,}404 & \cellcolor{yellow}16{,}673\\
& (266) & (261) & (230) & (231) & (231) & (227)\\\hline

\end{tabular}\\
\caption{Estimated $\mathbb{W}_\alpha(\pi _o)$ for different actual $\alpha$'s of interest and $\alpha'$'s for linear policy selection (policy class $\Pi_{\text{LES}}$). Standard errors are reported in parentheses, and all values are in USD. }
\label{Table: JTPA comparisons linear}
\end{table}
\vskip 0.2cm
\begin{table}[H]
\centering
\fontsize{8}{10}\selectfont
\begin{tabular}{|l||*{6}{c|}}\hline
\backslashbox{$\alpha$ of Interest}{$\alpha'$ for Policy Selection}
&\makebox[2.25em]{$0.25$}&\makebox[2.25em]{$0.3$}&\makebox[2.25em]{$0.4$}&\makebox[2.25em]{$0.5$}&\makebox[2.25em]{$0.8$}&\makebox[2.25em]{$1$}\\\hline\hline

\multirow{2}{*}{$0.25$} & \cellcolor{yellow}546 & 543 & 504 & 497 & 476 & 455\\
& (51) & (51) & (45) & (45) & (42) & (39)\\\hline

\multirow{2}{*}{$0.3$} & 917 & \cellcolor{yellow}918 & 911 & 897 & 872 & 846\\
& (68) & (72) & (63) & (62) & (62) & (58)\\\hline

\multirow{2}{*}{$0.4$} & 1{,}972 & 1{,}974 & \cellcolor{yellow}2{,}039 & 2{,}036 & 2{,}005 & 1{,}970\\
& (109) & (109) & (102) & (101) & (102) & (94)\\\hline

\multirow{2}{*}{$0.5$} & 3{,}365 & 3{,}369 & 3{,}526 & \cellcolor{yellow}3{,}527 & 3{,}510 & 3{,}469\\
& (144) & (146) & (130) & (132) & (135) & (122)\\\hline

\multirow{2}{*}{$0.8$} & 9{,}198 & 9{,}192 & 9{,}556 & 9{,}598 & \cellcolor{yellow}9{,}691 & 9{,}619\\
& (216) & (215) & (186) & (186) & (194) & (178)\\\hline

\multirow{2}{*}{$1$} & 15{,}816 & 15{,}810 & 16{,}243 & 16{,}284 & 16{,}482 & \cellcolor{yellow}16{,}678\\
& (267) & (267) & (231) & (231) & (244) & (227)\\\hline

\end{tabular}\\
\caption{Estimated $\mathbb{W}_\alpha(\pi _o)$ for different actual $\alpha$'s of interest and $\alpha'$'s for linear policy selection with $edu^2$ and $edu^3$ (policy class $\Pi^3_{\text{LES}}$).  Standard errors are reported in parentheses, and all values are in USD.}
\label{Table: JTPA comparisons cubic}
\end{table}

\subsection{Simulations Based on WGAN-Generated JTPA Data}
\label{Section: WGAN simulations}
We next present simulation results based on a superpopulation generated using Wasserstein Generative Adversarial Networks (WGANs) to evaluate the finite-sample performance of our debiased estimator. We focus on this simulation setup in the main text because the generated data more closely resembles real-world data distributions, making it more illustrative of practical applications. For comparison, we also conduct two additional simulation studies inspired by the DGPs in \cite{athey2021policy}, with adjustments that make the treatment assignment exogenous. Since the results across all three designs are qualitatively similar—our estimator consistently exhibits decreasing mean squared error as the sample size increases, and the coverage rates approach the nominal 95\% level in larger samples—we relegate the latter two studies to Section H.3 of the Online Supplement. 

In all three simulation setups, the propensity scores are assumed to be known, i.e., $\widehat{e}(\cdot) = e(\cdot)$. Cases with unknown propensity scores can be analyzed analogously using an estimator $\widehat{e}(\cdot)$ that satisfies Assumption \ref{assumption: nuisance parameter convergence rate}. Since uniform inference based on the multiplier bootstrap is computationally intensive, we report only the coverage rates based on confidence intervals constructed via Wald inference. We examine $\alpha\in\{0.25, 0.3, 0.4, 0.5, 0.8\}$.

We employ WGANs developed by \cite{athey2024using} to construct a hypothetical superpopulation, referred to as WGAN-JTPA, consisting of one million observations based on the JTPA data in Section \ref{Section: JTPA}. As mentioned by \cite{athey2024using}, a benefit of using WGAN-generated data for simulations is that this practice largely rules out the possibility for researchers to choose particular DGPs that favor their proposed methods. This subsection demonstrates robust performance of our debiased estimator even when the underlying superpopulation is built from real datasets like the JTPA, which has highly skewed outcome and covariate distributions. Section H.2 of the Online Supplement discusses the training process in more detail and presents some summary statistics. 

While technical details of WGANs can be found in \cite{athey2024using}, we highlight that to build the superpopulation, since we generate $X|A$ followed by $Y|(X,A)$ and apply the same generator on $(X, 1-A)$ to obtain $Y|(X, 1-A)$, both potential outcomes are available for each individual. As a result, we can directly compute the true expected welfare at any $\alpha$ induced by any policy, which is simply a tail average of post-treatment outcomes. For each $\alpha\in\mathcal{A}$, we run SA to find a linear policy $\pi_o\in\Pi_{\mathrm{LES}}$ (as defined in \eqref{JTPA LES policy class}) that maximizes $\mathbb{W}_\alpha(\pi)$ and treat the resulting optimum $\mathbb{W}_\alpha(\pi_o)$ as the population truth. 

\begin{figure}[t]
\centering
   \includegraphics[width=0.9\textwidth]{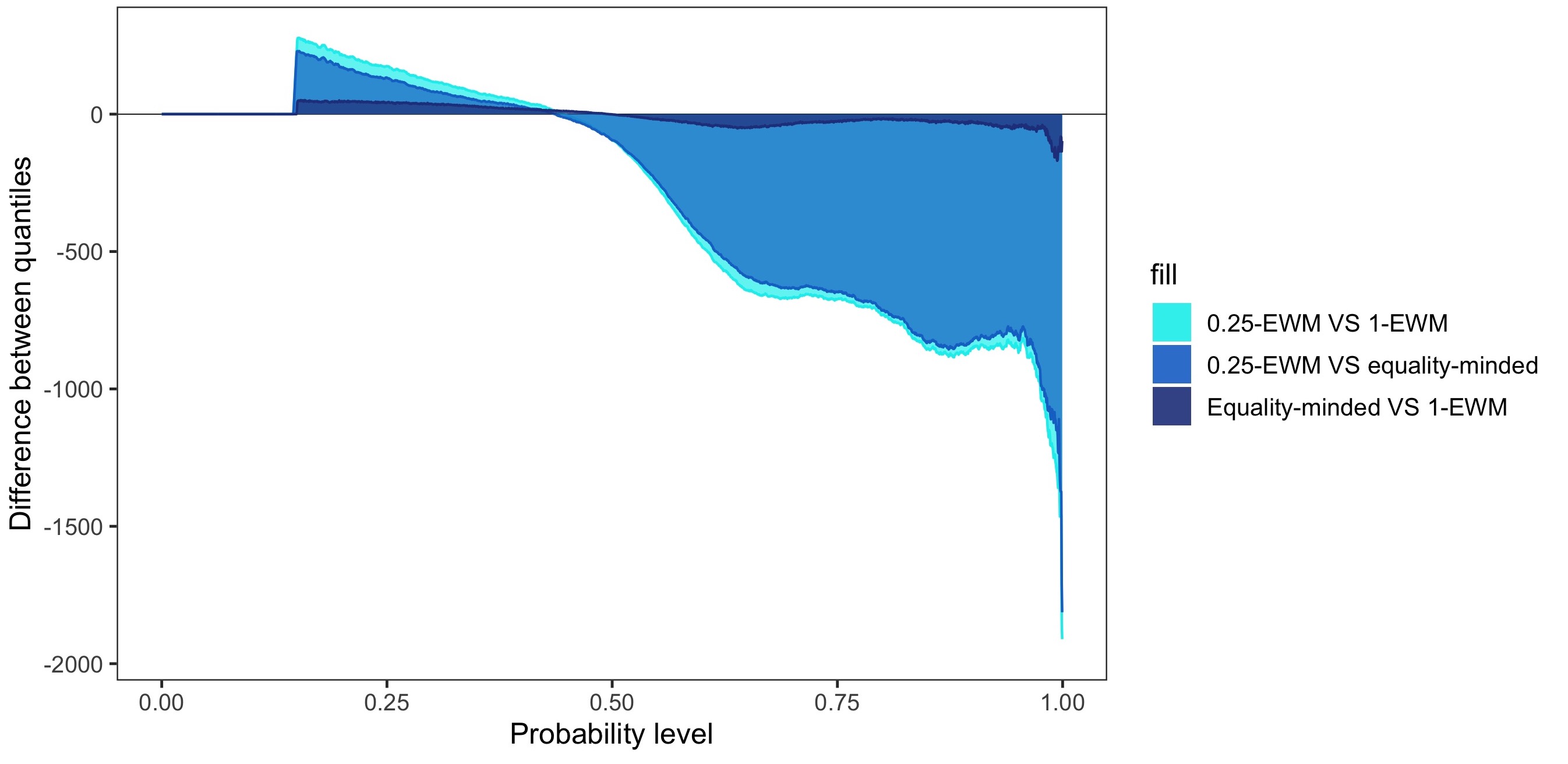}
   \caption{Between-quantile differences in outcomes for the 0.25-EWM, $1$-EWM, and equality-minded policies using the WGAN-JTPA data.}
   \label{Figure: WGAN-JTPA quantile comparisons all}
\end{figure}

As an illustration, we use WGAN-JTPA to compare the $0.25$-EWM policy with the $1$-EWM (mean-optimal) and equality-minded (standard Gini social welfare-optimal) policies. Inspired by Figure~3 in \cite{kitagawa2021equality}, Figure \ref{Figure: WGAN-JTPA quantile comparisons all} plots the between-quantile differences in post-treatment outcomes across these policies. The figure shows that both the $0.25$-EWM and equality-minded policies raise the welfare of lower-ranked individuals while lowering the welfare of higher-ranked individuals relative to the $1$-EWM policy at the population level, with the $0.25$-EWM policy placing much greater emphasis on these adjustments.

In the simulations, for each replicate, we draw a sample of size $n \in \{2{,}000, 5{,}000, 10{,}000\}$ without replacement from WGAN-JTPA. The propensity score is fixed at the population mean of $A$, which is approximately $0.66475$.\footnote{This is very close to the mean of $A$ in the actual JTPA data, $0.66497$. In the JTPA Study, treatment was randomized with probability $2/3$, and we assume randomized treatment in WGAN-JTPA as well.} For each $(n,\alpha)$, we conduct 1,000 Monte
Carlo replications using the estimation procedure described in
\cref{section: welfare_Debiased_Estimator} and Wald-type CIs in \cref{section: Wald CI}. The results are reported
in Table~\ref{Table: WGAN-JTPA}. As shown by the marginal histogram for \textit{earnings} in Figure H.1 in the Online Supplement, WGAN-JTPA inherits the high skewness present in the original JTPA data. Consequently, larger sample sizes are required to achieve satisfactory coverage. From Table \ref{Table: WGAN-JTPA}, our optimal welfare estimator achieves acceptable coverage when $n = 5{,}000$, which is a realistic sample size for both experimental and observational studies (for reference, the original JTPA sample used by \cite{kitagawa2018a} contains 9,223 observations).

\begin{table}[t] \centering
  \footnotesize{
\begin{tabular}
{@{\extracolsep{25pt}}lccc} \multicolumn{1}{l}{\textbf{Sample size}} & \textbf{2,000} & \textbf{5,000} & \textbf{10,000}\\
\\[-1.5ex]\hline
\multicolumn{4}{c}
{\cellcolor{blue!20}$\textbf{Panel 1: }\boldsymbol{\alpha}\boldsymbol{=0.25}, \textbf{truth}\boldsymbol{=1{,}119}$}
\vspace{.1cm}\\
\multicolumn{1}{l}{Avg. \% treated using $\widehat{\pi}_{n}$} & 53\% & 55\% & 55\%\\
\multicolumn{1}{l}{Bias} & 192 & 82 & 44\\
\multicolumn{1}{l}{Variance} & 46,486 & 18,779 & 9,050\\
\multicolumn{1}{l}{MSE} & 83,270 & 25,505 & 10,947\\
\multicolumn{1}{l}{95\% Coverage} & 93.1\% & 93.7\% & 94.9\%\\
\\[-2ex]\hline

\multicolumn{4}{c}
{\cellcolor{blue!20}$\textbf{Panel 2: }\boldsymbol{\alpha}\boldsymbol{=0.3}, \textbf{truth}\boldsymbol{=1{,}908}$}
\vspace{.1cm}\\
\multicolumn{1}{l}{Avg. \% treated using $\widehat{\pi}_{n}$} & 54\% & 54\% & 56\%\\
\multicolumn{1}{l}{Bias} & 207 & 96 & 49\\
\multicolumn{1}{l}{Variance} & 55,138 & 22,734 & 11,799\\
\multicolumn{1}{l}{MSE} & 97,934 & 32,002 & 14,166\\
\multicolumn{1}{l}{95\% Coverage} & 92.0\% & 94.3\% & 94.8\%\\
\\[-2ex]\hline

\multicolumn{4}{c}{\cellcolor{blue!20}$\textbf{Panel 3: }\boldsymbol{\alpha}\boldsymbol{=0.4}, \textbf{truth}\boldsymbol{=3{,}461}$}
\vspace{.1cm}\\
\multicolumn{1}{l}{Avg. \% treated using $\widehat{\pi}_{n}$} & 56\% & 57\% & 58\%\\
\multicolumn{1}{l}{Bias} & 229 & 101 & 48\\
\multicolumn{1}{l}{Variance} & 59,134 & 24,046 & 13,456\\
\multicolumn{1}{l}{MSE} & 111,699 & 34,292 & 15,763\\
\multicolumn{1}{l}{95\% Coverage} & 91.8\% & 93.9\% & 94.3\%\\
\\[-2ex]\hline

\multicolumn{4}{c}{\cellcolor{blue!20}$\textbf{Panel 4: }\boldsymbol{\alpha}\boldsymbol{=0.5}, \textbf{truth}\boldsymbol{=4{,}868}$}
\vspace{.1cm}\\
\multicolumn{1}{l}{Avg. \% treated using $\widehat{\pi}_{n}$} & 58\% & 60\% & 62\%\\
\multicolumn{1}{l}{Bias} & 205 & 96 & 50\\
\multicolumn{1}{l}{Variance} & 58,786 & 22,458 & 12,335\\
\multicolumn{1}{l}{MSE} & 100,721 & 31,742 & 14,810\\
\multicolumn{1}{l}{95\% Coverage} & 92.3\% & 94.4\% & 95.3\%\\
\\[-2ex]\hline

\multicolumn{4}{c}
{\cellcolor{blue!20}$\textbf{Panel 5: }\boldsymbol{\alpha}\boldsymbol{=0.8}, \textbf{truth}\boldsymbol{=9{,}475}$}
\vspace{.1cm}\\
\multicolumn{1}{l}{Avg. \% treated using $\widehat{\pi}_{n}$} & 76\% & 83\% & 88\%\\
\multicolumn{1}{l}{Bias} & 211 & 93 & 53\\
\multicolumn{1}{l}{Variance} & 80,007 & 33,275 & 16,695\\
\multicolumn{1}{l}{MSE} & 124,481 & 41,873 & 19,453\\
\multicolumn{1}{l}{95\% Coverage} & 93.5\% & 93.9\% & 95.3\%\\
\\[-2ex]\hline
\end{tabular}}
\caption{Simulation results based on WGAN-JTPA data (1,000 replications). All quantities inherit the units of USD from the empirical data; units are omitted for brevity. }
\label{Table: WGAN-JTPA}
\end{table}

\section{Concluding Remarks}
\label{Section: conclusion}

The $\alpha$-expected welfare function considered in this paper offers a flexible interpolation between the Rawlsian welfare ($\alpha\rightarrow 0$) and the empirical welfare maximization ($\alpha=1$) approach proposed by \cite{kitagawa2018a}. 
Like \cite{athey2021policy} for the empirical welfare maximization, our development of the doubly robust scores facilitates asymptotic inference for the optimal welfare and allows practitioners flexibility in how they estimate the nuisance parameters. Besides learning the optimal policies, our estimation strategy also enables more thorough policy evaluations by computing the average welfare of the worst-affected subpopulation of any size (fraction of the population). In addition to establishing regret bounds for the debiased estimator, we also develop inference for the optimal $\alpha$-expected welfare for any $\alpha \in (0,1)$. Results from extensive numerical studies based on both JTPA data and simulated data demonstrate the efficacy and practical value of policy learning through $\alpha$-EWM.

As pointed by an anonymous Associate Editor, $\alpha$-EWM objective for $\alpha\in (0,1)$ might come at the expense of the upper $(1-\alpha)$ quantile of the population and one approach to mitigate this is to solve the following constrained maximization problem: 
\begin{equation}\label{eq:Con_Max}
\max_{\pi \in \Pi_n} \left\{ \mathbb{W}_\alpha(\pi):   F_\pi(y) \leq F_0(y), \ \forall y \in \mathcal{Y} \right\},
\end{equation}
where the first-order stochastic dominance constraint between $F_{\pi}$ and $F_0$ is imposed. A different, but related question raised by an anonymous reviewer concerns the choice of an optimal policy when it is not unique. Inspired by the reviewer's comment, we recommend selecting the one that maximizes the expected welfare in this case. 
We will leave a thorough treatment of both in future work.

Several other extensions of this paper are worth investigating. Methodologically, it is important to develop statistical tests to compare whether one policy is superior to another. 
Practically, it would be beneficial to determine who is actually targeted by the optimal policy. For example, what characteristics do the worst-affected individuals have? Information like this could present a more comprehensive picture of the relevant population and promote the design of more equitable policies. 
\appendix

\section*{Appendix}

\section{First Best $\alpha$-EWM Policy}\label{section: First_best_policy}

It is insightful to compare the first-best (FB) policies based on expected welfare function and the AVaR welfare function for $\alpha\in (0,1)$. In EWM, the FB policy is 
\[
\mathds{1} \left\{ x \in \mathcal{X}:   \tau(x)> 0  \right\},
\]
where $\tau(x) =\mathbb{E}\left[Y_i(1) - Y_i(0) \mid X_i = x \right]$ is the CATE. We now provide a similar representation of the FB policy in our set-up.  The FB (optimal) policy is defined as 
\[ 
\pi_{\mathrm{FB}}^*\in\argmax_{ \pi \in \Pi_o  }\mathbb{W}_\alpha(\pi).
\] 
We assume the existence of $\pi_{\mathrm{FB}}^*$, which maximizes the average welfare of the size-$\alpha$ lowest-ranked subpopulation. 

Recall $\mu_1$, $\mu_0$, and $\tau$ defined in \cref{Section: debiased}. Under Assumption \ref{Assumption: Selection-on-observables}, for any given $\eta$, $\tau(x, \eta)$ is identified. Moreover, let $\chi_1 (\eta ) \equiv \mathbb{E}\left[ \mu_1(X_i, \eta) \mathds{1}\{ \tau(X_i, \eta) \geq 0 \} \right] $  and $\chi_0 (\eta ) \equiv \mathbb{E}\left[ \mu_0(X_i, \eta)  \mathds{1}\{ \tau(X_i, \eta) < 0 \} \right]$.   

\begin{lemma}\label{lemma: first best policy}
Suppose the functions $\chi_0(\cdot)$ and $\chi_1(\cdot)$ are continuous. Then, for each $\alpha \in (0,1]$, there is a constant $\eta^*_{\mathrm{FB}}$ depending on $\alpha$ such that the policy given by 
\[
\pi_{\mathrm{FB}}^*(x) = \mathds{1}\left\{ \tau\left(x,\eta^*_{\mathrm{FB}}  \right) >  0 \right \},
\]
maximizes $\mathbb{W}_\alpha(\pi)$ over $\pi \in \Pi_o$.
\end{lemma}

\begin{proof}
From Lemma \ref{lemma: AVaR dual} and  \cref{remark: compact feasible set}, it follows that
\begin{align}
\mathbb{W}_\alpha(\pi) = \mathrm{AVaR}_\alpha(Y_i(\pi))=\sup_{\eta \in \mathcal{B}_Y }\left\{\frac{1}{\alpha}\mathbb{E}\left[(Y_i(\pi)-\eta)_-\right]+\eta\right\}.
\end{align}
 Hence,
\[
\begin{aligned}
\sup_{\pi \in \Pi_o} \mathbb{W}_\alpha(\pi) & =  \sup_{\pi \in \Pi_o} \sup _{\eta \in \mathcal{B}_Y  }\left\{\frac{1}{\alpha} \mathbb{E}\left[\left(Y_i(\pi)-\eta\right)_{-}\right]+\eta\right\}  \\
& =  \sup _{\eta \in \mathcal{B}_Y  }   \sup_{\pi \in \Pi_o}   \left\{\frac{1}{\alpha} \mathbb{E}\left[\left(Y_i(\pi)-\eta\right)_{-}\right]+\eta\right\} .\\
\end{aligned}
\]
For a fixed $\eta \in \mathcal{B}_Y$, consider the following maximization:
\[
\sup_{\pi \in \Pi_o }   \frac{1}{\alpha} \mathbb{E}\left[ \left( Y_i(\pi) - \eta \right)_{-} \right] + \eta.
\]
An optimal solution to this problem is given by $\pi^*_\eta(x) = \mathds{1}\left\{ \tau(x,\eta) \geq  0 \right \}$. As a result, 
\[
\begin{aligned}
\sup_{\pi \in \Pi_o} \mathbb{W}_\alpha(\pi) & =  \sup_{\eta \in \mathcal{B}_Y}\left\{   \frac{1}{\alpha} \mathbb{E}\left[ \left( Y_i(\pi^*_\eta) - \eta \right)_{-} \right] + \eta \right\} \\
& = \sup_{\eta \in \mathcal{B}_Y  }\left\{   \frac{1}{\alpha} \mathbb{E}\left[ \left( Y_i(1) - \eta \right)_{-} \mathds{1}\{ \tau(X_i, \eta) \geq 0 \} + \left( Y_i(0) - \eta \right)_{-} \mathds{1}\{ \tau(X_i, \eta) < 0 \} \right] + \eta \right\} \\
& = \sup_{\eta \in \mathcal{B}_Y}\left\{   \frac{1}{\alpha} \mathbb{E}\left[ \mu_1(X_i, \eta) \mathds{1}\{ \tau(X_i, \eta) \geq 0 \} + \mu_0(X_i, \eta)  \mathds{1}\{ \tau(X_i, \eta) < 0 \} \right] + \eta \right\}.
\end{aligned}
\]
Since $\chi_0$ and $\chi_1$ are continuous, the function $ \frac{1}{\alpha} \mathbb{E}\left[ \left( Y_i(\pi^*_\eta) - \eta \right)_{-} \right] + \eta$ is also continuous
in $\eta$. Consequently, it attains its maximum at $\eta_{\mathrm{FB}}^*$ over the compact set $\mathcal{B}_Y$. Therefore, we conclude that $\mathds{1}\left\{ \tau (x,\eta^*_{\mathrm{FB}} ) \geq  0 \right \}$ is the FB policy.
\end{proof}

When $\alpha=1$, our FB policy $\pi_{\mathrm{FB}}^*(\cdot)$ reduces to $\mathds{1}\{\tau(x) \geq 0\}$, the FB policy under EWM. When $\alpha\in (0,1)$, $\pi_{\mathrm{FB}}^*(\cdot)$ depends on the distribution of post-treatment outcomes through the optimal cutoff $\eta^*_{\mathrm{FB}}$.  

\section{Improved Rate Under the Margin Assumption}
\label{section: improved rate}
In this section, we demonstrate that the asymptotic regret bound presented in \cref{theorem: regret bound with semiparametric efficiency score} can be further tightened under the margin assumption, a commonly adopted condition in the statistical learning literature. Throughout this subsection, we continue to uphold Assumption \ref{assumption: assumption for inference}.

\subsection{Curvature or Margin Assumption}
Since $\theta_o$ is the unique maximizer of $\mathbb{V}(\theta)$, the first order derivative of $\mathbb{V}(\theta)$ should vanish at $\theta_o$ and the second-order derivative should be negative definite. Motivated by this intuition, we introduce the following curvature (or margin) assumption.

\begin{assumption}[Curvature]
\label{assumption: V_theta smoothness_2}
Suppose there exist constants $\rho_o \geq 1$ and  $c_o >0$ such that for every $\theta$ in some nonempty neighborhood of $\theta_o$, the following inequality holds:
\[
\mathbb{V}(\theta_o) - \mathbb{V}(\theta)   \geq  c_o  \left\| \theta - \theta_o \right\|^{\rho_o}.
\]
\end{assumption}

Let $\check{\theta}_{n}$ denote a maximizer of the function $\mathbb{V}_{n} (\theta) = \mathbb{P}_n g_\theta$. Assumption \ref{assumption: V_theta smoothness_2} plays a pivotal role in establishing the convergence rate of $\check{\theta}_{n}$ as well as establishing the oracle regret bound, i.e., the convergence rate of $\mathbb{V}(\theta_o)-\mathbb{V}(\check{\theta}_{n} )$. The parameter 
$\rho_o$ is commonly referred to as the margin parameter in the statistical learning literature (see \cite{tsybakov2004optimal, scholkopf2002learning}). From this perspective, Assumption \ref{assumption: V_theta smoothness_2} serves as an analogue to the restricted eigenvalue condition in the Lasso framework. Let $\boldsymbol{X}$ be the design matrix, and let $\widehat{\beta}$ denote the Lasso estimator for $\beta$. The margin assumption helps to establish the relationship between the prediction error $\big \| \boldsymbol{X}^\prime( \widehat{\beta} - \beta) \big \|$ and the estimation error $\| \widehat{\beta} - \beta \|$.   In the context of Lasso estimation, $\rho_o$ is set to be one, whereas $\rho_o= 2$ in classical M-estimation theory (see \cite{vaart1998empirical, kosorok2008introduction}).

In the following example, we verify Assumption \ref{assumption: V_theta smoothness_2} for the linear rules introduced in \cref{example: Linear Rules}.

\begin{example}[Linear Rules]
We consider the policy class $\Pi = \left \{ \pi_\beta = \mathds{1}\{x^\prime \beta > 0\} : \| \beta \| = 1 \right \}$.\footnote{The restriction $\|\beta \| = 1$ ensures that if  $\beta_1 \neq \beta_2$ with $\|\beta_1 \| =\|\beta_2 \| = 1$ then $\pi_{\beta_1} \neq \pi_{\beta_2}$. } 
To verify Assumption \ref{assumption: V_theta smoothness_2}, we apply the primitive conditions stated in Assumption \ref{assumption: g_smoothness}.  With a slight abuse of notation, we write $\theta = (\beta, \eta)$ and $g_\theta = g_{(\pi_\beta, \eta)}$, and let $\theta_o = (\beta_o, \eta_o)$ denote the maximizer of the function $(\beta,\eta) \mapsto \mathbb{V}(\pi_\beta, \eta)$.

\begin{assumption}[Curvature Assumption for Linear Rules]
\label{assumption: g_smoothness}
\begin{enumerate}
    \item[(1)]  The function $\theta  \mapsto \mathbb{V}(\theta)$ is twice continuously differentiable in a neighborhood of $\theta_o$ with a negative definite Hessian matrix $\nabla_\theta^2 \mathbb{V}(\theta)$ evaluated at $\theta = \theta_o$. 
    \item[(2)] {\it Margin Assumption}:  There are $t_* >0$ and $\rho \geq 1$ such that  $P ( 0 < | X_i^\prime \beta_o  | \leq  t ) \lesssim t^{\rho}$ for all  $t \in (0, t_*)$.
    \item[(3)] The support $\mathcal{X}$ of $X_i$ is bounded.
\end{enumerate}
\end{assumption} 

Assumption \ref{assumption: g_smoothness} (1) is a standard assumption in parametric M-estimation (see \cite{vaart1998empirical, van2000asymptotic, kosorok2008introduction, kim1990cube, shi2018massive}). In contrast, Assumption \ref{assumption: g_smoothness} (2) is widely used in statistical and policy learning, as noted by \cite{kitagawa2018a, luedtke2020performance, tsybakov2004optimal, zhao2023semiparametric}. It is straightforward to see that Assumption \ref{assumption: g_smoothness} (2) and (3) together imply  $\left\| \pi_\beta - \pi_{\beta_o} \right \|_{L^2(P)} \lesssim \| \beta - \beta_o \| ^{\rho / 2}$. As a result, Assumption \ref{assumption: g_smoothness} provides the necessary conditions to verify Assumption \ref{assumption: V_theta smoothness_2} for linear policies, which can be established via a Taylor expansion:
\[
\begin{aligned}
\mathbb{V}(\theta) - \mathbb{V}(\theta_o) & < - c_o \left( \| \beta - \beta_o \|^2  + | \eta - \eta_o|^2 \right) \leq -c_o \left(  \left\| \pi_\beta - \pi_{\beta_o} \right \|_{L^2(P)}^{4/\rho} + |\eta- \eta_o|^{2} \right)\\
& \leq - c_o \left[ \left\| \pi_\beta - \pi_{\beta_o} \right \|_{L^2(P)}+ |\eta- \eta_o|  \right]^{\left(4/\rho\right) \wedge 2 } ,
\end{aligned}
\]
where $c_o >0$ is a constant  not depending on $\theta = (\beta, \eta)$. 
\end{example}

\subsection{Faster Rate}
In this subsection, we derive a sharper oracle regret bound than the one presented in \cref{theorem: regret bound with semiparametric efficiency score}. For illustrative purposes, this subsection focuses on the oracle regret bound based on the true influence scores $g_\theta(\cdot)$.\footnote{Based on Eq. (D.2) in Section D of the Online Supplement, the regret bound can be upper-bounded by the sum of the oracle regret bound combined with the nuisance parameter estimation error or the uniform coupling error, as established in \cref{Lemma: M_error}.   }
Fundamentally, the convergence rate of the regret $\mathbb{V}(\theta_o) - \mathbb{V}(\check{\theta}_n)$ is largely determined by the modulus of continuity of the empirical process $\sqrt{n} (\mathbb{P}_n - P) g_{\theta}$, indexed by $\theta$. This can be effectively controlled using maximal inequalities under uniform entropy conditions, see \cite{vaart1998empirical,van2011local,chernozhukov2014gaussian}. 

To establish the improved oracle regret bound rate under Assumption \ref{Assumption: V_theta smoothness (1)}, we introduce the following technical assumption. This helps circumvent measurability issues and enables the use of Talagrand's inequality to control the local empirical process effectively.

\begin{assumption}\label{Assumption: countable approximation}
There is a countable subset $\Theta^\prime$ of $\Theta$ satisfying that for any  $\theta \in \Theta$, there is a sequence  $(\theta_k)_{k=1}^\infty$
in $\Theta^\prime$ such that 
 $\lim_{k \rightarrow\infty}g_{\theta_k}(z)  = g_{\theta}(z)$ for $P$-a.s. $z \in \mathcal{Z}$.
\end{assumption} 

\begin{theorem}
Suppose that Assumptions \ref{assumption: nuisance parameter convergence rate}, \ref{Assumption: Bounded support},  \ref{Assumption: V_theta smoothness (1)}, \ref{assumption: V_theta smoothness_2}, and \ref{Assumption: countable approximation} hold. If $\tau(x, \eta)$ is uniformly bounded, i.e., $\sup_{x, \eta}\left|\tau(x, \eta) \right| < \infty$,  then there is a universal constant $c_o >0$ not depending on $n$ such that
\[
\mathbb{E}_{P} \left[  \mathrm{Reg}( \check{\theta}_{n}  ) \right] \leq c_o  \left( 
 \mathrm{VC}(\Pi)/ n \right)^{\rho_o / (2 \rho_o-1) },\quad \forall n \in \mathbb{N}^+.
\]
\end{theorem}

\begin{remark}
Let us analyze the role of the margin parameter $\rho_o$. If we remove the assumption on the margin parameter (i.e., letting $\rho_o\rightarrow \infty$), the regret convergence rate becomes $O(\sqrt{\mathrm{VC}(\Pi)/n})$, identical to the rate in \cref{theorem: regret bound with semiparametric efficiency score}, and independent of $\rho_o$. Notably, the knowledge of the margin parameter $\rho_o$ is not required, as it neither needs to be estimated nor plays a role in constructing the optimal policy.
\end{remark}

\section{Uniform Inference for the Optimal Welfare}
\label{section: inference for the optimal value}

In this section, we apply
the numerical delta method introduced by Hong and Li (2018) to develop inference for the optimal welfare without Assumption \ref{Assumption: V_theta smoothness (1)}. It improves upon the inference proposed in Appendix B of the supplemental material to \cite{kitagawa2018a}. Throughout this section, we assume that $\mathrm{VC}(\Pi)$ is finite, i.e., Assumption \ref{assumption: assumption for inference} is satisfied.

To develop uniform inference, we define a distance 
 $d_{\Pi}$ to measure the dissimilarity between policies in  $\Pi$, independent of the underlying distribution $P$. To do so, let $\nu$ be a finite product measure such that the distribution
of $X$ admits a uniformly bounded density with respect to $\nu$. The semimetric $d_\Pi$ is defined as
\[
d_{\Pi } (\pi, \tilde{\pi} ) = \left[\int_{\mathbb{R}^p} \left|\pi-\tilde{\pi}\right|^2 \mathrm{d} \nu \right]^{1/2}, \quad \forall  \pi, \tilde{\pi} \in \Pi.
\]
 Moreover, we introduce a semimetric $d_{\Theta}$ on $\Theta$, defined by $d_{\Theta} (\theta, \tilde{\theta}) = d_{\Pi} (\pi, \tilde{\pi}) + |\eta - \tilde{\eta}|$ for all $\theta = (\pi, \eta)$ and $\tilde{\theta} = ( \tilde{\pi}, \tilde{\eta})$.  Furthermore, the estimated functions $\widehat{e}\left(\cdot\right)$ and $\widehat{\mu}_a\left(\cdot,\cdot\right)$ need to satisfy Assumption \ref{assumption: nuisance parameter convergence rate} uniformly across a collection of distributions $P \in \mathcal{P}_n$. This, in turn, requires the nonparametric/ML models used to estimate $e_o(\cdot)$ and $\mu_a(\cdot,\cdot)$ to be not excessively complex. To formalize this condition, let $\Delta_n\searrow 0, \delta_n\searrow 0$, and $\tau_n \searrow 0$  be sequences that approach zero from above at a rate no faster than polynomial in $n$ (e.g. $\Delta_n > n^{-c}$ for some $c>0$). Let $\mathcal{M}_{n,a}$ and $\mathcal{D}_n$ denote the classes of measurable functions $\check{\mu}_a, \check{e}$ such that $\|\check{\mu}_a -  \mu_a  \|_{P,2}\leq \tau_n /2$ and $ \| \check{e} - e_o \|_{P,2} \leq \tau_n /2$.  Finally, let
 \[
\mathcal{F}_n = \left\{ g_\theta\left(\cdot; \check{\mu}, \check{e} \right)  :  \theta \in \Theta, \check{\mu}_a\in \mathcal{M}_{n,a},  \check{e} \in \mathcal{D}_n \right\},
\]
where $g_\theta( z; \check{\mu}, \check{e} )$ is defined in \cref{equation: g_theta}. We impose the following regularity conditions.

\begin{assumption}\label{assumption: Estimation of Nuisance Functions}
There exists $n_o \in \mathbb{N}^+$ and  a constant $c_o >0$ such that the following conditions hold for all $n \geq n_0$ and $P \in \mathcal{P}_n$.
\begin{assumpenum}
\item $|Y_i| \leq c_o$ $P$-a.s. and Assumption \ref{Assumption: Selection-on-observables} holds.
\item \label{assumption: covariate density}
 $X \in \mathbb{R}^p$ has density $f_P: \mathcal{X} \rightarrow \mathbb{R}_+$ such that $\| f_P\|_\infty \leq c_o$, with respect to $\nu$.

\item Suppose $\tau_n^2 \sqrt{n} \leq \delta_n$, and the estimated functions  $\widehat{\mu}_a(\cdot,\cdot) \in \mathcal{M}_{n,a}$ and $\widehat{e}\left(\cdot\right) \in \mathcal{D}_n$, with probability at least $1- \Delta_n$.  Let $a_n \geq n \vee e$ and $s_n \geq 1$ be two sequences such that
\[
\begin{aligned}
& n^{-1/2} \left( \sqrt{s_n \log a_n }  + n^{-1/4}  s_n  \log a_n  \right) \leq \tau_n \quad \text{and} \\
& \tau_n^{1/2} \sqrt{s_n \log a_n} + s_n n^{- 1/4} \log a_n \cdot \log n \leq \delta_n.
\end{aligned}
\]
The function class $\mathcal{F}_n$  is suitably measurable and its uniform covering entropy satisfies:
\[
\sup_{Q} \log N\left( \epsilon \| F_1 \|_{Q,2} , \mathcal{F}_n, \| \cdot \|_{Q,2}   \right) \leq  s_n \log \left(a_n/ \epsilon\right) \vee 0,
\]
where $F_1$ is an envelope for $\mathcal{F}_n$ with $\|F_1\|_\infty \leq C$ for all $n$.
\end{assumpenum}
\end{assumption}

Define the supremum functional  $\psi: \ell^{\infty} (\Theta) \rightarrow \mathbb{R}$ as $\psi: h \mapsto \sup_{\theta \in \Theta}  h (\theta)$. We can verify that $\psi$ is Hadamard directionally differentiable at $\mathbb{V}_P$ tangentially to $C_{b}( \Theta)$, which allows the application of generalized delta method, see \cite{belloni2017program,fang2019inference,hong2018numerical}.
 Let $\Theta^\star_P \defeq \arg \max_{\theta \in \Theta} \mathbb{V}_P( \theta)$. It is known that the directional derivative of $\psi$  at $\mathbb{V}_P$ is $\psi^{\prime}_{P} : C_b(\Theta) \rightarrow \mathbb{R}$ as $\psi^{\prime}_{P}(h) = \sup_{\theta \in \Theta^\star_P} h(\theta)$. 

We proceed in three steps. In the first step, we establish uniform weak convergence of the empirical process
$\sqrt{n} (\widehat{\mathbb{V}}_n - \mathbb{V})$ to a Gaussian process in Lemma F.1 in Section F of the Online Supplement; in the second step,  we apply the delta method to the supremum functional, validated by Lemma F.2: 
$
\sqrt{n} \left[ \sup_{\theta \in \Theta }  \widehat{ \mathbb{V}} _{n}   (\theta) -  \sup_{\theta \in \Theta } \mathbb{V} (\theta) \right]
$
to derive its limiting distribution in \cref{theorem: delta method for optimal value}; finally we estimate the limiting distribution by the numerical delta method introduced by \cite{hong2018numerical}, see Lemma F.3 and Lemma F.4 in Section F.

\begin{theorem}\label{theorem: delta method for optimal value}
Suppose $\mathrm{VC}(\Pi)< \infty$ and Assumption \ref{assumption: Estimation of Nuisance Functions} holds. Then 
\[
\sqrt{n} \left[\psi  \big(\widehat{\mathbb{V}}_n\big) -\psi \left( \mathbb{V}_P \right) \right]  \rightsquigarrow  \psi_P^{\prime}(\mathbb{G}_P) = \sup_{ \theta \in \Theta^\star_P  } \mathbb{G}_P(\theta) ,
\]
  where $\mathbb{G}_P: \theta \mapsto \mathbb{G}_P g_\theta$ is a mean-zero tight Gaussian process on $\ell^{\infty} (\Theta)$ with covariance function 
\[
\mathrm{Cov}_{P}(\theta_1, \theta_2) = \mathbb{E} \left[ \mathbb{G}_P(\theta_1) \mathbb{G}_P(\theta_2) \right].
\]
Moreover, the paths $\theta \mapsto \mathbb{G}_P(\theta)$ are a.s. uniformly continuous on $(\Theta, d_\Theta)$, satisfying the following conditions:
\[
\sup_{P \in \mathcal{P}_n}  \mathbb{E}_P \left[ \sup_{\theta \in \Theta} | \mathbb{G}_{P} |  \right ] < \infty \quad  \text{and} \quad \lim_{\delta \searrow 0} \sup_{P \in \mathcal{P}_n} \mathbb{E}_P \left[ \sup_{d_{\Theta}  (\theta, \bar{\theta} ) \leq \delta } \left|   \mathbb{G}_{P} (\theta) -  \mathbb{G}_{P}( \bar{\theta} ) \right| \right]= 0. 
\]
\end{theorem}

When the maximizer of $\mathbb{V}_P$ is unique, i.e., $\Theta^\star_P = \{ \theta_o \}$ is a singleton, \cref{theorem: delta method for optimal value} implies that $\sqrt{n} (\psi  \big(\widehat{\mathbb{V}}_n\big) -\psi \left( \mathbb{V}_P \right))$ weakly converges to the normal distribution defined in \cref{Theorem: approximation of DML M_n}. When $\Theta^\star_P$ is not a singleton, $\sqrt{n}(\psi  \big(\widehat{\mathbb{V}}_n\big) -\psi \left( \mathbb{V}_P \right) )$ no longer converges weakly to normal distribution.
Although \cref{theorem: delta method for optimal value} establishes the asymptotic distribution of the estimator for the optimal welfare, conducting valid inference still requires information on the distribution of $\mathbb{G}_P$ and the directional derivative $\psi_P^\prime$. We utilize the bootstrap approach to approximate the distribution of $\mathbb{G}_P$. In particular, we consider the multiplier bootstrap $ \widehat{\mathbb{G}}_{n}^*: \Theta \rightarrow \mathbb{R}$ defined as 
\begin{equation*}
\widehat{\mathbb{G}}_{n}^*: \theta \mapsto  n^{-1/2} \sum_{i=1}^n  \xi_i \left [ \widehat{g}_{\theta}(Z_i) - \widehat{\mathbb{V}}_n (\theta) \right ],    
\end{equation*}
where $\{\xi_i\}_{i=1}^n$ are i.i.d. random variables independent of $(Z_i)_{i=1}^n$, with $\mathbb{E}( \xi_i) = 0$, $\mathbb{E}(\xi_i^2) =1$ and $\mathbb{E}\left[ \exp |\xi_i| \right] < \infty$. 
We apply the numerical delta method proposed by \cite{hong2018numerical} to estimate the directional derivative $\psi_P^\prime(\mathbb{G}_P)$.\footnote{
Alternative estimators of $\psi'_P(\mathbb G_P)$ are available;
see, e.g., \cite{firpo2023uniform}.
}  This is justified by Theorem 3.1 in \cite{hong2018numerical} or Lemma F.2 in Section F of the Online Supplement.
For given $\epsilon_n = o(1)$ with $n^{1/2} \epsilon_n \rightarrow \infty$, we estimate  $\psi_P^\prime( \mathbb{G}_P )$  using the distribution of the random variable:
\begin{equation}
\begin{aligned}
\widehat{\psi}_n^{\prime} ( \widehat{\mathbb{G}}_n^* )  &= \frac{ \psi \big(\widehat{\mathbb{V}}_n + \epsilon_n  \widehat{\mathbb{G}}_n^* \big )  - \psi(\widehat{\mathbb{V}}_n )    }{  \epsilon_n }.
\end{aligned}
\end{equation}

\bibliographystyle{apalike}
\bibliography{reference.bib} 

@article{wang2024robust,
  title={Robust Network Targeting with Multiple Nash Equilibria},
  author={Wang, Guanyi},
  journal={arXiv preprint arXiv:2410.20860},
  year={2024}
}

@article{kallus2021minimax,
  title={Minimax-optimal policy learning under unobserved confounding},
  author={Kallus, Nathan and Zhou, Angela},
  journal={Management Science},
  volume={67},
  number={5},
  pages={2870--2890},
  year={2021},
  publisher={INFORMS}
}

@article{kido2022distributionally,
  title={Distributionally robust policy learning with wasserstein distance},
  author={Kido, Daido},
  journal={arXiv preprint arXiv:2205.04637},
  year={2022}
}

@article{si2023distributionally,
  title={Distributionally robust batch contextual bandits},
  author={Si, Nian and Zhang, Fan and Zhou, Zhengyuan and Blanchet, Jose},
  journal={Management Science},
  volume={69},
  number={10},
  pages={5772--5793},
  year={2023},
  publisher={INFORMS}
}

@article{van2011local,
  title={A local maximal inequality under uniform entropy},
  author={van der Vaart, Aad W and Wellner, Jon A},
  journal={Electronic Journal of Statistics},
  volume={5},
  number={2011},
  pages={192},
  year={2011},
  publisher={NIH Public Access}
}

@article{chernozhukov2014gaussian,
  title={GAUSSIAN APPROXIMATION OF SUPREMA OF EMPIRICAL PROCESSES},
  author={Chernozhukov, Victor and Chetverikov, Denis and Kato, Kengo},
  journal={The Annals of Statistics},
  pages={1564--1597},
   volume = {42},
 year = {2014}
}

@article{geman1984stochastic,
  title={Stochastic Relaxation, Gibbs Distributions, and the Bayesian Restoration of Images},
  author={Geman, Stuart and Geman, Donald},
  journal={IEEE Transactions on Pattern Analysis and Machine Intelligence},
  volume={6},
  number={6},
  pages={721--741},
  year={1984},
  publisher={IEEE Computer Society Washington, DC, USA}
}

@article{kennedy2016semiparametric,
  title={Semiparametric Theory and Empirical Processes in Causal Inference},
  author={Kennedy, Edward H},
  journal={Statistical Causal Inferences and Their Applications in Public Health Research},
  pages={141},
  year={2016},
  publisher={Springer}
}

@article{fang2019inference,
  title={Inference on directionally differentiable functions},
  author={Fang, Zheng and Santos, Andres},
  journal={The Review of Economic Studies},
  volume={86},
  number={1},
  pages={377--412},
  year={2019},
  publisher={Oxford University Press}
}

@article{hong2018numerical,
  title={The numerical delta method},
  author={Hong, Han and Li, Jessie},
  journal={Journal of Econometrics},
  volume={206},
  number={2},
  pages={379--394},
  year={2018},
  publisher={Elsevier}
}

@article{belloni2017program,
  title={Program evaluation and causal inference with high-dimensional data},
  author={Belloni, Alexandre and Chernozhukov, Victor and Fernandez-Val, Ivan and Hansen, Christian},
  journal={Econometrica},
  volume={85},
  number={1},
  pages={233--298},
  year={2017},
  publisher={Wiley Online Library}
}

@book{gine2021mathematical,
  title={Mathematical Foundations of Infinite-Dimensional Statistical Models},
  author={Gin{\'e}, Evarist and Nickl, Richard},
  year={2021},
  publisher={Cambridge university press}
}

@article{wang2018quantile,
  title={Quantile-optimal treatment regimes},
  author={Wang, Lan and Zhou, Yu and Song, Rui and Sherwood, Ben},
  journal={Journal of the American Statistical Association},
  volume={113},
  number={523},
  pages={1243--1254},
  year={2018},
  publisher={Taylor \& Francis}
}

@article{shi2018massive,
  title={A massive data framework for M-estimators with cubic-rate},
  author={Shi, Chengchun and Lu, Wenbin and Song, Rui},
  journal={Journal of the American Statistical Association},
  volume={113},
  number={524},
  pages={1698--1709},
  year={2018},
  publisher={Taylor \& Francis}
}

@article{zhao2023semiparametric,
  title={A Semiparametric Instrumented Difference-in-Differences Approach to Policy Learning},
  author={Zhao, Pan and Cui, Yifan},
  journal={arXiv preprint arXiv:2310.09545},
  year={2023}
}

@book{kosorok2008introduction,
  title={Introduction to Empirical Processes and Semiparametric Inference},
  author={Kosorok, Michael R},
  year={2008},
  publisher={Springer}
}

@book{shapiro2021lectures,
  title={Lectures on Stochastic Programming: Modeling and Theory},
  author={Shapiro, Alexander and Dentcheva, Darinka and Ruszczynski, Andrzej},
  year={2021},
  publisher={SIAM}
}

@article{chernozhukov2022locally,
  title={Locally robust semiparametric estimation},
  author={Chernozhukov, Victor and Escanciano, Juan Carlos and Ichimura, Hidehiko and Newey, Whitney K and Robins, James M},
  journal={Econometrica},
  volume={90},
  number={4},
  pages={1501--1535},
  year={2022},
  publisher={Wiley Online Library}
}

@article{duchi2023distributionally,
  title={Distributionally robust losses for latent covariate mixtures},
  author={Duchi, John and Hashimoto, Tatsunori and Namkoong, Hongseok},
  journal={Operations Research},
  volume={71},
  number={2},
  pages={649--664},
  year={2023},
  publisher={INFORMS}
}

@article{luedtke2020performance,
  title={Performance guarantees for policy learning},
  author={Luedtke, Alex and Chambaz, Antoine},
  journal={Annales de L'Institut Henri Poincare Section (B) Probability and Statistics},
  volume={56},
  number={3},
  pages={2162--2188},
  year={2020}
}

@article{viviano2024fair,
  title={Fair policy targeting},
  author={Viviano, Davide and Bradic, Jelena},
  journal={Journal of the American Statistical Association},
  volume={119},
  number={545},
  pages={730--743},
  year={2024},
}

@article{lei2023policy,
  title={Policy Learning under Biased Sample Selection},
  author={Lei, Lihua and Sahoo, Roshni and Wager, Stefan},
  journal={arXiv preprint arXiv:2304.11735},
  year={2023}
}

@article{rai2018statistical,
  title={Statistical inference for treatment assignment policies},
  author={Rai, Yoshiyasu},
  journal={Unpublished Manuscript},
  year={2018}
}

@article{fan2023quantifying,
  title={Quantifying Distributional Model Risk in Marginal Problems via Optimal Transport},
  author={Fan, Yanqin and Park, Hyeonseok and Xu, Gaoqian},
  journal={arXiv preprint arXiv:2307.00779},
  year={2023}
}

@article{adjaho2022externally,
  title={Externally valid treatment choice},
  author={Adjaho, Christopher and Christensen, Timothy},
  journal={arXiv preprint arXiv:2205.05561},
  volume={1},
  year={2022}
}

@article{luedtke2016statistical,
  title={Statistical inference for the mean outcome under a possibly non-unique optimal treatment strategy},
  author={Luedtke, Alexander R and van der Laan, Mark J},
  journal={The Annals of Statistics},
  volume={44},
  number={2},
  pages={713--742},
  year={2016}
}

@article{luedtke2018parametric,
  title={Parametric-rate inference for one-sided differentiable parameters},
  author={Luedtke, Alexander R and van der Laan, Mark J},
  journal={Journal of the American Statistical Association},
  volume={113},
  number={522},
  pages={780--788},
  year={2018},
  publisher={Taylor \& Francis}
}

@article{qi2023robustness,
  title={On robustness of individualized decision rules},
  author={Qi, Zhengling and Pang, Jong-Shi and Liu, Yufeng},
  journal={Journal of the American Statistical Association},
  volume={118},
  number={543},
  pages={2143--2157},
  year={2023},
  publisher={Taylor \& Francis}
}

@article{kohler2021rate,
  title={On the rate of convergence of fully connected deep neural network regression estimates},
  author={Kohler, Michael and Langer, Sophie},
  journal={The Annals of Statistics},
  volume={49},
  number={4},
  pages={2231--2249},
  year={2021},
  publisher={Institute of Mathematical Statistics}
}

@article{shi2020breaking,
  title={Breaking the Curse of Nonregularity with Subagging---Inference of the Mean Outcome under Optimal Treatment Regimes},
  author={Shi, Chengchun and Lu, Wenbin and Song, Rui},
  journal={Journal of Machine Learning Research},
  volume={21},
  number={176},
  pages={1--67},
  year={2020}
}

@article{belloni2015some,
  title={Some new asymptotic theory for least squares series: Pointwise and uniform results},
  author={Belloni, Alexandre and Chernozhukov, Victor and Chetverikov, Denis and Kato, Kengo},
  journal={Journal of Econometrics},
  volume={186},
  number={2},
  pages={345--366},
  year={2015},
  publisher={Elsevier}
}

@article{chen2015optimal,
  title={Optimal uniform convergence rates and asymptotic normality for series estimators under weak dependence and weak conditions},
  author={Chen, Xiaohong and Christensen, Timothy M},
  journal={Journal of Econometrics},
  volume={188},
  number={2},
  pages={447--465},
  year={2015},
  publisher={Elsevier}
}

@article{bertsimas2017optimal,
  title={Optimal classification trees},
  author={Bertsimas, Dimitris and Dunn, Jack},
  journal={Machine Learning},
  volume={106},
  pages={1039--1082},
  year={2017},
  publisher={Springer}
}

@article{chernozhukov2018double,
  title={Double/debiased machine learning for treatment and structural parameters},
  author={Chernozhukov, Victor and Chetverikov, Denis and Demirer, Mert and Duflo, Esther and Hansen, Christian and Newey, Whitney and Robins, James},
  journal={The Econometrics Journal},
  volume={21},
  number={1},
  pages={C1--C68},
  year={2018},
  publisher={Oxford University Press (OUP)}
}

@article{kim2023fair,
	author = {Kim, Kwangho and Zubizarreta, Jos{\'e}R},
	journal = {arXiv preprint arXiv:2306.03625},
	title = {Fair and Robust Estimation of Heterogeneous Treatment Effects for Policy Learning},
	year = {2023}}

@article{kitagawa2021equality,
  title={Equality-minded treatment choice},
  author={Kitagawa, Toru and Tetenov, Aleksey},
  journal={Journal of Business \& Economic Statistics},
  volume={39},
  number={2},
  pages={561--574},
  year={2021},
  publisher={Taylor \& Francis}
}

@book{scholkopf2002learning,
  title={Learning with Kernels: Support Vector Machines, Regularization, Optimization, and Beyond},
  author={Sch{\"o}lkopf, Bernhard and Smola, Alexander J},
  year={2002},
  publisher={MIT press}
}

@article{athey2021policy,
  title={Policy learning with observational data},
  author={Athey, Susan and Wager, Stefan},
  journal={Econometrica},
  volume={89},
  number={1},
  pages={133--161},
  year={2021}
}

@article{firpo2023uniform,
  title={Uniform inference for value functions},
  author={Firpo, Sergio and Galvao, Antonio F and Parker, Thomas},
  journal={Journal of Econometrics},
  volume={235},
  number={2},
  pages={1680--1699},
  year={2023},
  publisher={Elsevier}
}

@article{tsybakov2004optimal,
  title={Optimal aggregation of classifiers in statistical learning},
  author={Tsybakov, Alexander B},
  journal={The Annals of Statistics},
  volume={32},
  number={1},
  pages={135--166},
  year={2004},
  publisher={Institute of Mathematical Statistics}
}

@article{robins1994estimation,
  title={Estimation of regression coefficients when some regressors are not always observed},
  author={Robins, James M and Rotnitzky, Andrea and Zhao, Lue Ping},
  journal={Journal of the American statistical Association},
  volume={89},
  number={427},
  pages={846--866},
  year={1994},
  publisher={Taylor \& Francis}
}

@book{vaart1998empirical,
  title={Weak Convergence and Empirical Processes: {With} Applications to Statistics},
  author={van der Vaart, Aad W and Wellner, Jon A},
  pages={127--384},
  year={1996},
  publisher={Springer}
}

@book{van2000asymptotic,
  title={Asymptotic Statistics},
  author={van der Vaart, Aad W},
  year={2000},
  publisher={Cambridge University Press}
}

@article{newey1994asymptotic,
  title={The Asymptotic Variance of Semiparametric Estimators},
  author={Newey, Whitney},
  journal={Econometrica},
  volume={62},
  number={6},
  pages={1349--82},
  year={1994},
  publisher={Econometric Society}
}

@article{rubin1990comment,
  title={Comment: Neyman (1923) and causal inference in experiments and observational studies},
  author={Rubin, Donald B},
  journal={Statistical Science},
  volume={5},
  number={4},
  pages={472--480},
  year={1990}
}

@article{fang2023fairness,
  title={Fairness-oriented learning for optimal individualized treatment rules},
  author={Fang, Ethan X and Wang, Zhaoran and Wang, Lan},
  journal={Journal of the American Statistical Association},
  volume={118},
  number={543},
  pages={1733--1746},
  year={2023},
  publisher={Taylor \& Francis}
}

@article{bloom1997benefits,
  title={The benefits and costs of JTPA Title II-A programs: Key findings from the national Job Training Partnership Act study},
  author={Bloom, Howard S and Orr, Larry L and Bell, Stephen H and Cave, George and Doolittle, Fred and Lin, Winston and Bos, Johannes M},
  journal={Journal of Human Resources},
  volume={32},
  number={3},
  pages={549--576},
  year={1997},
  publisher={University of Wisconsin Press}
}

@article{athey2024using,
  title={Using wasserstein generative adversarial networks for the design of monte carlo simulations},
  author={Athey, Susan and Imbens, Guido W and Metzger, Jonas and Munro, Evan},
  journal={Journal of Econometrics},
  volume={240},
  number={2},
  pages={105076},
  year={2024},
  publisher={Elsevier}
}

@book{41549aa6-42b1-36d1-ab6a-84fdd10b1f93,
 ISBN = {9780674005105},
 URL = {http://www.jstor.org/stable/j.ctv31xf5v0},
 author = {John Rawls},
 publisher = {Harvard University Press},
 title = {Justice as Fairness: A Restatement},
 urldate = {2024-02-17},
 year = {2001}
}

@article{greselin2018classical,
  title={From the classical Gini index of income inequality to a new Zenga-type relative measure of risk: A modeller’s perspective},
  author={Greselin, Francesca and Zitikis, Ri{\v{c}}ardas},
  journal={Econometrics},
  volume={6},
  number={1},
  pages={4},
  year={2018},
  publisher={MDPI}
}

@article{schmidt2020nonparametric,
  title={Nonparametric regression using deep neural networks with ReLU activation function},
  author={Schmidt-Hieber, Johannes},
  journal={The Annals of Statistics},
  volume={48},
  number={4},
  pages={1875},
  year={2020},
  publisher={Institute of Mathematical Statistics}
}

@article{rubin1978bayesian,
  title={Bayesian Inference for Causal Effects: The Role of Randomization},
  author={Rubin, Donald B},
  journal={The Annals of statistics},
  volume={6},
  number={1},
  pages={34--58},
  year={1978}
}

@article{qian2011performance,
author = {Min Qian and Susan A. Murphy},
title = {{Performance guarantees for individualized treatment rules}},
volume = {39},
journal = {The Annals of Statistics},
number = {2},
publisher = {Institute of Mathematical Statistics},
pages = {1180-1210},
year = {2011},
doi = {10.1214/10-AOS864},
URL = {https://doi.org/10.1214/10-AOS864}
}

@article{rockafellar2000optimization,
  title={Optimization of conditional value-at-risk},
  author={Rockafellar, R Tyrrell and Uryasev, Stanislav and others},
  journal={Journal of risk},
  volume={2},
  pages={21--42},
  year={2000},
  publisher={Citeseer}
}

@article{zhou2023offline,
  title={Offline multi-action policy learning: Generalization and optimization},
  author={Zhou, Zhengyuan and Athey, Susan and Wager, Stefan},
  journal={Operations Research},
  volume={71},
  number={1},
  pages={148--183},
  year={2023},
  publisher={INFORMS}
}

@inproceedings{kallus2018confounding,
  title={Confounding-robust policy improvement},
  author={Kallus, Nathan and Zhou, Angela},
  booktitle={Proceedings of the 32nd International Conference on Neural Information Processing Systems},
  pages={9289--9299},
  year={2018}
}

@article{zhang2012estimating,
  title={Estimating optimal treatment regimes from a classification perspective},
  author={Zhang, Baqun and Tsiatis, Anastasios A and Davidian, Marie and Zhang, Min and Laber, Eric},
  journal={Stat},
  volume={1},
  number={1},
  pages={103--114},
  year={2012},
  publisher={Wiley Online Library}
}

@article{bhattacharya2012inferring,
  title={Inferring welfare maximizing treatment assignment under budget constraints},
  author={Bhattacharya, Debopam and Dupas, Pascaline},
  journal={Journal of Econometrics},
  volume={167},
  number={1},
  pages={168--196},
  year={2012},
  publisher={Elsevier}
}

@article{zhao2012estimating,
  title={Estimating individualized treatment rules using outcome weighted learning},
  author={Zhao, Yingqi and Zeng, Donglin and Rush, A John and Kosorok, Michael R},
  journal={Journal of the American Statistical Association},
  volume={107},
  number={499},
  pages={1106--1118},
  year={2012},
  publisher={Taylor \& Francis}
}

@article{shorrocks1983ranking,
  title={Ranking income distributions},
  author={Shorrocks, Anthony F},
  journal={Economica},
  volume={50},
  number={197},
  pages={3--17},
  year={1983},
  publisher={JSTOR}
}

@article{robins1995analysis,
  title={Analysis of semiparametric regression models for repeated outcomes in the presence of missing data},
  author={Robins, James M and Rotnitzky, Andrea and Zhao, Lue Ping},
  journal={Journal of the American Statistical Association},
  volume={90},
  number={429},
  pages={106--121},
  year={1995}
}

@article{rockafellar2002deviation,
  title={Deviation measures in risk analysis and optimization},
  author={Rockafellar, R Tyrrell and Uryasev, Stanislav P and Zabarankin, Michael},
  journal={University of Florida, Department of Industrial \& Systems Engineering Working Paper},
  number={2002-7},
  year={2002}
}

@article{athey2019generalized,
  title={GENERALIZED RANDOM FORESTS},
  author={Athey, Susan and Tibshirani, Julie and Wager, Stefan},
  journal={The Annals of Statistics},
  volume={47},
  number={2},
  pages={1148--1178},
  year={2019}
}

@article{kim1990cube,
  title={Cube Root Asymptotics},
  author={Kim, Jeankyung and Pollard, David},
  journal={The Annals of Statistics},
  volume={18},
  number={1},
  pages={191--219},
  year={1990}
}

@article{kitagawa2018a,
  title = {Who should be treated? Empirical welfare maximization methods for treatment choice},
  author = {Kitagawa, Toru and Tetenov, Aleksey},
  journal = {Econometrica},
  volume = {86},
  number = {2},
  pages = {591--616},
  year = {2018},
  publisher = {Wiley Online Library},
  note = {Supplementary materials and online appendix available at \url{https://doi.org/10.3982/ECTA13288}}
}

@article{fan2025policy_supplement,
  title={{Supplement} to ``{Policy} Learning with {$\alpha$}-Expected Welfare''},
  author={Fan, Yanqin and Qi, Yuan and Xu, Gaoqian},
  journal={arXiv preprint arXiv:2505.00256},
  year={2025}
}

@article{kirkpatrick1983optimization,
  title={Optimization by simulated annealing},
  author={Kirkpatrick, Scott and Gelatt Jr, C Daniel and Vecchi, Mario P},
  journal={Science},
  volume={220},
  number={4598},
  pages={671--680},
  year={1983},
  publisher={American association for the advancement of science}
}

@article{husmann2017r,
  title={The R package optimization: Flexible global optimization with simulated-annealing},
  author={Husmann, Kai and Lange, Alexander and Spiegel, Elmar},
  journal={CRAN citation},
  year={2017}
}

@inproceedings{kallus2018balanced,
  title={Balanced policy evaluation and learning},
  author={Kallus, Nathan},
  booktitle={Proceedings of the 32nd International Conference on Neural Information Processing Systems},
  pages={8909--8920},
  year={2018}
}

@article{gine2002rates,
  title={Rates of strong uniform consistency for multivariate kernel density estimators},
  author={Gin{\'e}, Evarist and Guillou, Armelle},
  journal={Annales de l'Institut Henri Poincare (B) Probability and Statistics},
  volume={38},
  number={6},
  pages={907--921},
  year={2002},
  organization={Elsevier}
}
\end{document}